\documentclass[a4paper,11pt]{article}
\pdfoutput=1 

\usepackage{jheppub} 

\usepackage[T1]{fontenc} 

\preprint{KIAS-P17001}

\title{Fundamental Vortices, Wall-Crossing, and Particle-Vortex Duality}

\author[a]{Chiung Hwang,}
\author[a]{Piljin Yi}
\author[b]{and Yutaka Yoshida}

\affiliation[a]{School of Physics, Korea Institute for Advanced Study, \\
Seoul 02455, Korea}
\affiliation[b]{Research Institute for Mathematical Sciences, Kyoto University, \\
Kyoto 606-8502, Japan}

\emailAdd{chwang@kias.re.kr}
\emailAdd{piljin@kias.re.kr}
\emailAdd{yyoshida@kurims.kyoto-u.ac.jp}

\abstract{We explore 1d vortex dynamics of 3d supersymmetric Yang-Mills
theories, as inferred from factorization of exact
partition functions. Under Seiberg-like dualities, the 3d
partition function must remain invariant, yet it is not a priori
clear what should happen to the vortex dynamics. We observe
that the 1d quivers for the vortices remain the same, and the
net effect of the 3d duality map manifests as 1d Wall-Crossing
phenomenon; Although the vortex number can shift along such
duality maps, the ranks of the 1d quiver theory are unaffected,
leading to a notion of fundamental vortices as basic building
blocks for topological sectors. For Aharony-type duality, in particular,
where one must supply extra chiral fields to couple with monopole
operators on the dual side, 1d wall-crossings of an infinite
number of vortex quiver theories are neatly and collectively
encoded by 3d determinant of such extra chiral fields. As such,
1d wall-crossing of the vortex theory encodes the particle-vortex
duality embedded in the 3d Seiberg-like duality. For
$\mathcal N = 4$, the D-brane picture is used to motivate this
3d/1d connection, while, for $\mathcal N = 2$, this 3d/1d
connection is used to fine-tune otherwise ambiguous vortex
dynamics. We also prove some identities of 3d supersymmetric partition functions for the Aharony duality using this vortex wall-crossing interpretation.}

\begin{document}
\maketitle
\flushbottom

\section{Introduction}

In recent years, a plethora of exact partition functions became
available for supersymmetric gauge theories. The localization method,
responsible for these, is powerful and universal but such
universality comes with costs. Much of the dynamics is lost,
as the end result depends only on handful of UV information,
such as field contents and their representation under the
gauge and the global symmetries. This should be hardly surprising.
When the spacetime that admits a circle, for example, the
supersymmetric partition function can be regarded as a refined
index, well-known to be robust under continuous deformations.
Despite this ultraviolet nature of the computation, these
partition functions proved to quite useful, for example as a
litmus test for various dualities.
For dimensions less than three also, where there is no notion of
vacuum expectation value of moduli, a UV theory often flows
down to a unique theory in IR. As such, the partition functions
in such low dimensions contain more useful information
than one may generally hope for. The Gromov-Witten invariants
cleverly embedded \cite{Jockers:2012dk,Gomis:2012wy}
in the $S^2$ partition functions \cite{Benini:2012ui,Doroud:2012xw}
of $d=2$ GLSM are the prime example of this, while the
$d=2$ elliptic genus \cite{Benini:2013xpa} and $d=1$  refined
Witten index \cite{Hori:2014tda} are more obvious ones.

A trick of convenience involved in the localization
computations, which lifts flat directions as much as possible,
is to introduce chemical potentials and other susy-preserving
masses. For $d=3$ theories, in particular, one
turns on real masses associated with flavor symmetries, which
generically simplify the vacuum structures to those of isolated
ones. Note that not all theories
admit such computations. When the theory must involve
superpotentials, for example, the number of the available
flavor symmetries get reduced. Also, given a reduced flavor
symmetry that allows some superpotential, the computation will
tend to compute the partition function for theories with
generic superpotential consistent with the flavor charge
assignment. The usual mantra that the localization is
insensitive to the details of the superpotential must be
taken with such genericity presumed. In this note, we will
be considering theories where all matter fields acquire
real masses, independent of one another, which means that
superpotential is turned off by imposing global symmetries.

When the number of matter multiplets and accompanying flavor
symmetry are sufficiently large, quantum vacua then tend to
be isolated \cite{Intriligator:2013lca}. One type, which we
refer to as the Higgs vacua, is such that chiral fields are
turned on to cancel FI constants with the Coulombic vev's
pinned at some of the real masses. Another type, more
characteristic of $d=3$ and called the topological vacua
by Intriligator and Seiberg,
achieves the vacuum condition entirely along the Coulomb branch
with all chiral fields turned off. Although the total number
of vacua is invariant under continuous deformations of
the theory, this split of vacua between the Higgs type and
the topological type is not robust, and in particular affected
by signs of 3d FI constants. One interesting class of theories
is where one can tune the 3d FI constant so that all vacua are of Higgs
variety. In such theories, the exact partition functions
on $S^1$ fibred over $S^2$ are known to admit the so-called
factorization \cite{Krattenthaler:2011da,Dimofte:2011ju,Pasquetti:2011fj,Beem:2012mb,Hwang:2012jh,Taki:2013opa,Cecotti:2013mba,Fujitsuka:2013fga,Benini:2013yva,Benini:2015noa} where the partition function can be rewritten
as a sum over product of three multiplicative pieces: vortex
contributions at the north pole, anti-vortex contributions
at the south pole, and the perturbative 1-loop from fields
not involved in the vortex construction.

When a partition function is thus factorizable, one has
a good glimpse into the supersymmetric vortices. In an isolated
Higgs vacuum, one chiral field gets a vev on top of
each Cartan vev $\sigma_a$ pinned at a real mass $m_i$, and
form ${\mathbb P}^0$ after the gauge identification. Each
such chiral field can acquire the winding number and the associated
quantized magnetic flux $2\pi n_a$. Due to the chemical potential
associated with $R+2 J_3$ where $R$ is an R-symmetry generator and $J_3$ a
rotation generator on $S^2$, the (anti-)vortices are pushed into
the north(south)-pole, much like the $\Omega$-deformation on
${\mathbb R}^4$ pushing instantons to the origin \cite{Nekrasov:2002qd}.

As such, contributions of vortices and of anti-vortices
to the partition function can be read off from the factorization,
and in turn one can ask what low energy dynamics of vortices
is capable of producing such contributions. This gives us
an indirect way to explore $d=1$ low energy dynamics of vortices,
from 3d exact partition functions. The latter is
reliably and universally computed via the Coulombic localization,
which  has no knowledge of the Higgs vacua or of vortices. Yet,
it can be used to extract vortex dynamics, once factorized.

The vortex theory found this way is typically a quiver quantum
mechanics with either ${\cal N}=4$ or ${\cal N}=2$ supercharges,
respectively, for 3d ${\cal N}=4$ or ${\cal N}=2$ gauge theories.
Since the partition function
itself carries limited information about the 3d field
theory, one should not really expect the exact low energy
dynamics of the vortices \cite{Hanany:2003hp};  Rather, we are
interested in more robust (hence more coarse) aspects of the
low energy dynamics, as much as can be encoded in the 1d twisted
partition functions of vortices. The main question we ask
is how the 3d Seiberg-like dualities manifest in the
1d vortex theory, and what else we learn from such
investigations.

Since Seiberg-like dualities change the UV data of 3d
theory entirely, it is not clear whether there is a simple
meaningful action of this duality on vortices themselves.
It does preserve the partition function, yet we are asking
about individual vortex sector contributions, an infinite number
of which must be combined to contribute to a single 3d
partition function. Also the rank of the 3d theory changes
and vortices are naturally associated with the Cartan
part of the gauge group, so the vortex number must change
upon duality, even if somehow the contributions remain
intact sector by sector. This seems a contradiction in
itself, as we usually interpret the vortex contribution
as coming from the low energy quantum mechanics of the
relevant topological sector, which is in turn closely
related to the Cartan $U(1)$'s of the gauge group.

A useful analogy can be learned from BPS monopoles in $d=4$ \cite{Weinberg:1982ev,Weinberg:2006rq}.
For a non-Abelian Yang-Mills broken to the Cartan by a single
adjoint Higgs field, there can be as many ``unit'' magnetic
monopole solutions as the number of positive roots; Along
each positive root, one can embed $SU(2)$ and a unit
magnetic monopole solution thereof. Upon a closer look, however,
one quickly realizes that most of such monopoles are
composite states of more than one ``fundamental'' monopoles \cite{Weinberg:1982ev}.
Recall that
the fundamental monopole makes sense when the Yang-Mills group is
broken to the Cartan, with help of real adjoint Higgs vev $h$.
The latter defines the positivity in the dual root space, which
in turn divides monopoles into BPS and anti-BPS. A BPS monopole,
associated with positive dual root $\alpha^*$, is then generally
written as
$$\alpha^*= \sum n_a \beta^*_a$$
with nonnegative integers $n_a$, which defines the fundamental
monopole charges $\beta^*_a$. It is easy to see that the
collection $\{\beta_a\}$ can be regarded as a collection of
simple roots, with the positivity defined by $h$, and mass of the
monopole is built additively from masses of these
fundamental monopoles. Such a monopole is separable into $\sum_a n_a$
distinct fundamental monopoles in real space.

Similarly, we will find that a notion of ``fundamental'' vortices
emerges quite naturally. For vortices in $\mathcal N=2$ theories,
the 3d FI constants $\xi$, regarded as a vector in the gauged
Cartan subalgebra, will play the analog of $h$ while the roles
of magnetic charges (such as $\alpha^*$ and $\beta_a^*$) are
played by the Chern numbers $q$ of the vortices. As with the
monopole analogy, the fundamental vortices are those with
``minimal'' positive masses $\xi\cdot q_i > 0$, so that general
positive vortices are constructed as
$$q= \sum k_i q_i$$
with nonnegative integer $k_i$'s. For 3d $\mathcal N=4$ theories,
for which $\xi$'s are naturally triplets, we must also address
why it makes sense to pick out one out of three possible
directions for $\xi$, which will be addressed later.

Then, what does happen to the theory of such fundamental vortices
when 3d duality is performed? Much like ADHM of
instantons, the effective quantum mechanics of vortices,
whenever known, is of quiver type. Although this
representation is not accurate enough in dynamical sense,
it appears to be good enough for some supersymmetric observables
such as the twisted partition function that enters the 3d
partition functions. Our finding shows the following:
the quiver theory for vortices remains invariant, quite
surprisingly, upon Seiberg-like duality. Not only the
quiver itself is invariant, so are the rank vectors.
Clearly, this 1d quiver theory describes
the dynamics of the fundamental vortices, and the invariance
implies that the notion of these fundamental vortices is
also robust under the 3d duality as long as we correctly
keep track of signs of 3d FI constants along the Seiberg-like
duality. The rule for the latter can be naturally inferred
from quiver mutation mechanism  along with their underlying
D-brane pictures.

Instead, the 1d vortex theory reacts to 3d duality by
flipping sign(s) of some 1d FI parameters $\zeta$.
Recall that in supersymmetric gauged quantum mechanics,
the sign flip of an FI parameter often causes a wall-crossing \cite{Denef:2002ru}.
At the level of computation via JK-residue sum, the sign flip
implies that the list of contributing residues changes.
Depending on details of the quiver theory, this may or may not
translate to different twisted partition functions in the end \cite{Hori:2014tda}.
When the result remains unchanged after the residue sum, our
assertion implies that not only the vortex quiver theory but
the twisted partition function contributing to 3d partition functions
remain exactly the same despite the Seiberg-duality.
Since the vortex numbers in 3d sense differ between dual pairs,
we will also identify a canonical 3d theory in a given duality
chain where the fundamental vortices are identified by
``unit'' Chern numbers and where the resulting vortex quiver
theory is most naturally identified.

When 1d twisted partition functions do change, signalling
nontrivial wall-crossing in the vortex quiver theory, a new
issue arises. Since such a wall-crossing tends to be universal
for all rank vectors, the discrepancies exist in an infinite number
of vortex sectors. On the other hand, 3d partition function
itself has to be invariant under the duality, so there has to be
something else that corrects this change of vortex contributions.
The answer to this is also simple and elegant: Such discrepancies
exist if and only if the Seiberg-like duality becomes an Aharony
type \cite{Aharony:1997gp}, where one must insert extra neutral
chiral multiplets on the dual side, usually coupled to monopole
operators via F-term. Furthermore, an infinite number of
discrepancies sum up exactly to the simple perturbative
contributions from such extra chiral multiplets
on the dual side, compensating the difference neatly.
In effect, whenever the 1d vortex theory undergoes
a wall-crossing, we are witnessing a particle-vortex duality
embedded in the 3d Seiberg-like duality.

On the other hand, we must  caution the readers against taking the
low energy theory of vortices too literally. What they keep
track of is topological information rather than dynamical one.
For instance there are cases where, even though the vortex quiver
theory looks nontrivial, it has no supersymmetric vacua and thus
offers no accompanying vortex contribution to the 3d partition
function. This commonly indicates that the vortex in question
cannot be quantized preserving supersymmetry, but could also
happen simply because one side of a 3d dual pair has no gauge group.
Clearly, we cannot really speak of vortex dynamics in any dynamical
sense on the side with no 3d gauge group, yet, amazingly, the
common 1d vortex quiver theory, naturally derived on the gauged
side, correctly keeps track of ``vortex'' contributions to both sides.
In the absence of 1d wall-crossing, this would tell us the gauged
side has no supersymmetric and quantum vortex even though such a
solution is possible classically; in the presence of 1d wall-crossing,
this would indicate that a complete vortex-particle duality has occurred. In a way, such versatility of these 1d fundamental
vortices is both puzzling and intriguing.

In this note, we study and explore such relations between
3d Seiberg-like dualities and 1d wall-crossings with
emphasis on concrete examples.
For $d=3$ ${\cal N}=4$ theories, the 3d/1d relation can be
recovered quite straightforwardly from Hanany-Witten D-brane
realizations which also tell us much about the 1d quiver
theory of vortices.  For 3d ${\cal N}=2$
theories, things are no longer so simple. Among various
complications are the rampant Chern-Simons terms and how they
modify the 1d quiver theory. Although the general answer
to this can be seen to be supersymmetry-preserving Wilson lines
in the vortex quantum mechanics, we find the details of how this
is realized are actually ambiguous if one cares only about
the final twisted partition functions, chamber by chamber.
In contrast, our main observation, which gives an unambiguous
rule for connecting different 1d chambers and thus related
3d Seiberg-dual theories, will be used to resolve such ambiguities.

In section \ref{sec:review}, we review some background materials for 3d supersymmetric gauge theories and 1d gauged linear sigma model.
In section \ref{sec:vortex QM}, we discuss the quantum mechanics descriptions for half-BPS vortices in 3d $\mathcal N = 4, 2$ linear quiver gauge theories and compute their refined Witten indices.
In section \ref{sec:SD}, we examine the connection between vortex quantum mechanics and 3d Seiberg-like dualities, in particular, for $\mathcal N = 2, 4$ SQCD-like theories with (anti-)fundamental matters and linear quiver theories with bi-fundamental matters.
Section \ref{sec:summary} is a summary of the note.
We also work out the factorization of the topologically twisted index on $S^2 \times S^1$ in Appendix \ref{sec:fact} and prove some identities of 3d supersymmetric partition functions for the Aharony duality in Appendix \ref{sec:identity}.
\\

\section{Review}
\label{sec:review}

\subsection{3d $\mathcal N = 2$ gauge theories and Seiberg-like dualities}
\label{sec:3d}

In this section we review some basic properties of 3d $\mathcal N = 2$ supersymmetric gauge theories, which will be relevant in the subsequent sections. As a concrete example, we consider a 3d $\mathcal N = 2$ theory with gauge group $U(N_c)$. The theory contains an $\mathcal N = 2$ vector multiplet, which consists of a gauge field $A$, a real scalar $\sigma$, a two-component Dirac fermion, called a gaugino, $\lambda$ and an auxiliary real scalar $D$ in the adjoint representation of the gauge group $U(N_c)$. One can also introduce a chiral multiplet, which consists of a complex scalar $\phi$, a two-component Dirac fermion $\psi$ and an auxiliary complex scalar $F$. We here consider $N_f$ chiral multiplets $Q_a^i$ in the fundamental representation and $N_a$ chiral multiplets $\tilde Q_i^{\tilde b}$ in the anti-fundamental representation where $a = 1,\ldots,N_f$, $\tilde b = 1,\ldots,N_a$ and $i = 1,\ldots,N_c$. A holomorphic function $W(Q,\tilde Q)$ of those chiral multiplets defines the superpotential, which yields interaction terms of the theory.

Another type of an interaction that a 3d $\mathcal N = 2$ theory has is the Chern-Simons interaction. For the gauge group $U(N_c)$, one can include the CS interaction of level $\kappa$:
\begin{align}
\mathcal L_\text{CS}^{\mathcal N = 2} = \frac{\kappa}{4 \pi} \mathrm{Tr} \left(\epsilon^{\mu \nu \rho} \left( A_\mu \partial_\nu A_\rho- \frac{2 i}{3} A_\mu A_\nu A_\rho \right)+2 D \sigma+\overline \lambda \lambda\right)
\end{align}
where $\kappa$ should satisfy $\kappa+\frac{N_f-N_a}{2} \in \mathbb Z$ due to the so-called parity anomaly. Moreover, one can also turn on the additional CS interaction for the $U(1)$ factor of the gauge group. This $U(1)$ CS interaction can be generalized such that one can consider a CS-like interaction between different $U(1)$ symmetries:
\begin{align}
\mathcal L_\text{BF}^{\mathcal N = 2} = \frac{1}{2 \pi} \left(\epsilon^{\mu \nu \rho} A^{(1)}_\mu \partial_\nu A^{(2)}_\rho+D^{(1)} \sigma^{(2)}+D^{(2)} \sigma^{(1)}+\frac{1}{2} \left(\overline \lambda^{(1)} \lambda^{(2)}+\overline \lambda^{(2)} \lambda^{(1)}\right)\right).
\end{align}
This is called a mixed CS interaction, or a BF interaction because it is a coupling between field strength $F$ of gauge field $A^{(1)}$ and another gauge field $B = A^{(2)}$. In particular there is a special kind of a BF interaction which corresponds to the Fayet-Iliopoulos term. In 3d, a $U(N_c)$ gauge theory has a global symmetry whose conserved current is defined as follow:
\begin{align}
\label{eq:tcurrent}
j = \frac{1}{2 \pi} * \mathrm{Tr} F.
\end{align}
This is usually called the topological symmetry, which we will denote by $U(1)_T$. If we consider the BF interaction between this $U(1)_T$ global symmetry and the $U(1)$ factor of the gauge symmetry, it gives rise to the following Lagrangian, which is the same as the FI term:
\begin{align}
\mathcal L_\text{FI}^{\mathcal N = 2} = \frac{1}{2 \pi} D \xi
\end{align}
where $\xi$ is the real scalar in the background vector multiplet for the $U(1)_T$ symmetry.

The theory has the $U(1)_R$ R-symmetry as well as other global symmetries $SU(N_f) \times SU(N_a) \times U(1)_A  \times U(1)_T$. The charges of the fundamental and anti-fundamental chiral multiplets $Q, \tilde Q$ are summarized in table \ref{tab:AGK}.
\begin{table}[tbp]
\centering
\begin{tabular}{|c|cccccc|}
\hline
 & $U(N_c)$ & $SU(N_f)$ & $SU(N_a)$ & $U(1)_A$ & $U(1)_T$ & $U(1)_R$ \\
\hline
$Q$ & $\mathbf{N_c}$ & $\mathbf{\overline{N_f}}$ & $\mathbf 1$ & $1$ & $0$ & $0$ \\
$\tilde Q$ & $\mathbf{\overline{N_c}}$ & $\mathbf 1$ & $\mathbf{N_a}$ & $1$ & $0$ & $0$ \\
\hline
$q$ & $\mathbf{N_c}$ & $\mathbf 1$ & $\mathbf{\overline{N_a}}$ & $-1$ & $0$ & $1$ \\
$\tilde q$ & $\mathbf{\overline{N_c}}$ & $\mathbf{N_f}$ & $\mathbf 1$ & $-1$ & $0$ & $1$ \\
$M$ & $\mathbf 1$ & $\mathbf{\overline{N_f}}$ & $\mathbf{N_a}$ & $2$ & $0$ & $0$ \\
$V_\pm$ & $\mathbf 1$ & $\mathbf 1$ & $\mathbf 1$ & $-\frac{N_f+N_a}{2}$ & $\pm 1$ & $\frac{N_f+N_a}{2}-N_c+1$ \\
\hline
\end{tabular}
\caption{\label{tab:AGK} The symmetry charges of the chiral multiplets we introduce. Note that $V_\pm$ exist only when $\kappa \pm \frac{N_f-N_a}{2} = 0$ respectively. The $U(1)_R$ charges are UV values we conventionally choose. Their IR superconformal values are given by combinations of the UV R-charge and other global $U(1)$ charges, which can be determined by $Z$-extremization \cite{Jafferis:2010un}.}
\end{table}
\\

Without the superpotential, the theory has supersymmetric vacua when $\textrm{max}\left(\kappa,\frac{|N_f-N_a|}{2}\right)+\frac{N_f+N_a}{2} \geq N_c$. The analysis of those supersymmetric vacua can be found in, e.g., \cite{Aharony:1997bx,Giveon:2008zn,Intriligator:2013lca}.\footnote{Also note that the algebraic structure of the vacuum moduli space, which is captured by the Hilbert series, is examined in \cite{Hanany:2015via,Cremonesi:2015dja,Cremonesi:2016nbo}.} Here we briefly summarize the analysis of \cite{Intriligator:2013lca} for a $U(1)$ theory.

After integrating out the auxiliary fields, one has the following semi-classical effective potential:
\begin{align}
V = \frac{e_\text{eff}^2}{32 \pi^2} \left(\sum_i 2 \pi n_i |Q_i|^2-\xi_\text{eff}-\kappa_\text{eff} \sigma\right)^2+\sum_i (m_i+n_i \sigma)^2 |Q_i|^2
\end{align}
where $Q_i$ is the scalar in the $i$-th chiral multiplet of charge $n_i$ and real mass $m_i$. $e_\text{eff}$ is the effective gauge coupling. The quantum corrected FI parameter and CS level, $\xi_\text{eff}$ and $\kappa_\text{eff}$, are given by
\begin{align}
\xi_\text{eff} &= \xi+\frac{1}{2} \sum_i n_i m_i ~ \textrm{sgn} (m_i+n_i \sigma), \\
\kappa_\text{eff} &= \kappa+\frac{1}{2} \sum_i n_i^2 ~ \textrm{sgn} (m_i+n_i \sigma).
\end{align}
The semi-classical vacua are given by the solutions of $V = 0$, or equivalently
\begin{gather}
F(\sigma) = \sum_i 2 \pi n_i |Q_i|^2, \label{eq:vac eq 1}\\
(m_i+n_i \sigma) Q_i = 0 \label{eq:vac eq 2}
\end{gather}
where we have defined
\begin{align}
F(\sigma) &\equiv \xi_\text{eff}+\kappa_\text{eff} \sigma \\
&= \xi+\kappa \sigma+\frac{1}{2} \sum_i n_i |m_i+n_i \sigma|. \label{eq:F}
\end{align}

Equations \eqref{eq:vac eq 1} and \eqref{eq:vac eq 2} allow three types of solutions: Higgs, Coulomb and topological vacua. A Higgs vacuum is a solution with the nonzero vacuum expectation value $\left<Q_i\right>$. Nonzero $\left<Q_i\right>$, from \eqref{eq:vac eq 2}, requires the vanishing effective real mass, $m_i+n_i \sigma = 0$. For generic real masses, the Higgs vacua are isolated while for special values of real masses, they can have a continuous moduli space called a Higgs branch.

A Coulomb vacuum is a solution with $\xi_\text{eff} = \kappa_\text{eff} = 0$, which implies that $F(\sigma) = \sum_i 2 \pi n_i |Q_i|^2 = 0$. There is a continuous moduli space of Coulomb vacua, which is parameterized by $\sigma$ in a range keeping $\xi_\text{eff} = \kappa_\text{eff} = 0$. This is called a Coulomb branch.

A topological vacuum is a solution with $F(\sigma) = \sum_i 2 \pi n_i |Q_i|^2 = 0$ but with nonzero $\xi_\text{eff}$ and $\kappa_\text{eff}$. Unlike the Coulomb vacuum, a topological vacuum is isolated, which reflects the fact that there is no massless degree of freedom at a topological vacuum. Indeed, the low-energy effective theory at a topological vacuum is the $\mathcal N = 2$ CS theory of level $\kappa_\text{eff}$. A classical topological vacuum obtained from $F(\sigma) = 0$ acquires the topological multiplicity $|\kappa_\text{eff}|$ \cite{Witten:1999ds,Intriligator:2013lca}.
\\

A 3d $\mathcal N = 2$ $U(N_c)$ theory is known to have a dual gauge description. More precisely, the 3d $\mathcal N = 2$ $U(N_c)_\kappa$ gauge theory with $N_f$ fundamental and $N_a$ anti-fundamental chiral multiplets flows to the same IR fixed point as the $U(N_c^D)_{-\kappa}$ theory with $N_a$ fundamental and $N_f$ anti-fundamental chiral multiplets (and additional gauge singlet chiral multiplets we explain shortly) flows to. The dual gauge rank, $N_c^D$, is determined as follows:
\begin{align}
\label{eq:dual rank}
N_c^D = \left\{\begin{array}{cc}
\mathrm{max}(N_f,N_a)-N_c, \qquad & |\kappa| \leq \frac{|N_f-N_a|}{2}, \\
|\kappa|+\frac{N_f+N_a}{2}-N_c, \qquad & |\kappa| > \frac{|N_f-N_a|}{2}. \\
\end{array}\right.
\end{align}
Originally the duality is proposed by Aharony for $\kappa = N_f-N_a = 0$ \cite{Aharony:1997gp} and by Giveon and Kutasov for $\kappa \neq N_f-N_a = 0$ \cite{Giveon:2008zn}. Later those are generalized to arbitrary $N_f$, $N_a$ and $\kappa$ \cite{Benini:2011mf}.

The dual theory contains additional gauge singlet chiral multiplets which couple to the gauge theory via the superpotential. There are $N_f N_a$ chiral multiplets $M_a^{\tilde b}$ coupling to the dual fundamental and anti-fundamental chiral multiplets as follows:
\begin{align}
W = M_a^{\tilde b} \tilde q^i_{\tilde b} q^a_i.
\end{align}
Those $M_a^{\tilde b}$ correspond to the meson operators in the original theory, $Q_a^i \tilde Q_i^{\tilde b}$, whose vacuum expectation values parameterize the Higgs branch of the moduli space. Due to the superpotential, the dual mesons $q^a_i q^i_{\tilde b}$ cannot have the vacuum expectation values.

In addition, there is another chiral multiplet $V_\pm$ if $\kappa = \mp\frac{N_f-N_a}{2}$ respectively. For each case, the dual theory has a gauge invariant monopole operator $v_\mp$, which couples to $V_\pm$ as follows:
\begin{align}
W = \delta_{2 \kappa,N_a-N_f} V_+ v_- + \delta_{2 \kappa,N_f-N_a} V_- v_+.
\end{align}
Due to this superpotential, $v_\pm$ cannot have the vacuum expectation values while the vacuum expectation values of $V_\pm$ parameterize the Coulomb branches of the moduli space. The charges of those extra chiral multiplets are summarized in table \ref{tab:AGK}.

Note that the duality patterns for $N_f \neq N_a$ are distinguished into two classes: $|\kappa| < \frac{|N_f-N_a|}{2}$ and $|\kappa| > \frac{|N_f-N_a|}{2}$. In \cite{Benini:2011mf} the former is called maximally chiral, whose duality pattern resembles the Aharony duality, while the latter is called minimally chiral, whose duality pattern resembles the Giveon-Kutasov duality. In this note, for brevity, we call the Seiberg-like dualities for $|\kappa| \leq \frac{|N_f-N_a|}{2}$ Aharony dualities and the dualities for $|\kappa| > \frac{|N_f-N_a|}{2}$ Giveon-Kutasov dualities.
\\

\begin{figure}[tbp]
\centering 
\includegraphics[height=.3\textheight]{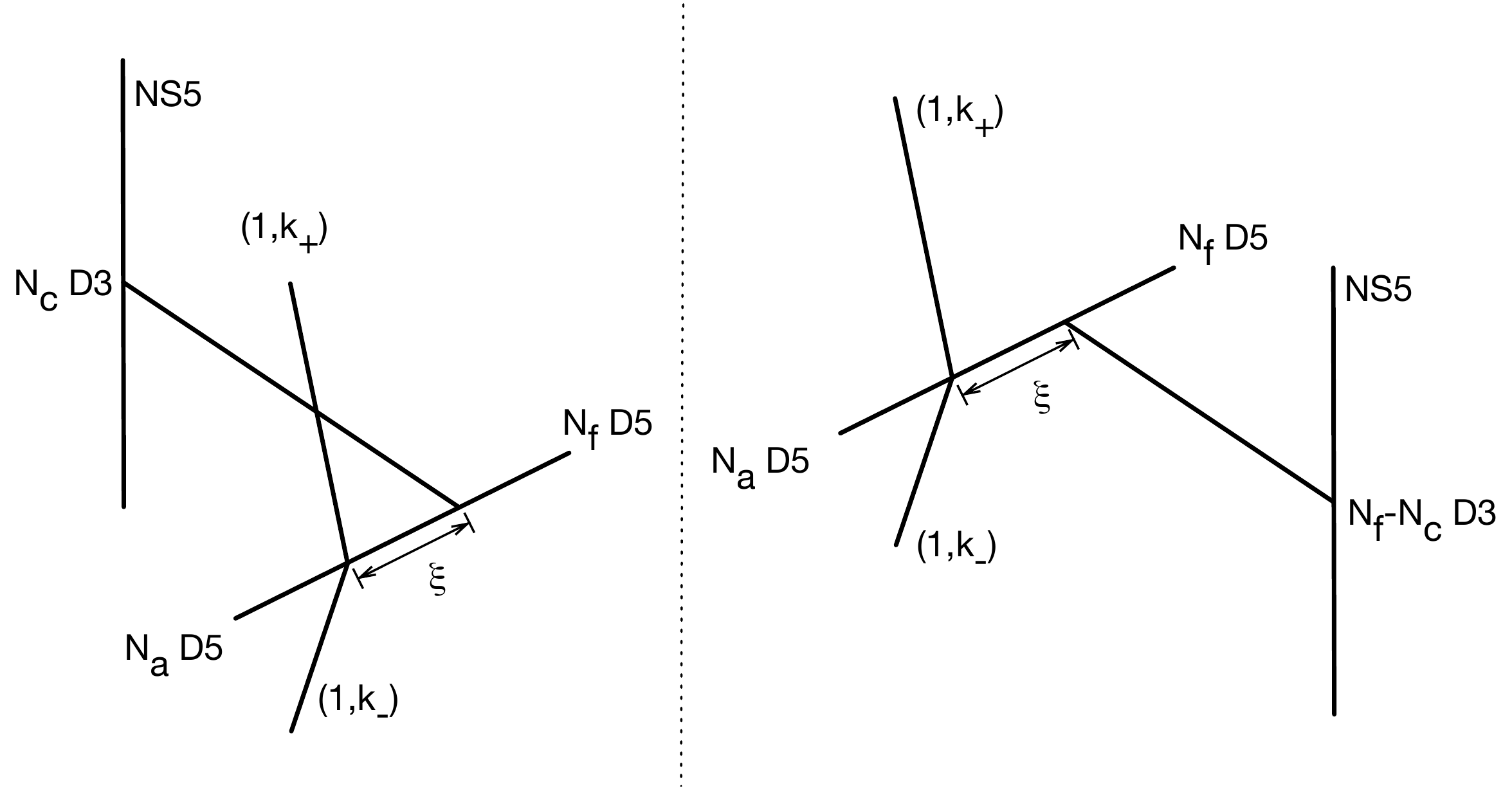}
\caption{\label{fig:AD} The brane motion representing the Aharony duality. The spacetime directions occupied by each brane are summarized in table \ref{tab:AGK branes}. Due to the Hanany-Witten effect, the number of the D3-branes changes from $N_c$ to $N_f-N_c$ when the NS5-brane passes through D5-branes.}
\end{figure}

\begin{figure}[tbp]
\centering 
\includegraphics[height=.3\textheight]{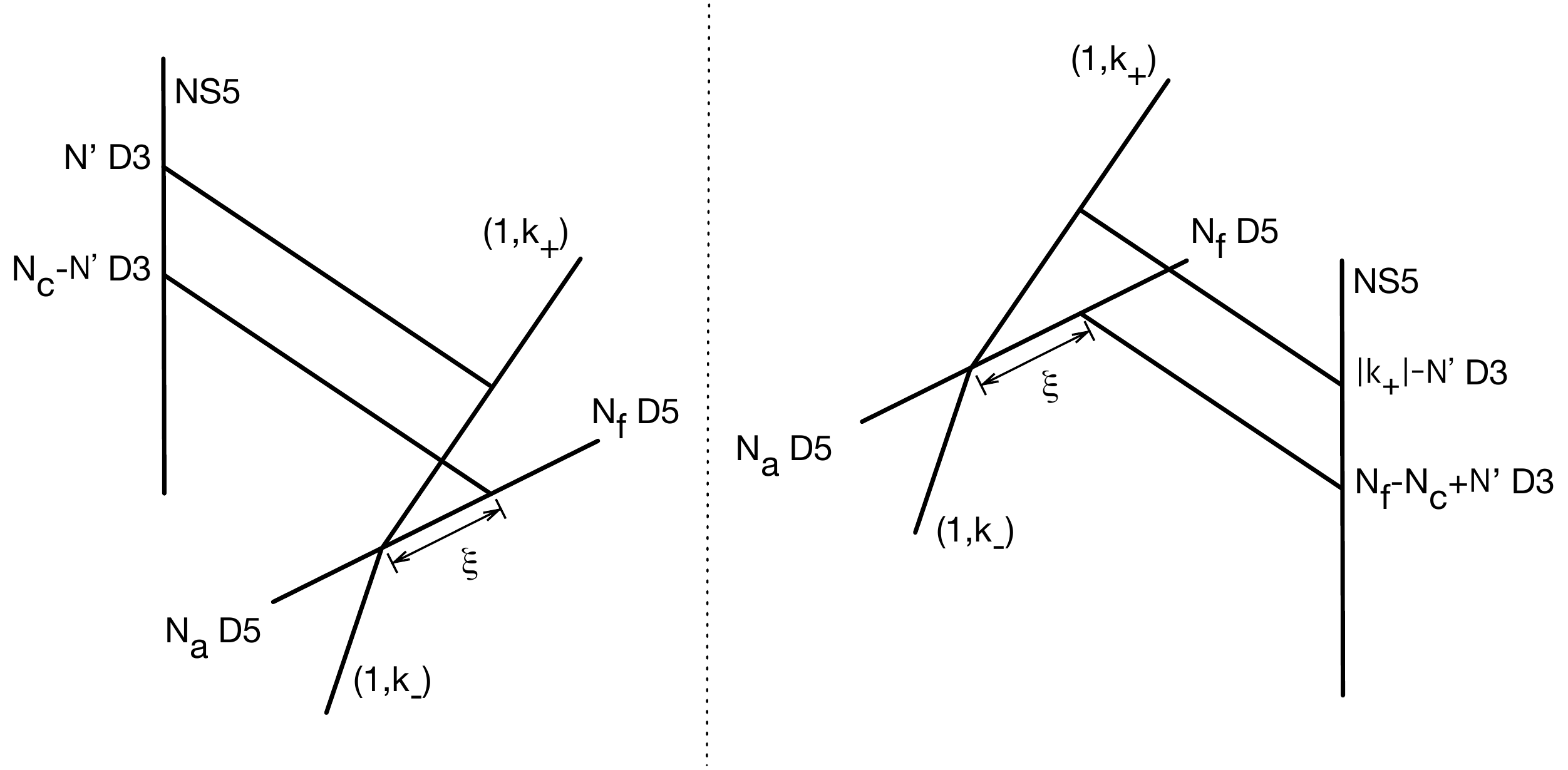}
\caption{\label{fig:GKD} The brane motion representing the Giveon-Kutasov duality. The spacetime directions occupied by each brane are summarized in table \ref{tab:AGK branes}. For positive FI parameter $\xi > 0$, one end of a D3-brane is attached either to D5-brane or to $(1,k_+)$-brane. In the figure, $N'$ D3-branes are ending on the $(1,k_+)$-brane. Recall that $k_+ = \kappa+\frac{N_f-N_a}{2}$, which is negative in the figure. After the NS5-brane passes through D5-branes, (and the $(1,k_+)$-brane,) there are $|k_+|+N_f-N_c = |\kappa|+\frac{N_f+N_a}{2}-N_c$ D3-branes due to the Hanany-Witten effect.}
\end{figure}

The Aharony duality and the Giveon-Kutasov duality are inferred from the Hanany-Witten transitions \cite{Hanany:1996ie} between the brane setups illustrated in figure \ref{fig:AD} and figure \ref{fig:GKD}.
\begin{table}[tbp]
\centering
\begin{tabular}{|c|cccccccccc|}
\hline
Branes & 0 & 1 & 2 & 3 & 4 & 5 & 6 & 7 & 8 & 9 \\
\hline
NS5 & $\times$ & $\times$ & $\times$ & $\times$ & $\times$ & $\times$ & & & & \\
$(1,k_\pm)$ & $\times$ & $\times$ & $\times$ & & & $\cdot$ & $\cdot$ & $\times$ & $\times$ & \\
D5 & $\times$ & $\times$ & $\times$ & & & & $\times$ & $\times$ & $\times$ & \\
D3 & $\times$ & $\times$ & $\times$ & & & & & & & $\times$ \\
\hline
\end{tabular}
\caption{\label{tab:AGK branes} The spacetime directions occupied by the branes in figure \ref{fig:AD} and figure \ref{fig:GKD} are marked by $\times$. The $(1,k_\pm)$-branes occupy 1-dimensional subspaces in the 56-plane.}
\end{table}
In the brane picture, those two dualities can be distinguished by whether $(1,k_+)$-brane and $(1,k_-)$-brane belong to the same side or not with respect to $x_6 = 0$. $k_+$ and $k_-$ are the effective CS levels for $\sigma \rightarrow \infty$ and $\sigma \rightarrow -\infty$ respectively. They are determined by $k_\pm = \kappa+\frac{N_f-N_a}{2}$.

The original and dual 3d gauge theories are realized as the effective theories on D3-branes. Since the D3-branes are stretched along a finite interval in the 9-direction, the effective theory is three dimensional. In figure \ref{fig:AD}, due to the Hanany-Witten effect, the number of the D3-branes changes from $N_c$ to $N_f-N_c$ when the NS5-brane passes through D5-branes. In figure \ref{fig:GKD}, on the other hand, the NS5-brane also meets the $(1,k_+)$-brane such that there are $|k_+|+N_f-N_c = |\kappa|+\frac{N_f+N_a}{2}-N_c$ D3-branes after the transition. We have assumed positive FI parameter $\xi > 0$. Those transitions of the number of the D3-branes reflect the rank of the dual gauge group shown in \eqref{eq:dual rank}.

In figure \ref{fig:AD} a D3-brane is attached to a D5-brane while in figure \ref{fig:GKD} a D3-brane is attached either to a D5-brane or to a $(1,k_+)$-brane. It indicates that a theory exhibiting the Aharony duality, with a suitable FI parameter, has only Higgs vacua while a theory exhibiting the Giveon-Kutasov duality has both Higgs vacua and topological vacua. In this note, our main interests are vortex states sitting at Higgs vacua and their behavior under Seiberg-like dualities. For this reason, we will focus on theories with Higgs vacua only and vortices therein. Under Aharony dualities of those theories, vortex states, which are excited by monopole operators, are partly mapped to particle states excited by elementary fields in dual theories. This phenomenon is called the particle-vortex duality, which is generic for 3d dualities even without supersymmetry. Non-supersymmetric particle-vortex dualities \cite{Peskin1978122,PhysRevLett.47.1556,Son:2015xqa,2015PhRvX...5d1031W,Metlitski:2015eka} and their connections to supersymmetric dualities such as the 3d mirror symmetry \cite{Intriligator:1996ex,Aharony:1997bx,Kapustin:1999ha} have been discussed recently \cite{Karch:2016sxi,Murugan:2016zal,Seiberg:2016gmd,Kachru:2016rui,Kachru:2016aon}. Indeed, the Aharony duality is also a type of the particle-vortex duality and should tell us something about the connection between vortex states and particle states. In section \ref{sec:SD}, we will see that this connection between vortex and particle states can be understood in the perspective of vortex quantum mechanics by examining the behavior of vortex quantum mechanics under the Aharony duality.
\\

Next we consider a 3d $\mathcal N = 4$ $U(N_c)$ gauge theory. A 3d $\mathcal N = 4$ theory has the $SO(4)_R = SU(2)_H \times SU(2)_C$ R-symmetry. An $\mathcal N = 4$ vector multiplet contains an $\mathcal N = 2$ vector multiplet and an $\mathcal N = 2$ chiral multiplet $\Phi$ in the adjoint representation of the gauge group. Three real scalars in the $\mathcal N = 4$ vector multiplet, one from the $\mathcal N = 2$ vector and two from the adjoint chiral multiplet, form a triplet of $SU(2)_C$. Another $\mathcal N = 4$ multiplet is a hypermultiplet. It consists of a pair of $\mathcal N = 2$ chiral multiplets $Q,\tilde Q$ whose scalars are organized to form a doublet of $SU(2)_H$. We consider $N_f$ hypermultiplets in the fundamental representation of the gauge group. Thus, the theory has $SU(N_f) \times U(1)_T$ global symmetries where $U(1)_T$ is the topological symmetry defined by the current \eqref{eq:tcurrent}. The theory has the superpotential $W = \tilde Q \Phi Q$ in $\mathcal N = 2$ language.
\\

3d $\mathcal N = 4$ $U(N_c)$ gauge theories with fundamental hypermultiplets are classified into three classes according to the number of the hypermultiplets: good, ugly and bad \cite{Gaiotto:2008ak,Kapustin:2010mh}. When $N_f > 2 N_c-1$, the theory is called good. The gauge invariant monopole operators of a good theory has the conformal dimensions larger than 1/2, which are required for the unitarity at the interacting IR fixed point.

When $N_f = 2 N_c-1$, the theory is called ugly. There is a gauge invariant monopole operator having conformal dimension 1/2. Since an operator of conformal dimension 1/2 in a 3d superconformal theory is free, this monopole operator decouples from the interacting IR fixed point theory. Apart from this decoupled monopole operator, which is described by a free twisted hypermultiplet,\footnote{Two scalars in a hypermultiplet form a doublet of $SU(2)_H$ while two scalars in a twisted hypermultiplet form a doublet of $SU(2)_C$.} the interacting IR fixed point allows another UV description, the $U(N_c-1)$ gauge theory with $N_f$ fundamental hypermultiplets \cite{Gaiotto:2008ak,Kapustin:2010mh}.

When $N_c \leq N_f < 2 N_c-1$, the theory is called bad. There are gauge invariant monopole operators having the UV R-charges less than (or equal to) 1/2. If those UV R-charges are maintained in the IR limit, the naive conformal dimensions, which are the same as the R-charges, break the unitarity of the theory. However, it is argued that a bad theory has accidental IR symmetries. The UV R-charges are corrected by those accidental IR symmetry charges such that the IR R-charges of the monopole operators, and accordingly their conformal dimensions, become 1/2. Thus, those monopole operators decouple from the interacting IR fixed point theory. Again this interacting IR fixed point allows dual UV description, the $U(N_f-N_c)$ theory with $N_f$ fundamental hypermultiplets \cite{Kim:2012uz,Yaakov:2013fza,Gaiotto:2013bwa}. The decoupled monopole operators are described by $2 N_c-N_f$ free twisted hypermultiplets.

\begin{figure}[tbp]
\centering 
\includegraphics[height=.3\textheight]{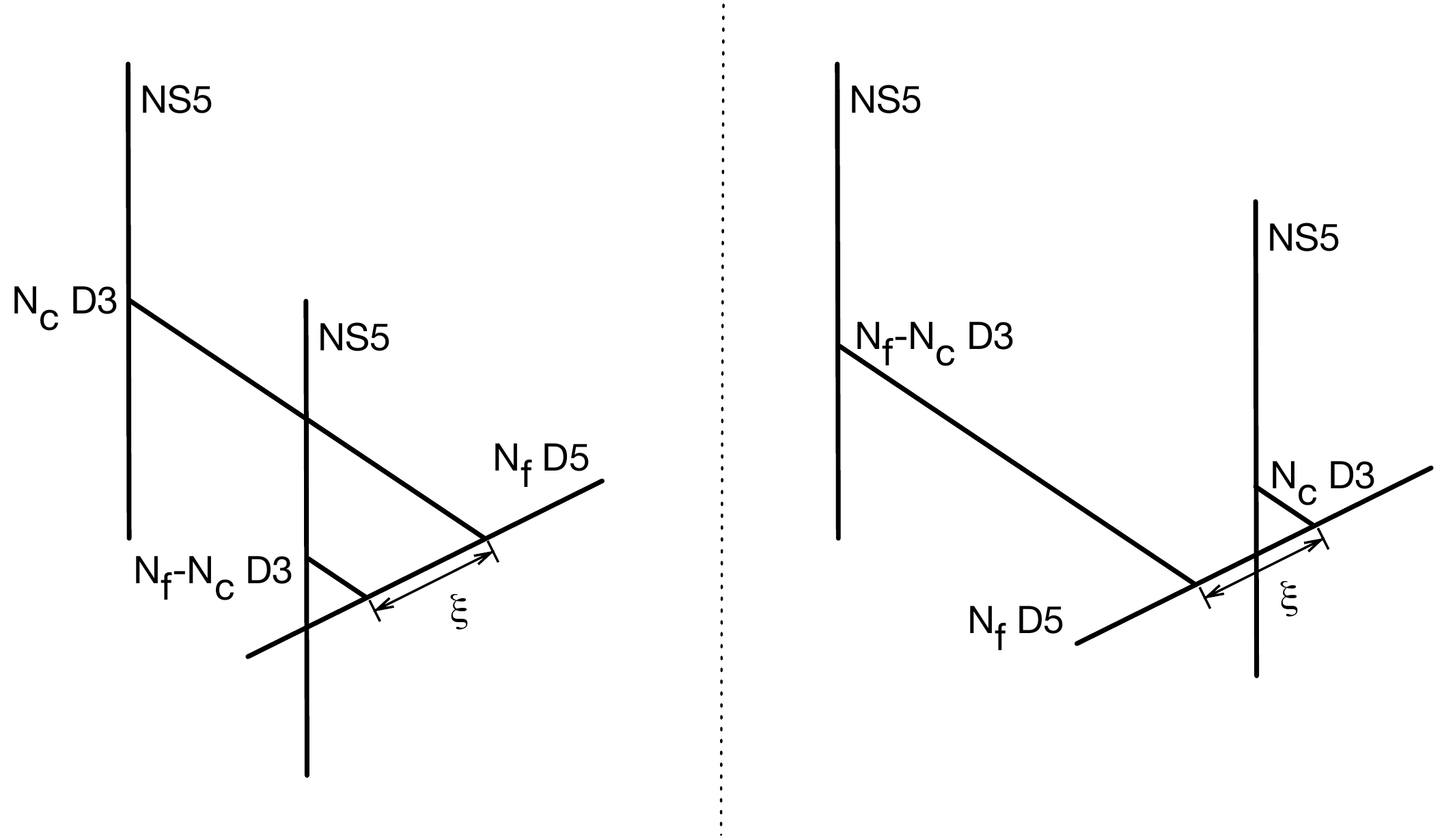}
\caption{\label{fig:N=4 SD} The brane motion representing the Seiberg-like duality of a 3d $\mathcal N = 4$ SQCD. The spacetime directions occupied by each brane are summarized in table \ref{tab:N=4 SD}.}
\end{figure}

In conclusion, the $\mathcal N = 4$ $U(N_c)$ theory with $N_f$ fundamental hypermultiplets has the Seiberg-like dual theory, which is given by the $\mathcal N = 4$ $U(N_f-N_c)$ theory with $N_f$ hypermultiplets and $2 N_c-N_f$ decoupled free twisted hypermultiplets. This duality is realized as the Hanany-Witten transition illustrated in figure \ref{fig:N=4 SD}.
\begin{table}[tbp]
\centering
\begin{tabular}{|c|cccccccccc|}
\hline
Branes & 0 & 1 & 2 & 3 & 4 & 5 & 6 & 7 & 8 & 9 \\
\hline
NS5 & $\times$ & $\times$ & $\times$ & $\times$ & $\times$ & $\times$ & & & & \\
D5 & $\times$ & $\times$ & $\times$ & & & & $\times$ & $\times$ & $\times$ & \\
D3 & $\times$ & $\times$ & $\times$ & & & & & & & $\times$ \\
\hline
\end{tabular}
\caption{\label{tab:N=4 SD} The spacetime directions occupied by the branes in figure \ref{fig:N=4 SD} are marked by $\times$.}
\end{table}
\\

\subsection{1d $\mathcal N = 2$ GLSMs and the refined Witten indices}
\label{sec:1d}

Next let us briefly review the properties of a 1d $\mathcal N = 2$ supersymmetric gauged linear sigma model with gauge group $G$. The theory contains a 1d $\mathcal N = 2$ vector multiplet, which consists of a gauge field $v_t$, a real scalar $\sigma$, a gaugino $\lambda$ and an auxiliary real scalar $D$ in the adjoint representation of $G$. One can introduce a 1d $\mathcal N = 2$ chiral multiplet as well, which consists of a scalar $\phi$ and a fermion $\psi$ in representation $V_\text{chiral}$ of $G$. In addition, there is another supersymmetric multiplet not appearing in higher dimensions: a fermi multiplet. An $\mathcal N = 2$ fermi multiplet consists of a fermion $\eta$ and an auxiliary scalar $F$ in representation $V_\text{fermi}$ of $G$.
\\

One should note that the supersymmetry transformation of the fermi multiplet $(\eta,F)$ is determined by a $G$-equivariant holomorphic map $E:V_\text{chiral} \rightarrow V_\text{fermi}$, which satisfies $E(g \phi) = g E(\phi)$:
\begin{align}
\delta \eta &= \epsilon F+\overline \epsilon E(\phi), \\
\delta F &= \overline \epsilon \left(-i D_t^{(+)} \eta+\psi^i \partial_i E(\phi) \right)
\end{align}
where $D_t^{(+)} = \partial_t+i v_t+i \sigma$. $(\phi,\psi)$ is a chiral multiplet in representation $V_\text{chiral}$. The supersymmetric kinetic term of the fermi multiplet is thus given by
\begin{align}
L_\text{fermi} = i \overline \eta D_t^{(+)} \eta+\overline F F-\overline{E(\phi)} E(\phi)-\overline \eta \partial_i E(\phi) \psi^i-\overline \psi^{\bar i} \partial_{\bar i} \overline{E(\phi)} \eta,
\end{align}
which includes the interaction terms associated to $E(\phi)$.

There is a different type of an interaction associated to another $G$-equivariant holomorphic map $J:V_\text{chiral} \rightarrow V_\text{fermi}^*$ satisfying $J(\phi) E(\phi) = 0$. Given such a map $J(\phi)$, one can turn on the following supersymmetric interaction:
\begin{align}
L_J = \psi^i \partial_i J(\phi) \eta-J(\phi) F+c.c.,
\end{align}
which is called the $F$-term associated with the superpotential $\mathfrak W = J(\phi) \eta$.

In addition, a 1d $\mathcal N = 2$ GLSM can include the Fayet-Iliopoulos term as well as the supersymmetric Wilson line. If $G$ contains $U(1)$ factors, there is an adjoint invariant linear form $\zeta: i \mathfrak g \rightarrow \mathbb R$, which defines the FI interaction term:
\begin{align}
L_\text{FI} = -\zeta(D).
\end{align}
Given a $\mathbb Z_2$ graded vector space $M$ with a hermitian inner product, the $\mathcal N = 2$ supersymmetric Wilson line is defined by
\begin{gather}
\mathrm{P exp} \left(-i \int \mathcal A_t dt\right), \\
\mathcal A_t = \rho(v_t+\sigma)-\psi^i \partial_i Q(\phi)+\overline \psi^{\bar i} \partial_{\bar i} Q(\phi)^\dagger+\{Q(\phi),Q(\phi)^\dagger\}
\end{gather}
with $\rho:G \rightarrow U(M)$, a unitary representation of $G$ on $M$ and $Q:V_\text{chiral} \rightarrow \mathrm{End}^\text{od} (M)$, a $G$-equivariant holomorphic map satisfying $Q(\phi)^2 = 0$.

With the canonical interaction terms in the supersymmetric kinetic terms of the vector multiplet and the chiral multiplet, those supersymmetric interaction terms determine the interactions of a 1d $\mathcal N = 2$ GLSM. More details about 1d $\mathcal N = 2$ GLSMs can be found in, e.g., \cite{Hori:2014tda}.
\\

One can define the refined Witten index \cite{Witten:1982df,AlvarezGaume:1986nm} of a 1d $\mathcal N = 2$ GLSM with flavor twists as follows:
\begin{align}
I = \mathrm{Tr} \left[(-1)^F e^{-\beta H} x^{G_F}\right]
\end{align}
where we collectively denote the flavor symmetry generators by $G_F$. For a compact theory,\footnote{What we mean by the refined Witten index in this note is, strictly speaking, the twisted partition function computed by the localization
procedure. For theories with non-compact low energy dynamics in the limit of vanishing chemical potentials, there
 are many subtleties in relating the two objects. For instance, the Witten index, usually understood to be defined with the $L^2$ boundary condition, needs not be computable as a limit of the twisted partition function. See \cite{Lee:2016dbm} for more discussions. Nevertheless, we stick to the former nomenclature in this note since it is not clear, a priori, whether the vortex theory in question makes the usual
sense in the vanishing limit of 3d real masses.} the refined Witten index can be computed as the twisted partition function on $S^1$, whose path integral in the end is reduced to a finite matrix integral over bosonic zero modes $u \in (\mathbb C^*)^r$ where $r$ is the rank of $G$ \cite{Hwang:2014uwa,Cordova:2014oxa,Hori:2014tda}.

This bosonic zero mode integration is encapsulated in the Jeffrey-Kirwan residue \cite{1993alg.geom..7001J} as follows:
\begin{align}
\label{eq:JK}
I = \frac{1}{|\mathsf W|} \sum_{u_* \in \mathfrak M_\text{sing}^*} \text{JK-Res}_{u = u_*} (Q(u_*),\eta) \left[g(u) d^r u\right].
\end{align}
$|\mathsf W|$ is the Weyl group order of the gauge group. The integrand $g(u)$ is determined by the Wilson line contribution and the 1-loop determinant of each multiplet:
\begin{align}
\begin{aligned}
\label{eq:g(u)}
W(u) &= \sum_{k \in K} (-1)^{r_k} e^{q_k(u)+q_k^F(\mu)}, \\
g_\text{vector}(u) &= \prod_{\alpha \in \Delta_G} 2 \sinh \frac{-\alpha \cdot u}{2}, \\
g_\text{chiral}(u) &= \prod_a \prod_{\rho \in R_a} \frac{1}{2 \sinh \frac{\rho \cdot u+F_a \cdot \mu}{2}}, \\
g_\text{fermi}(u) &= \prod_b \prod_{\rho \in R_b} 2 \sinh \frac{-\rho \cdot u-F_a \cdot \mu}{2}
\end{aligned}
\end{align}
where the flavor chemical potential is denoted by $x = e^\mu$. For the Wilson line contribution, we have adopted the weight decomposition of $M$,
\begin{align}
M = \oplus_{k \in K} \mathbb C \left(q_k,q_k^F\right) [r_k]
\end{align}
where the $\mathbb Z_2$-grade of $M$ is labeled by $r_k$ with $r_k = 0$ for the even part and $r_k = 1$ for the odd part. For the 1-loop determinants, $\Delta_G$ denotes the set of the roots of the gauge group $G$. $a$ and $b$ label chiral and fermi multiplets respectively, each of which has flavor charge $F_{a,b}$ and gauge representation $R_{a,b}$. The integrand $g(u)$ is given by the product of those factors in \eqref{eq:g(u)}.

Each factor in the denominator of $g(u)$ defines a hyperplane in $(\mathbb C^*)^r$. The set of the poles determined by intersections of such hyperplanes is denoted by $\mathfrak M_\text{sing}^*$. The charge vectors associated to $u_* \in \mathfrak M_\text{sing}^*$ is collectively denoted by $Q(u_*)$. $Q(u_*)$ is assumed to be projective for every $u_* \in \mathfrak M_\text{sing}^*$; i.e., every charge vector in $Q(u_*)$ belongs to the same half space.

For simplicity, let us assume $u_* = 0$; generic $u_*$ can be restored by a coordinate shift. When the pole at $u = 0$ is not degenerate, i.e., exactly $r$ linearly independent hyperplanes intersect at $u = 0$, the JK-residue with given $\eta$ is evaluated as follow \cite{1993alg.geom..7001J,1999math......3178B}:
\begin{align}
\text{JK-Res}_{u = 0}(Q(0),\eta) \frac{d^r u}{\prod_{p = 1}^r Q_{i_p} \cdot u} = \left\{\begin{array}{ll}
\frac{1}{|\mathrm{det}(Q_{i_1} \ldots Q_{i_r})|}, \qquad & \text{if $\eta \in \text{Cone}(Q_{i_1},\ldots,Q_{i_r})$,} \\
0, \qquad & \text{otherwise.}
\end{array}\right.
\end{align}
The auxiliary JK-vector $\eta$ determines which poles contribute to the result while the final result is independent of the choice of $\eta$. $\eta$ should be generic, i.e., it shouldn't be a linear combination of less than $r$ charge vectors. When the pole is degenerate, a constructive definition of the JK-residue can be used \cite{1999math......3178B,2004InMat.158..453S}, which is also reviewed in \cite{Benini:2013xpa}.

We also comment that a pole would be an intersection of $r-1$ hyperplanes and the asymptotic infinity where we associate the charge vector $-\zeta$ to the latter. In most cases, however, we choose $\eta$ belonging to the same chamber as $\zeta$ in the charge space such that the asymptotic poles do not participate in the actual computation.\footnote{Nevertheless, those asymptotic poles and their residues are important to understand the wall-crossing of a 1d GLSM. See \cite{Hori:2014tda} for related discussions.}
\\

\section{Vortex quantum mechanics}
\label{sec:vortex QM}
\subsection{$T_{\rho}[SU(N)]$}
\label{sec:T[SU(N)]}

\begin{figure}[tbp]
\begin{center}
\includegraphics[width=9cm]{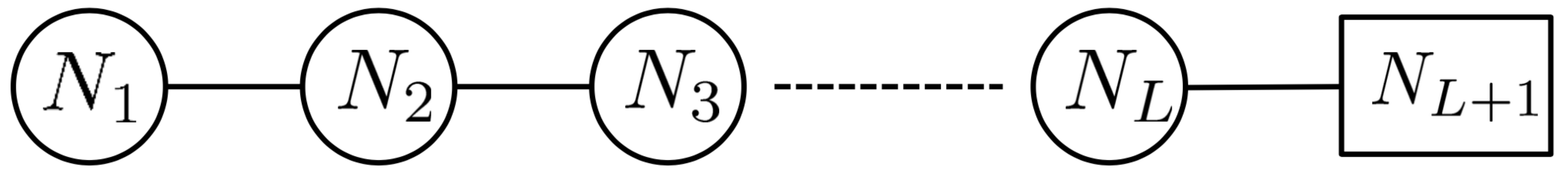}
\end{center}
\vspace{-0.5cm}
\caption{The quiver diagram for $T_\rho [SU(N)]$ with $\rho=(N_1,N_2-N_1, \cdots, N_{L+1}-N_{L})$ and $N:=N_{L+1}$. A circle with $N_l$ expresses the $\mathcal{N}=4$ $U(N_l)$ vector multiplet. A line attached between two circles with $N_l$ and $N_{l+1}$ expresses an $\mathcal{N}=4$ bi-fundamental hypermultiplet.
A line attached between the circle with $N_{L}$ and the box with $N_{L+1}$ represents $N_{L+1}$ $U(N_L)$ fundamental hypermultiplets.}
\label{quiver:fig}
\end{figure}

\begin{figure}[tbp]
\begin{center}
\includegraphics[width=11cm]{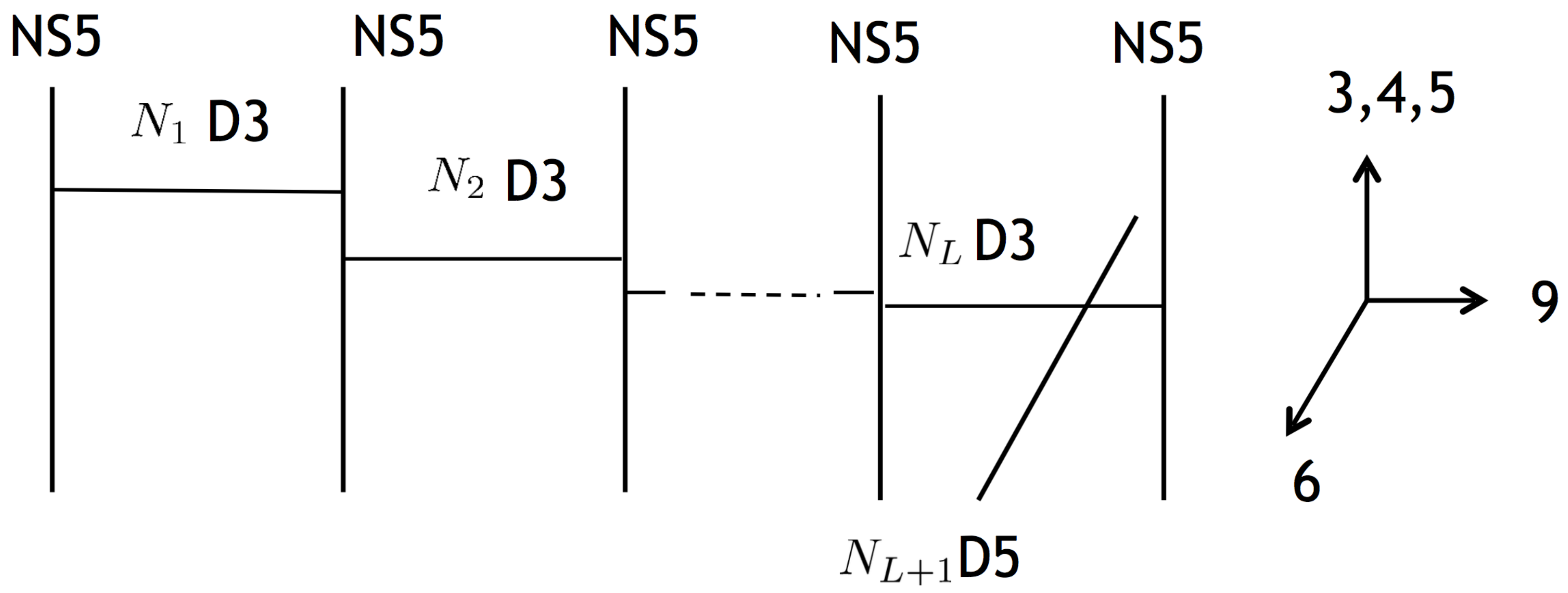}
\end{center}
\vspace{-0.5cm}
\caption{The brane configuration of $T_\rho [SU(N)]$. The vertical lines express NS5-branes and the horizontal lines express D3-branes.}
\label{fig:branequiver}
\end{figure}

First, we review brane constructions of vortex quantum mechanics \cite{Hanany:2003hp, Aganagic:2014oia, Bullimore:2016hdc} for 3d $\mathcal{N}=4$ linear quiver gauge theories called $T_{\rho} [SU(N)]$ (figure \ref{quiver:fig}).
The Hanany-Witten brane setup of the linear quiver gauge theory is shown in figure \ref{fig:branequiver}.
The $L+1$ NS5-branes extend along $012345$-directions with separations in the $9$-direction.
The $N_{l=1,2, \cdots, L}$ D3-branes extend along $0129$-directions and are stretched between two adjacent NS5-branes.
These D3-branes give rise to the $\mathcal{N}=4$ $\prod_{l=1}^L U(N_l)$ vector multiplets and
also $L-1$ bi-fundamental hypermultiplets.
The $N_{L+1}$ D5-branes extending along $012678$-directions give $N_{L+1}$ fundamental hypermultiplets of a gauge group $U(N_L)$. The theory has the flavor symmetry $SU(N_{L+1})$ as well as the R-symmetry $SU(2)_C \times SU(2)_H$.

One can introduce an FI term for each $U(1)$ factor of the gauge group. Each FI parameter is a triplet of $SU(2)_H$, which can be decomposed into one real and one complex FI parameters. Since we are interested in the half-BPS vortex solutions, which are in general allowed only for the vanishing complex FI parameters \cite{Bullimore:2016hdc}, we only turn on the real FI parameters $(\xi_1,\xi_2, \cdots, \xi_L)$ for the gauge group $\prod_{l=1}^L U(N_l)$. The R-symmetry group $SU(2)_C \times SU(2)_H$ is broken to $SU(2)_C \times U(1)_H$ in the presence of the FI term.

\begin{figure}[tbp]
\begin{center}
\includegraphics[width=9cm]{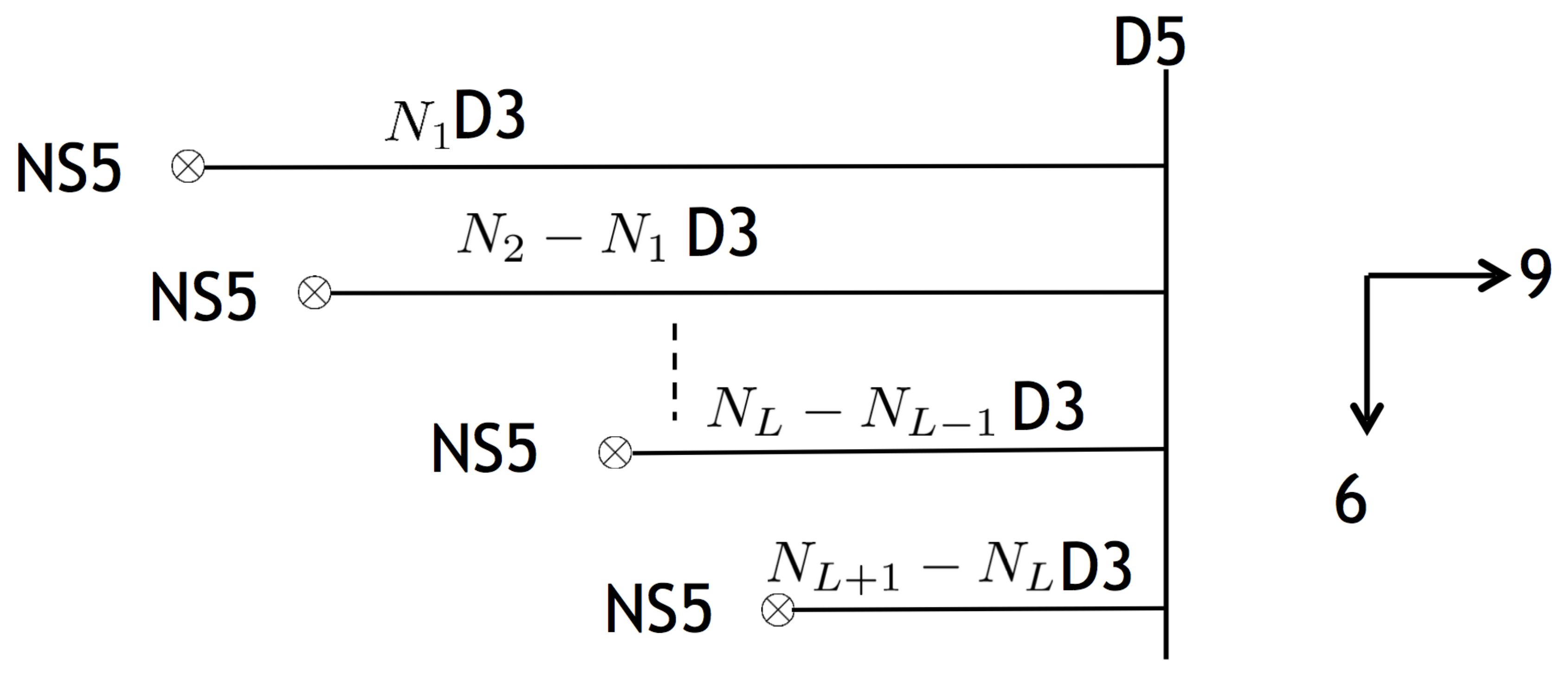}
\end{center}
\vspace{-0.5cm}
\caption{The brane configuration for a Higgs branch.
The horizontal lines are D3-branes suspended between a NS5-brane and D5-branes.
The distance between $N_{l+1}-N_l$ D3-branes and $N_{l}-N_{l-1}$ D3-branes along the $6$-direction is proportional to the magnitude of FI parameter $\xi_l$ of $U(N_l)$.}
\label{fig:Higgsbranch}
\end{figure}

In the brane setup, the nonzero real FI parameters are achieved by separations of the NS5-branes along the 6-direction. We also take the $N_{L+1}$ D5-branes to the right in the $9$-direction. When the D5-branes across the rightmost NS5-brane, the D3-brane annihilations and creations occur and $N_{L+1}-N_L$ D3-branes suspended between the rightmost NS5-brane and the D5-branes appear.
Then we obtain a brane configuration sketched in figure \ref{fig:Higgsbranch}. The magnitude of an FI parameter $\xi_l$ is proportional to the distance between $N_{l+1}-N_{l}$ D3-branes and $N_l-N_{l-1}$ D3-branes in the 6-direction.

\begin{figure}[tbp]
\begin{center}
\includegraphics[width=9cm]{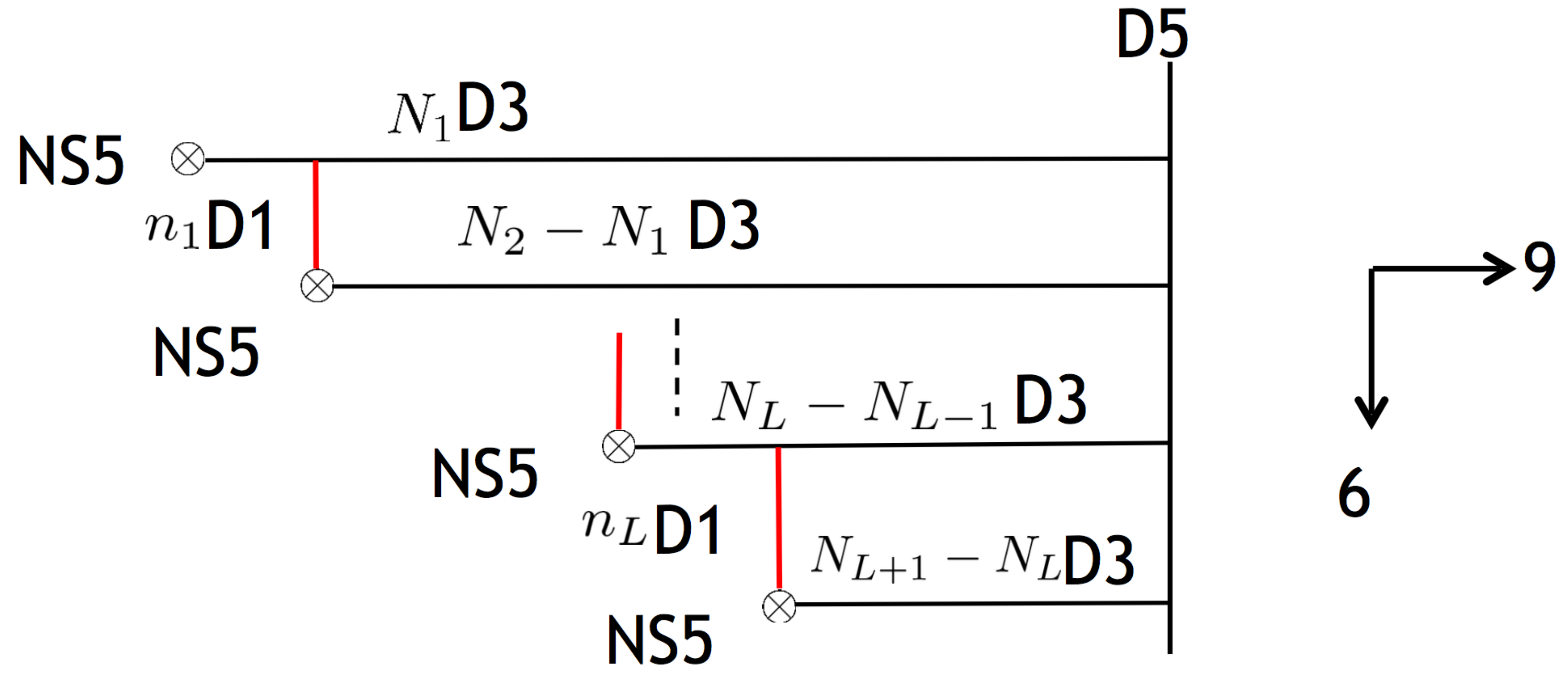}
\end{center}
\vspace{-0.5cm}
\caption{The brane configuration of the $(n_1,n_2, \cdots ,n_L)$ vortex. The  vertical red lines express D1-branes.}
\label{fig:vortex1}
\end{figure}

\begin{figure}[tbp]
\begin{center}
\includegraphics[width=9cm]{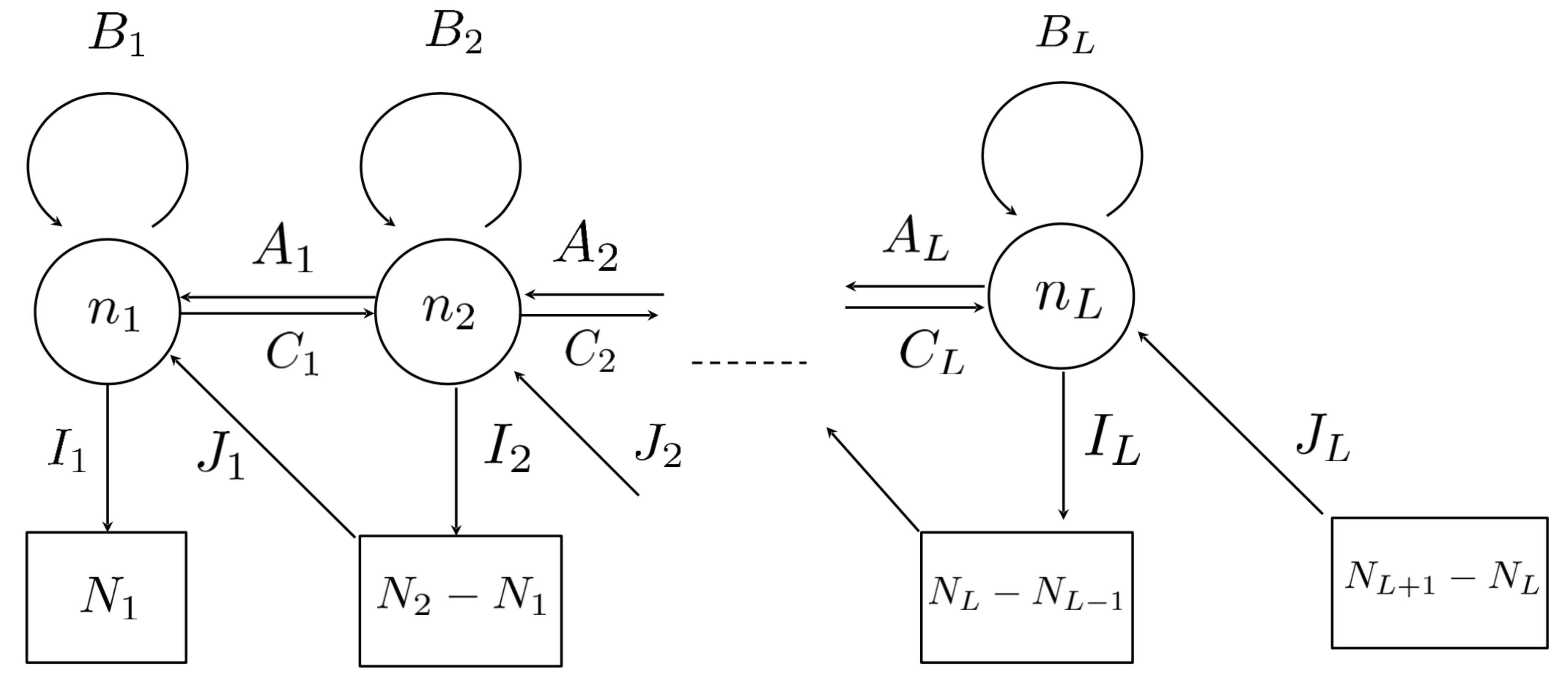}
\end{center}
\vspace{-0.5cm}
\caption{The quiver diagram for the $(n_1,n_2, \cdots, n_L)$ vortex world-line theory. The circle with $n_l$ expresses the 1d $\mathcal{N}=4$ $U(n_l)$ vector multiplet and the arrows express 1d $\mathcal{N}=4$ chiral multiplets. The box with $N_l-N_{l-1}$ expresses the number of chiral multiplets.}
\label{fig:handsaw}
\end{figure}

The half-BPS vortices are engineered by $n_{l=1, \cdots, L}$ D1-branes stretched between D3-branes as shown in figure \ref{fig:vortex1}, where the world-volume of D1-branes are $06$-directions.
By sending $N_{L+1}$ D5-branes to right infinity in the 9-direction, one can read off the world-volume theory of D1-branes, which is 1d $\mathcal{N}=4$ supersymmetric quiver quantum mechanics. The quiver diagram of the quantum mechanics is specified by figure \ref{fig:handsaw}.
The closed loops of arrows in figure \ref{fig:handsaw} correspond to the following superpotential terms:
\begin{eqnarray}
W=\sum_{l=1}^{L-1} J_l C_{l} I_{l+1}+\sum_{l=1}^{L-1} \mathrm{Tr} B_{l} C_l A_l- \sum_{l=1}^{L-1} \mathrm{Tr} B_{l+1} A_l C_l
\end{eqnarray}
where $I_l$, $J_l$ and $B_l$ are $N_{l}-N_{l-1}$ fundamental, $N_{l+1}-N_{l}$ anti-fundamental and an adjoint chiral multiplets of a gauge group $U(n_l)$, respectively. $A_{l}$ and $C_{l}$ are $U(n_l) \times U(n_{l+1})$ bi-fundamental chiral multiplets. The moduli space of vortices is given by the D-term and F-term solution of the quiver quantum mechanics.
The gauge coupling $e_l$ of 3d gauge group $U(N_l)$ is related to the FI parameter $\zeta_l$ of 1d gauge group $U(n_l)$ as
$\zeta_l ={2 \pi }/{e^{2}_l} $.

The global symmetry group of the 1d $\mathcal{N}=4$ quantum mechanics is $[\prod_{l=1}^{L+1} U(N_{l}-N_{l-1})]/U(1)^L \times SU(2)_C \times U(1)_H  \times U(1)_Z$, where $U(1)_{Z}$ is associated with the rotation in the 12-directions and $SU(2)_C \times U(1)_H$ is the R-symmetry group, which descends from the 3d R-symmetry. The diagonal combination $U(1)_{\epsilon}:=\mathrm{diag}(U(1)_H \times U(1)_Z)$ commutes with the 1d supersymmetry and acts on each multiplet as a flavor symmetry \cite{Bullimore:2016hdc}. The charge assignment is summarized in table \ref{tab:1dN4}.
\begin{table}[tbp]
\centering
\begin{tabular}{|c|cccccc|}
\hline
 & $U(n_l)$ & $U(n_{l+1})$ & $U(N_{l}-N_{l-1})$ & $U(N_{l+1}-N_{l})$ & $U(1)_H$ & $U(1)_{\epsilon}$ \\
\hline
$I_l$ & $\mathbf{n_l}$ & $\mathbf 1$ & $\mathbf{\overline{N_{l}-N_{l-1}}}$ & $\mathbf 1$  & $0$ & $-1$ \\
$J_l$ & $\mathbf{\overline{n_l}}$ & $\mathbf{1}$ & $\mathbf{1}$ & $\mathbf{N_{l+1}-N_{l}}$ & $0$ & $-1$ \\
$A_l$ & $\mathbf{\overline{n_l}}$ & $\mathbf{n_{l+1}}$ & $\mathbf{1}$ & $\mathbf 1$ & $0$ & $0$ \\
$B_l$ & $\mathbf{\mathrm{adj}}$ & $\mathbf 1$ & $\mathbf{1}$ & $\mathbf 1$ & $0$ & $-2$ \\
$C_l$ & $\mathbf{n_{l}}$ & $\mathbf{\overline{n_{l+1}}}$ & $\mathbf 1$ & $\mathbf 1$ & $2$ & $2$ \\
\hline
\end{tabular}
\caption{\label{tab:1dN4} The symmetry charges of the 1d $\mathcal{N}=4$ chiral multiplets we introduce. There is also the $SU(2)_C$ symmetry, which is a part of the 1d R-symmetry. All the multiplets in the table carry spin zero under this $SU(2)_C$, i.e., $J_3 = 0$.}
\end{table}

Now we would like to compute the index of this vortex quantum mechanics. The refined Witten index of $\mathcal N = 4$ handsaw quiver quantum mechanics is written as \cite{Hori:2014tda}
\begin{align}
I= \mathrm{Tr} \left[ (-1)^F y^{R-2 J_3} \prod_a x^{G^{(a)}_F}_a \right]
\end{align}
where $J_3$ is the Cartan generator of  $SU(2)_C$ and $R$ is the generator of $U(1)_H$.
$x_a$'s denote fugacities for the $[\prod_{l=1}^{L+1} U(N_{l}-N_{l-1})]/U(1)^L \times U(1)_{\epsilon}$ flavor symmetry.
$G^{(a)}_{F}$ is the corresponding flavor charge for $x_a$.
More explicitly, we set $y=e^{\mu}$ as well as the other flavor fugacities: $e^{\gamma}$ for $U(1)_{\epsilon}$ and $e^{m^{(l)}_a}$  $(a=1, \cdots, N_{l}-N_{l-1})$ for the Cartan part of $U(N_l-N_{l-1})$.

As discussed in the previous section, this refined Witten index is given by the following JK-residue:\footnote{``$\eta = \zeta$'' means that $\eta$ is generic but belongs to the same chamber as $(\zeta_1 \vec 1_{N_1},\ldots,\zeta_L \vec 1_{N_L})$ in the charge space.}
\begin{align}
\label{eq:T[SU(N)] ind}
I = \frac{1}{|\mathsf W|} \text{JK-Res}_{\eta = \zeta} \left[g(u) d^r u\right]
\end{align}
where
\begin{align}
g(u) &= \left(\frac{1}{2 \sinh \frac{-2 \mu}{2}}\right)^{\sum_{l = 1}^L n_l} \prod_{l = 1}^L \left(\prod_{i \neq j}^{n_l} \frac{\sinh \frac{u^{(l)}_i-u^{(l)}_j}{2}}{\sinh \frac{u^{(l)}_i-u^{(l)}_j-2 \mu}{2}}\right) \left(\prod_{i,j = 1}^{n_l} \frac{\sinh \frac{u^{(l)}_i-u^{(l)}_j-2 \mu-2 \gamma}{2}}{\sinh \frac{u^{(l)}_i-u^{(l)}_j-2 \gamma}{2}}\right) \nonumber \\
&\quad \times \left(\prod_{i = 1}^{n_{l+1}} \prod_{j = 1}^{n_l} \frac{\sinh \frac{u^{(l+1)}_i-u^{(l)}_j-2 \mu}{2}}{\sinh \frac{u^{(l+1)}_i-u^{(l)}_j}{2}}\right) \left(\prod_{i = 1}^{n_{l+1}} \prod_{j = 1}^{n_l} \frac{\sinh \frac{-u^{(l+1)}_i+u^{(l)}_j+2 \gamma}{2}}{\sinh \frac{-u^{(l+1)}_i+u^{(l)}_j+2 \mu+2 \gamma}{2}}\right) \nonumber \\
&\quad \times \left(\prod_{i = 1}^{n_l} \prod_{b = N_{l-1}}^{N_l} \frac{\sinh \frac{u^{(l)}_i-m_b-2 \mu-\gamma}{2}}{\sinh \frac{u^{(l)}_i-m_b-\gamma}{2}}\right) \left(\prod_{j = 1}^{n_l} \prod_{a = N_l}^{N_{l+1}} \frac{\sinh \frac{-u^{(l)}_j+m_a-2 \mu-\gamma}{2}}{\sinh \frac{-u^{(l)}_j+m_a-\gamma}{2}}\right).
\end{align}
We have defined $n_{L+1} = 0$. Since $g(u)$ has degenerate poles for general gauge ranks $n_l$, the JK-residue computation requires a constructive definition of the JK-residue called the flag method \cite{Benini:2013xpa}, which is very complicated to do analytically. Instead we adopt a prescription for the selection of the contributing poles and conduct tests for the prescription numerically. Motivated by the honest computation for $L = 1$ in section \ref{sec:N=4 SQCD}, we select the following poles:
\begin{align}
u^{(l)}_i = m_a+(2 k_a-1) \gamma, \qquad a \in \{1,\ldots,N_l\}, \quad k_a \in \{1,\ldots,n^{(l)}_a\}
\end{align}
where $(n^{(l)}_1,\ldots,n^{(l)}_{N_l})$ is a partition of $n_l$, i.e., an ordered set of $N_l$ nonnegative integers satisfying $\sum_{a = 1}^{N_l} n^{(l)}_a = n_l$. Every pair $(a,k_a)$ is assigned to one of $i = 1,\ldots,n_l$ exactly once. Evaluating the residue at each pole, we have
\begin{align}
\label{eq:T[SU(N)] res}
& I = \nonumber \\
%
%
& \prod_{l = 1}^{L} \left(\prod_{a = 1}^{N_{l+1}} \prod_{b = 1}^{N_l} \prod_{k_a = 1}^{n^{(l+1)}_a} \frac{\sinh \frac{m_a-m_b+2 (n^{(l+1)}_a-k_a) \gamma}{2}}{\sinh \frac{m_a-m_b-2 \mu+2 (n^{(l+1)}_a-k_a) \gamma}{2}} \prod_{a,b = 1}^{N_l} \prod_{k_a = 1}^{n^{(l)}_a} \frac{\sinh \frac{m_a-m_b-2 \mu+2 (k_a-n^{(l)}_b-1) \gamma}{2}}{\sinh \frac{m_a-m_b+2 (k_a-n^{(l)}_b-1) \gamma}{2}}\right)' \nonumber \\
&\quad \times \left(\prod_{a = 1}^{N_{l+1}} \prod_{b = 1}^{N_l} \prod_{k_b = 1}^{n^{(l)}_b} \frac{\sinh \frac{m_a-m_b-2 \mu+2 (n^{(l+1)}_a-k_b) \gamma}{2}}{\sinh \frac{m_a-m_b+2 (n^{(l+1)}_a-k_b) \gamma}{2}} \prod_{a,b = 1}^{N_l} \prod_{k_b = 1}^{n^{(l)}_b} \frac{\sinh \frac{m_a-m_b+2 (k_b-n^{(l)}_b-1) \gamma}{2}}{\sinh \frac{m_a-m_b-2 \mu+2 (k_b-n^{(l)}_b-1) \gamma}{2}}\right)'
\end{align}
where $'$ denotes that the vanishing factors are omitted. The permutations among $u^{(l)}_i$'s give rise to factor $n_l!$, which cancels the Weyl group factor $|\mathsf W|$. The detailed computation of \eqref{eq:T[SU(N)] res} is similar to that of the single gauge node case, which is explicitly described in section \ref{sec:N=4 SQCD}. Using
\begin{align}
(a;q)_{n-m} = \frac{(a q^{-m};q)_n}{(a q^{-m};q)_m},
\end{align}
\eqref{eq:T[SU(N)] res} is further simplified such that the final expression of the index is given by
\begin{align}
\label{eq:T[SU(N)] result}
& I = \nonumber \\
& \prod_{l = 1}^{L} \left(\prod_{a,b = 1}^{N_l} \prod_{k = 1}^{n^{(l)}_a-n^{(l)}_b} \frac{\sinh \frac{m_a-m_b-2 \mu+2 (k-1) \gamma}{2}}{\sinh \frac{m_a-m_b+2 (k-1) \gamma}{2}}\right) \left(\prod_{a = 1}^{N_{l+1}} \prod_{b = 1}^{N_l} \prod_{k = 1}^{n^{(l)}_b-n^{(l+1)}_a} \frac{\sinh \frac{m_a-m_b-2 \mu-2 k \gamma}{2}}{\sinh \frac{m_a-m_b-2 k \gamma}{2}} \right),
\end{align}
which agrees with the vortex partition function of $T_\rho [SU(N)]$ obtained in \cite{Bullimore:2014awa} using the factorization of the $S^3_b$ partition function. We also conduct the numerical computation of \eqref{eq:T[SU(N)] ind} using the flag method and confirm that it agrees with \eqref{eq:T[SU(N)] result}.
\\

\subsection{$\mathcal N = 2$ linear quiver gauge theories}
\label{sec:N=2}

\begin{figure}[tbp]
\centering 
\includegraphics[width=.5\textwidth]{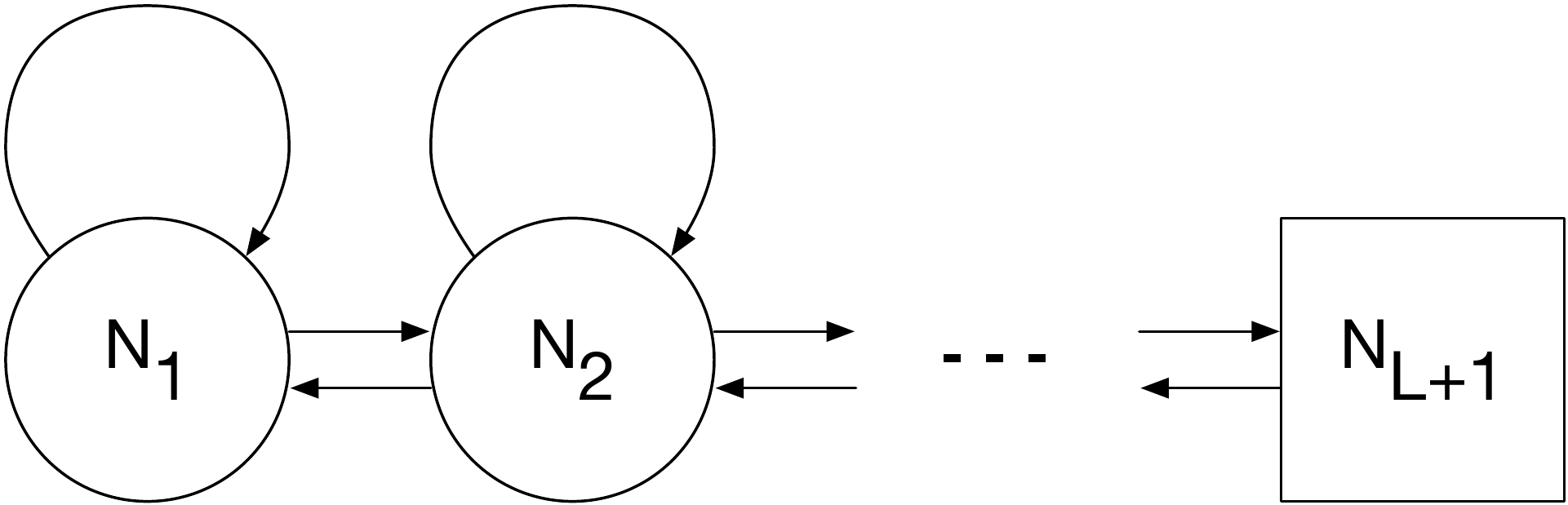}
\caption{\label{fig:N=2 T[SU(N)]} The quiver diagram representation of $T_\rho [SU(N)]$ is illustrated in terms of $N=2$ multiplets. A round node represents a gauge group factor with a corresponding $\mathcal N = 2$ vector multiplet. Each arrow represents an $\mathcal N = 2$ chiral multiplet in the bi-fundamental representation between the gauge group factors connected by the arrow. A square node represents flavor group factor.}
\end{figure}

\begin{figure}[tbp]
\centering 
\includegraphics[width=.5\textwidth]{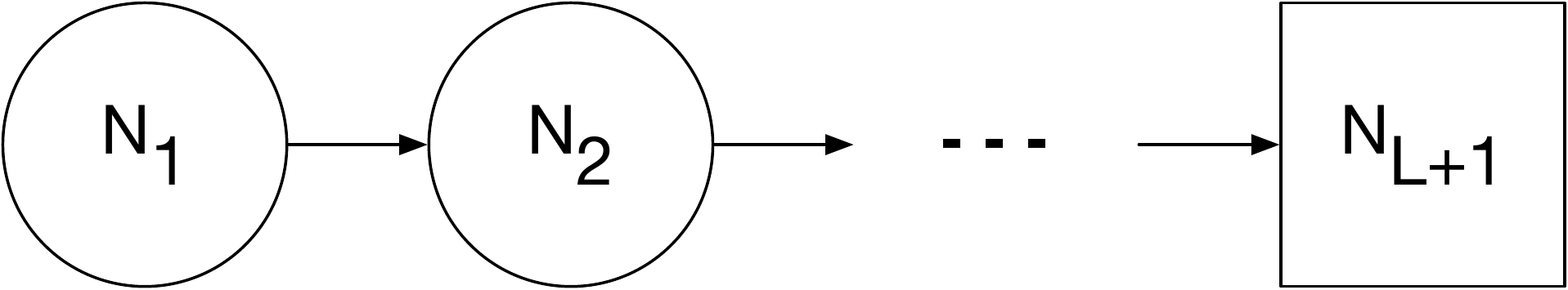}
\caption{\label{fig:def T[SU(N)]} The quiver diagram representation of the 3d $\mathcal N = 2$ theory deformed from $T_\rho [SU(N)]$. The deformed theory also includes various CS/BF interactions.}
\end{figure}

Now we would like to extend our discussion of vortex quantum mechanics to 3d $\mathcal N = 2$ linear quiver gauge theories. Unlike $T_\rho [SU(N)]$, the brane setup of an $\mathcal N = 2$ linear quiver theory is in general not known. Thus, one cannot directly read off vortex quantum mechanics from the brane setup. Here, instead, we take an indirect approach: we first consider the $\mathcal N = 2$ deformation of the previous $\mathcal N = 4$ $T_\rho [SU(N)]$ example. Figure \ref{fig:N=2 T[SU(N)]} represents $T_\rho [SU(N)]$ in terms of $\mathcal N = 2$ multiplets.
In general, one can deform a 3d theory by turning on real mass for a global symmetry. Here we turn on real mass for the $U(1)_A$ symmetry, which is the off-diagonal combination of two $SU(2)$ R-symmetries. Since this real mass breaks the R-symmetry $SU(2)^2$ to $U(1)_R \times U(1)_A$ where $U(1)_A$ is a non-$R$ global symmetry, the deformed theory only preserves $\mathcal N = 2$ supersymmetry. In addition, we also turn on the vacuum expectation values of vector multiplet scalars such that only the right-directed chiral multiplets in figure \ref{fig:N=2 T[SU(N)]} remain massless. Thus, the deformed theory is given by figure \ref{fig:def T[SU(N)]}.

We emphasize that this deformation of $T_\rho [SU(N)]$ incorporates various CS/BF interactions in the deformed $\mathcal N = 2$ theory. First, the fermions in the left-directed bi-fundamental chiral multiplets, which are integrated out, leave their remnants as the CS/BF interactions of levels
\begin{gather}
\begin{gathered}
\label{eq:CS bifund}
\kappa^{(l)} = \frac{N_{l-1}+N_{l+1}}{2}, \\
\kappa_{U(1)}^{(l,l+1)} = -\frac{1}{2}
\end{gathered}
\end{gather}
where we have defined $N_0 = 0$. $\kappa^{(l)}$ is the CS level for the $l$-th gauge node and $\kappa_{U(1)}^{(l,l+1)}$ is the BF level between the $U(1)$ factors of the $l$-th and $(l+1)$-th gauge nodes. The $U(1)$ BF level is normalized such that the corresponding Lagrangian term is given by
\begin{align}
\frac{\kappa_{U(1)}^{(l,l+1)}}{2 \pi} \mathrm{Tr} A^{(l)} d \mathrm{Tr} A^{(l+1)}.
\end{align}
Second, the  fermions in adjoint chiral multiplets, which are parts of the $\mathcal N = 4$ vector multiplets, also give CS interactions but only for the $SU(N)$ parts:
\begin{align}
\label{eq:CS adj}
\kappa^{(l)}_{SU(N)} = -N_l.
\end{align}
Combining \eqref{eq:CS bifund} and \eqref{eq:CS adj}, the induced CS/BF levels by the deformation are as follows:
\begin{gather}
\begin{gathered}
\label{eq:def levels}
\kappa^{(l)} = \frac{N_{l-1}+N_{l+1}}{2}-N_l, \\
\Delta \kappa_{U(1)}^{(l)} = 1, \\
\kappa_{U(1)}^{(l,l+1)} = -\frac{1}{2}
\end{gathered}
\end{gather}
with $N_0 = 0$. For later convenience, we organize the $SU(N)$ and $U(1)$ CS levels such that they are given by $U(N)$ CS levels, $\kappa^{(l)}$, and additional level shifts for the $U(1)$ parts, $\Delta \kappa_{U(1)}^{(l)}$. Again $\kappa_{U(1)}^{(l,l+1)}$ is the BF level between the $l$-th and $(l+1)$-th gauge nodes.
\\

\begin{figure}[tbp]
\centering 
\includegraphics[height=.2\textheight]{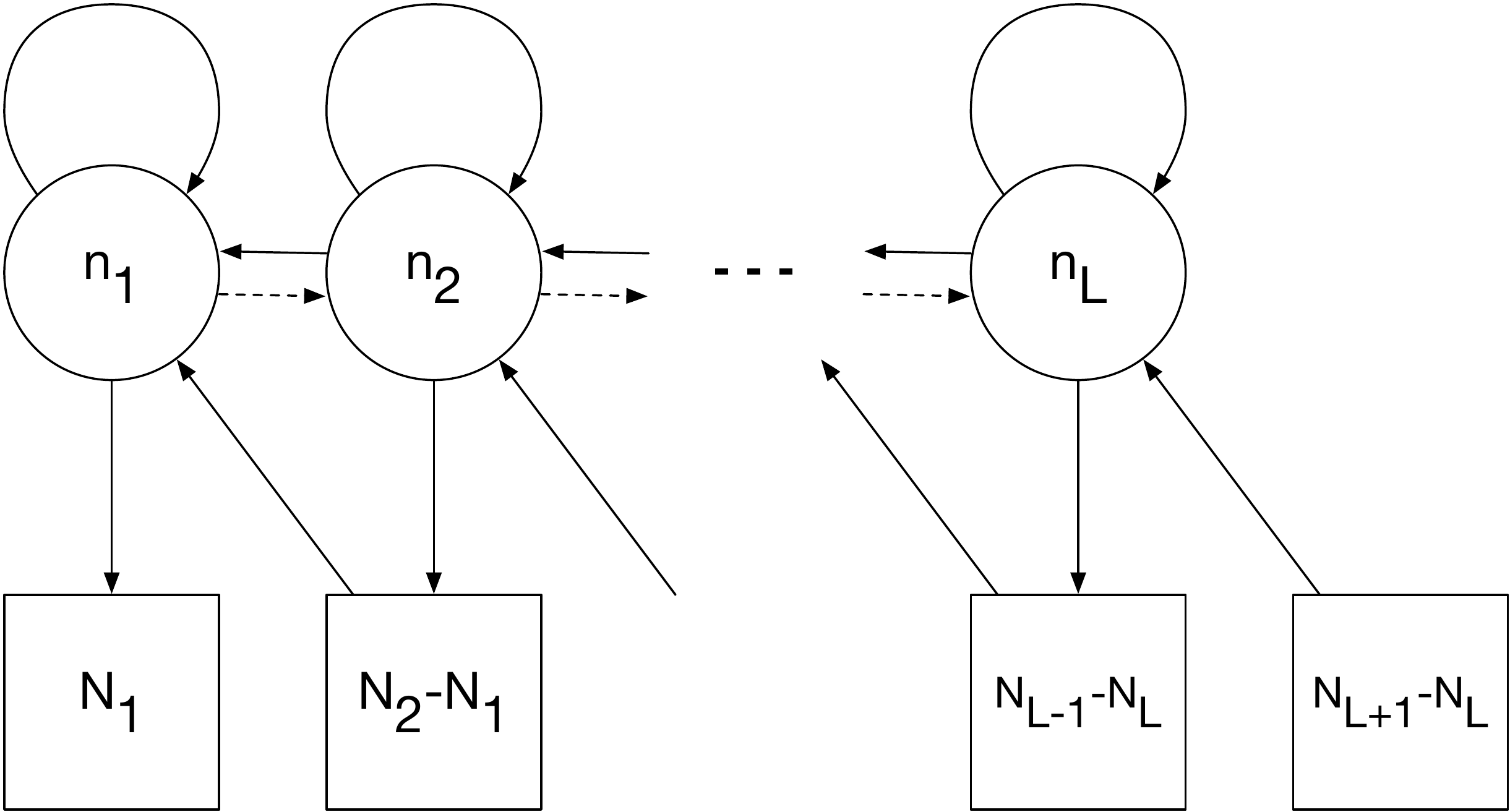}
\caption{\label{fig:def vortex} The quiver diagram representation of vortex quantum mechanics for the $\mathcal N = 2$ deformed theory of $T_\rho [SU(N)]$. }
\end{figure}

We have realized the $\mathcal N = 2$ deformation of $T_\rho [SU(N)]$ by turning on real mass associated with the $U(1)_A$ symmetry, which is the off-diagonal combination of $SU(2)_C \times SU(2)_H$. The 3d $SU(2)_C \times SU(2)_H$ R-symmetry is broken to $SU(2)_C \times U(1)_H$ for nonzero 3d FI parameters and descends down to the 1d R-symmetry of vortex quantum mechanics. We have denoted the (Cartan) generators of 1d $SU(2)_C \times U(1)_H$ R-symmetry by $J_3$ and $R$. The 1d version of the $U(1)_A$ symmetry is thus generated by $R-2 J_3$, whose mass parameter is denoted by $\mu$. Recall that the charges of the 1d multiplets are summarized in table \ref{tab:1dN4}. Each $\mathcal N = 4$ chiral multiplet of $R-2 J_3 = r$ is decomposed into an $\mathcal N = 2$ chiral of $R-2 J_3 = r$ and an $\mathcal N = 2$ fermi of $R-2 J_3 = r-2$. Thus, one can read off the $\mathcal N = 2$ chiral/fermi multiplets charged under $U(1)_A$, which become massive under the deformation $\mu \rightarrow \infty$. After integrating them out, the remaining quantum mechanics is given by figure \ref{fig:def vortex}.

Note that the integrated out 1d multiplets leave their remnants as Wilson lines in quantum mechanics. First, the fundamental and anti-fundamental fermi multiplets induce the following gauge/global Wilson lines:\footnote{\eqref{eq:CS Wilson lines} is the Wilson line value at a saddle point of the localization. One can easily restore their full Lagrangian expressions.}
\begin{align}
\label{eq:CS Wilson lines}
\prod_{l = 1}^L e^{q_l \sum_{i = 1}^{n_l} u^{(l)}_i+\frac{1}{2} n_l \sum_{a = N_{l-1}+1}^{N_l} m_a-\frac{1}{2} n_l \sum_{a = N_l+1}^{N_{l+1}} m_a}
\end{align}
where the gauge Wilson line charge,
\begin{align}
q_l = \frac{N_{l-1}+N_{l+1}}{2}-N_l,
\end{align}
coincides with the CS level $\kappa^{(l)}$ of the 3d theory. Second, the bi-fundamental chiral and fermi multiplets that are integrated out induce the following global Wilson lines:
\begin{align}
\prod_{l = 1}^{L-1} e^{-n_l n_{l+1} \gamma}.
\end{align}
Last, the adjoint chiral and fermi multiplets that are integrated out induce global Wilson lines
\begin{align}
\prod_{l = 1}^L e^{n_{l}^2 \gamma}.
\end{align}
Those gauge/global Wilson lines should be the quantum mechanics counterparts of the CS/BF interactions in the deformed 3d theory. By comparing them with the result from the factorization of the 3d topologically twisted index, which is shown in appendix \ref{sec:fact}, one can identify the 3d origin of each Wilson line. It turns out that the 3d CS/BF interactions of the levels \eqref{eq:def levels} have their Wilson line counterparts in quantum mechanics as follows:
\begin{gather}
\begin{gathered}
\label{eq:Wilson lines}
\prod_{l = 1}^L e^{\kappa^{(l)} \sum_{i = 1}^{n_l} u^{(l)}_i}, \\
\prod_{l = 1}^L e^{\Delta \kappa_{U(1)}^{(l)} (n_l^2 \gamma+n_l \sum_{a = 1}^{N_l} m_a)}, \\
\prod_{l = 1}^{L-1} e^{\kappa_{U(1)}^{(l,l+1)} (2 n_l n_{l+1} \gamma+n_l \sum_{a = 1}^{N_{l+1}} m_a+n_{l+1} \sum_{a = 1}^{N_l} m_a)}
\end{gathered}
\end{gather}
where we have used $\sum_{a = 1}^{N_{L+1}} m_a = 0$, the traceless condition for $SU(N_{L+1})$. Indeed, from the factorization result, we expect that \eqref{eq:Wilson lines} is not restricted to the specific levels in \eqref{eq:def levels} but is generally applicable. Thus, we conclude that a 3d $\mathcal N = 2$ linear quiver gauge theory of figure \ref{fig:def T[SU(N)]} with CS/BF levels $\kappa^{(l)}$, $\Delta \kappa_{U(1)}^{(l)}$ and $\kappa_{U(1)}^{(l,l+1)}$ has vortex quantum mechanics of figure \ref{fig:def vortex} with the Wilson lines \eqref{eq:Wilson lines}. Those Wilson lines will play important roles when we discuss Seiberg-like dualities for $\mathcal N = 2$ linear quiver theories in the next section.
\\

Next we move on to the index of this vortex quantum mechanics. The refined Witten index is given by
\begin{align}
\label{eq:N=2 ind}
I = \frac{1}{|\mathsf W|} \text{JK-Res}_{\eta = \zeta} \left[g(u) d^r u\right]
\end{align}
where $g(u)$ is now written as
\begin{align}
\label{eq:N=2 1-loop}
g(u) &= W(u) \prod_{l = 1}^L \frac{\left(\prod_{i \neq j}^{n_l} \sinh \frac{u^{(l)}_i-u^{(l)}_j}{2}\right) \left(\prod_{i = 1}^{n_{l+1}} \prod_{j = 1}^{n_l} \sinh \frac{u^{(l+1)}_i-u^{(l)}_j-2 \gamma}{2}\right)}{\left(\prod_{i,j = 1}^{n_l} \sinh \frac{u^{(l)}_i-u^{(l)}_j-2 \gamma}{2}\right) \left(\prod_{i = 1}^{n_{l+1}} \prod_{j = 1}^{n_l} \sinh \frac{u^{(l+1)}_i-u^{(l)}_j}{2}\right)} \nonumber \\
&\qquad \times \frac{1}{\left(\prod_{i = 1}^{n_l} \prod_{b = N_{l-1}+1}^{N_l} \sinh \frac{u^{(l)}_i-m_b-\gamma}{2}\right) \left(\prod_{j = 1}^{n_l} \prod_{a = N_l+1}^{N_{l+1}} \sinh \frac{-u^{(l)}_j+m_a-\gamma}{2}\right)}
\end{align}
with $n_{L+1} = 0$. $W(u)$ is the Wilson line contribution given by the product of the three factors in \eqref{eq:Wilson lines}. We adopt the same prescription as in $T_\rho [SU(N)]$ which selects the following poles contributing to the index:
\begin{align}
u^{(l)}_i = m_a+(2 k_a-1) \gamma, \qquad a \in \{1,\ldots,N_l\}, \quad k_a \in \{1,\ldots,n^{(l)}_a\}.
\end{align}
$(n^{(l)}_1,\ldots,n^{(l)}_{N_l})$ is a partition of $n_l$ into $N_l$ nonnegative integers. Again every pair $(a,k_a)$ is assigned to one of $i = 1,\ldots,n_l$ exactly once.
Along the same computations as in $T_\rho [SU(N)]$, the final expression of the index is obtained as follows:
\begin{align}
\label{eq:N=2 result}
I &= W \prod_{l = 1}^L \left(\prod_{a \neq b}^{N_l} \prod_{k = 1}^{n^{(l)}_a-n^{(l)}_b} \sinh \frac{m_a-m_b+2 (k-1) \gamma}{2}\right)^{-1} \nonumber \\
&\qquad \times \left(\prod_{a = 1}^{N_{l+1}} \prod_{b = 1}^{N_l} \prod_{k = 1}^{n^{(l)}_b-n^{(l+1)}_a} \sinh \frac{m_a-m_b-2 k \gamma}{2}\right)^{-1}
\end{align}
where the Wilson line contribution $W$ is given by
\begin{align}
W &= \prod_{l = 1}^L e^{\kappa^{(l)} \sum_{a = 1}^{N_l} (n^{(l)}_a m_a+n^{(l)}_a{}^2 \gamma)} e^{\Delta \kappa_{U(1)}^{(l)} (n_l^2 \gamma+n_l \sum_{a = 1}^{N_l} m_a)} \nonumber \\
&\quad \times e^{\kappa_{U(1)}^{(l,l+1)} (2 n_l n_{l+1} \gamma+n_l \sum_{a = 1}^{N_{l+1}} m_a+n_{l+1} \sum_{a = 1}^{N_l} m_a)}.
\end{align}
We also conduct the numerical computation of \eqref{eq:N=2 ind} using the flag method \cite{Benini:2013xpa} and confirm that it agrees with \eqref{eq:N=2 result}. Indeed, \eqref{eq:N=2 result} agrees with \eqref{eq:vort}, the vortex partition function obtained from the factorization of the 3d topologically twisted index.
\\

We have seen that the index of quantum mechanics in figure \ref{fig:def vortex} gives the vortex partition function of the 3d linear quiver gauge theory in figure \ref{fig:def T[SU(N)]}, which is also obtained from the factorization of a 3d supersymmetric partition function. Introducing 2d parameters $m^\text{2d}_a =  r^{-1} m_a, \gamma^\text{2d} = r^{-1} \gamma$, one can also consider the 2d reduction of the 3d vortex partition function, which is equivalent to \eqref{eq:N=2 result}, by taking the limit $r \rightarrow 0$. We observe that the 2d reduction of \eqref{eq:N=2 result} correctly reproduces the known 2d vortex partition function \cite{Nawata:2014nca}. This is another evidence that quantum mechanics in figure \ref{fig:def vortex} describes vortices of the 3d theory in figure \ref{fig:def T[SU(N)]}. Moreover, one should note that nontrivial Wilson lines are allowed in the quantum mechanics description. We have fine-tuned those Wilson lines so that they correctly reflect CS/BF interactions of the parent 3d theory.

We comment on mathematical aspects of the world volume theory of vortices. The type of a quiver in figure \ref{fig:def vortex} is called a `hand-saw' quiver, which is isomorphic
to a parabolic Laumon space \cite{Finkelberg2014}. The parabolic Laumon space  coincides with  the moduli space of based quasi maps $\mathbf P^1$ into the flag variety.
The precise relation between quasimaps and the moduli space of vortex equation was studied in \cite{Venugopalan2013}.
The equivariant integrations over the based quasi map spaces give the equivariant J-function of the flag variety, which is the 2d reduction of  vortex partition function. Then
 our construction of vortex quantum mechanics is regarded as K-theoretic uplift.\footnote{The relation between K-theoretic J-function \cite{Givental2003} and the vortex partition function in three dimensions  was first pointed out in \cite{Dimofte:2010tz}.}  The index of vortex quantum mechanics with a particular choice of Wilson lines  reproduces the K-theoretic J-function.
\\

\section{Vortices and Seiberg-like dualities}
\label{sec:SD}
\subsection{$\mathcal N = 2$ SQCDs}
\label{sec:Aharony}

We have constructed 1d quantum mechanical systems which describe the low energy dynamics of vortices in 3d linear quiver theories. The moduli space of vortices is given by the Higgs branch of such vortex quantum mechanics. We have also computed the refined Witten indices of vortex quantum mechanics, which can be identified as the partition functions of vortices on $\Omega$-deformed $\mathbb R^2_\Omega \times S^1$. In this section, using these vortex partition functions, we examine how vortex quantum mechanics behave under 3d Seiberg-like dualities we reviewed in section \ref{sec:3d}.
\\

The example we discuss in this section is the $\mathcal N = 2$ $U(N_c)_\kappa$ gauge theory with $N_f$ fundamental chiral multiplets and $N_a$ anti-fundamental chiral multiplets. We include the Chern-Simons interaction with level $|\kappa| \leq \frac{|N_f-N_a|}{2}$ and FI parameter $\xi$. We assume $N_f \geq N_a$ and $\xi > 0$. The ranges of $\kappa$ and $\xi$ are restricted such that the theory only has Higgs vacua and avoids topological vacua as we discuss in section \ref{sec:3d}. The original Aharony duality and its generalizations tell us that this theory has a dual description, $U(N_f-N_c)_{-\kappa}$ gauge theory with $(N_a,N_f)$ flavors and extra gauge singlet fields described in section \ref{sec:3d} \cite{Aharony:1997gp,Benini:2011mf}.

\begin{figure}[tbp]
\centering 
\includegraphics[height=.3\textheight]{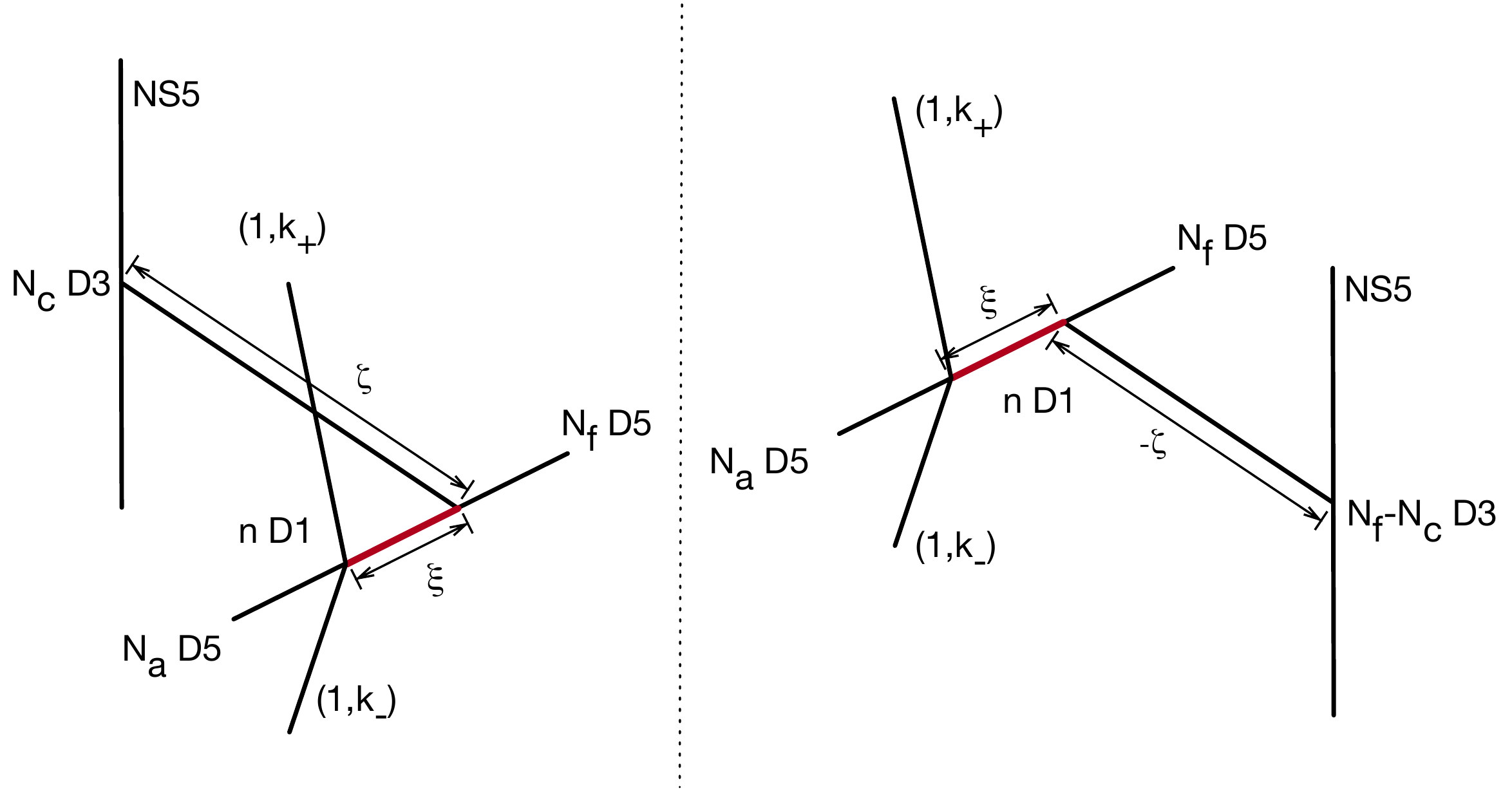}
\caption{\label{fig:Aharony duality} The brane motion representing the Aharony duality in the presence of vortices. Vortices are realized as D1-branes denoted by the red line in the figure. The length of the D1-branes corresponds to 3d FI parameter $\xi$ while the length of the D3-branes corresponds to 1d FI parameter $\zeta$. The spacetime directions occupied by each brane are summarized in table \ref{tab:Aharony branes}.}
\end{figure}

In order to understand the effect of the Aharony duality on vortex quantum mechanics, we first consult the brane picture. Recall the brane setup and the motion representing the Aharony duality, which are illustrated in figure \ref{fig:AD}. Now we insert additional D1-branes ending on D3-branes, which correspond to the presence of vortices in the 3d theory. The brane motion in the presence of D1-branes is illustrated in figure \ref{fig:Aharony duality}.
\begin{table}[tbp]
\centering
\begin{tabular}{|c|cccccccccc|}
\hline
Branes & 0 & 1 & 2 & 3 & 4 & 5 & 6 & 7 & 8 & 9 \\
\hline
NS5 & $\times$ & $\times$ & $\times$ & $\times$ & $\times$ & $\times$ & & & & \\
$(1,k_\pm)$ & $\times$ & $\times$ & $\times$ & & & $\cdot$ & $\cdot$ & $\times$ & $\times$ & \\
D5 & $\times$ & $\times$ & $\times$ & & & & $\times$ & $\times$ & $\times$ & \\
D3 & $\times$ & $\times$ & $\times$ & & & & & & & $\times$ \\
D1 & $\times$ & & & & & & $\times$ & & & \\
\hline
\end{tabular}
\caption{\label{tab:Aharony branes} The spacetime directions occupied by the branes in figure \ref{fig:Aharony duality} are marked by $\times$. The $(1,k_\pm)$-branes occupy 1-dimensional subspaces in the 56-plane.}
\end{table}
One can see that the brane motion is controlled by the relative distance between the NS5-brane and the $(1,k_\pm)$-branes along the 9-direction. In the 3d theory point of view, this distance is proportional to the inverse of the gauge coupling squared, $1/g^2$. The position that the NS5-brane and the $(1,k_\pm)$-branes are exchanged is therefore the infinite-coupling point, which is consistent with the fact the Aharony duality is an IR duality where 3d theories strongly interact.

On the other hand, in the vortex quantum mechanics point of view, that distance corresponds to Fayet-Iliopoulos parameter $\zeta$. Especially, the position exchanging the NS5-brane and the $(1,k_\pm)$-branes corresponds to $\zeta = 0$ where a non-compact Coulomb branch of the quantum mechanics can appear depending on the values of $k_\pm = \kappa \pm \frac{N_f-N_a}{2}$. In figure \ref{fig:Aharony duality}, the NS5-brane and the $(1,k_\pm)$-branes have the common 9-direction coordinate when $\zeta = 0$. Thus, D1-branes can be suspended between them. When either $k_\pm = 0$, we have two NS5-branes sharing a (semi-)infinite parallel direction, which allows the D1-branes to move along that direction. Thus, the D1-brane theory has a flat direction at $\zeta = 0$, which we call a Coulomb branch. Due to the appearance of the flat direction, some states of vortex quantum mechanics can escape through the flat direction such that a jump of the spectrum can happen at $\zeta = 0$. This phenomenon is called the wall-crossing. The important point is that the wall-crossing of vortex quantum mechanics and the Aharony duality of the 3d theory are inferred from the same brane motion.
\\

\begin{figure}[tbp]
\centering 
\includegraphics[height=.2\textheight]{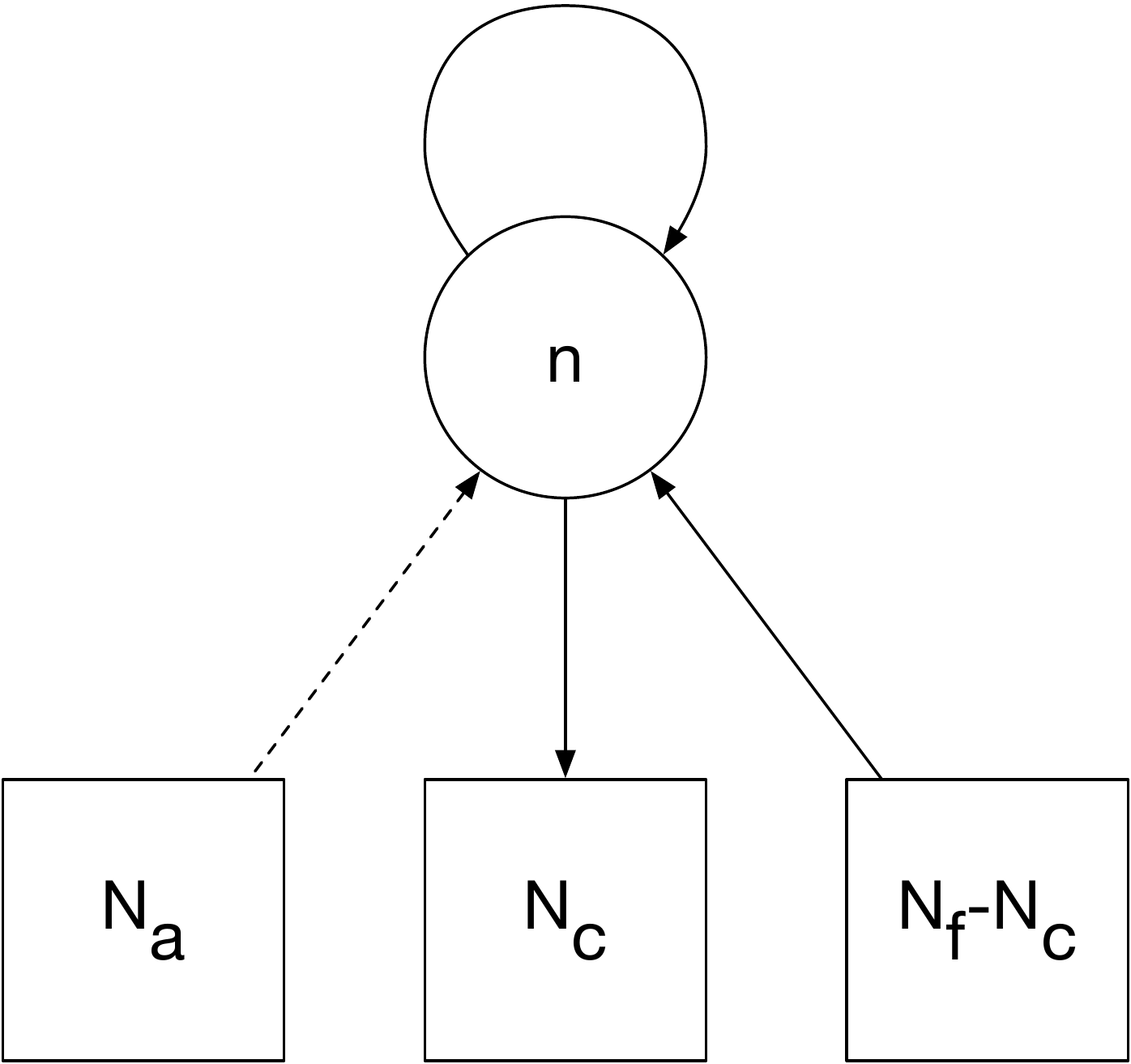}
\caption{\label{fig:Aharony vortex} The quiver diagram representation of vortex quantum mechanics for the 3d $\mathcal N = 2$ $U(N_c)_\kappa$ theory with $(N_f,N_a)$ flavors. A solid arrow denotes a 1d $\mathcal N = 2$ chiral multiplet while a dashed arrow denotes a 1d $\mathcal N = 2$ fermi multiplet.}
\end{figure}

From the brane picture, we expect that vortex quantum mechanics experiences the shift of the FI parameter from $\zeta > 0$ to $\zeta < 0$, and possibly the nontrivial wall-crossing at $\zeta = 0$, under the Seiberg-like duality of the parent 3d theory. We now validate this expectation by the explicit computations of the quantum mechanics indices for different 1d FI parameters. For the 3d $\mathcal N = 2$ $U(N_c)_\kappa$ theory with $(N_f,N_a)$ flavors, the moduli space of $n$ vortices is described by the $\mathcal N = 2$ gauged quantum mechanics illustrated in figure \ref{fig:Aharony vortex} \cite{Fujitsuka:2013fga}.
The refined Witten index of this quantum mechanics is given by
\begin{align}
\label{eq:Aharony ind}
I^n = \frac{1}{|\mathsf W|} \text{JK-Res}_{\eta = \zeta} \left[g^n(u) d^n u\right]
\end{align}
where $|\mathsf W| = n!$ is the Weyl group order of the gauge group $U(n)$ and
\begin{align}
\label{eq:Aharony 1-loop}
& g^n(u) = \nonumber \\
& \frac{e^{\kappa \sum_{i = 1}^n u_i} \left(\prod_{i \neq j}^n \sinh \frac{u_i-u_j}{2}\right) \left(\prod_{j = 1}^n \prod_{a = 1}^{N_a} \sinh \frac{-u_j+\tilde m_a-\mu+\gamma}{2}\right)}{\left(\prod_{i,j = 1}^n \sinh \frac{u_i-u_j-2 \gamma}{2}\right) \left(\prod_{i = 1}^n \prod_{b = 1}^{N_c} \sinh \frac{u_i-m_b-\mu-\gamma}{2}\right) \left(\prod_{j = 1}^n \prod_{a = N_c+1}^{N_f} \sinh \frac{-u_j+m_a+\mu-\gamma}{2}\right)}
\end{align}
is the integrand given by the classical action and the 1-loop determinant. We have made shifts of mass parameters $m \rightarrow m+\mu$, $\tilde m \rightarrow \tilde m-\mu$ where $\mu$ is associated with the $U(1)_A$ symmetry rotating the 3d fundamental and anti-fundamental fields simultaneously. Note that we have chosen the auxiliary JK-vector $\eta = \zeta$ such that asymptotic poles do not participate.\footnote{Again the meaning of ``$\eta = \zeta$'' is that $\eta$ is generic but belongs to the same chamber as $(\zeta_1 \vec 1_{N_1},\ldots,\zeta_L \vec 1_{N_L})$ in the charge space.} We want to compare the indices of this vortex quantum mechanics for different FI parameters: $\zeta > 0$ and $\zeta < 0$.

Let us consider the $\zeta > 0$ case first. The JK-residue rule chooses sets of linearly independent hyperplanes in $\mathbb R^n$ such that a chosen set of hyperplanes determine each $u_i$ as follows:
\begin{align}
u_i = \left\{\begin{array}{l}
m_a+\mu+\gamma, \qquad a \in \{1,\ldots,N_c\} \\
u_j+2 \gamma.
\end{array}\right.
\end{align}
The pole determined by the intersection of those hyperplanes is given by
\begin{align}
\label{eq:Aharony pole}
u_i = m_a+\mu+(2 k_a-1) \gamma, \qquad a \in \{1,\ldots,N_c\}, \quad k_a \in \{1,\ldots,n_a\}
\end{align}
where $(n_1,\ldots,n_{N_c})$ is a partition of $n$, i.e., an ordered set of $N_c$ nonnegative integers satisfying $\sum_{a = 1}^{N_c} n_a = n$. Every pair $(a,k_a)$ is assigned to one of $i = 1,\ldots,n$ exactly once. Evaluating the residue at this pole, we have
\begin{align}
\label{eq:Aharony res}
&\frac{e^{\kappa \sum_{a = 1}^{N_c} \sum_{k_a = 1}^{n_a} (m_a+\mu+(2 k_a-1) \gamma)} \left(\prod_{a,b = 1}^{N_c} \prod_{k_a = 1}^{n_a} \prod_{k_b = 1}^{n_b} \sinh \frac{m_a-m_b+2 (k_a-k_b) \gamma}{2}\right)'}{\left(\prod_{a,b = 1}^{N_c} \prod_{k_a = 1}^{n_a} \prod_{k_b = 1}^{n_b} \sinh \frac{m_a-m_b+2 (k_a-k_b-1) \gamma}{2}\right)' \left(\prod_{a = 1}^{N_c} \prod_{k_a = 1}^{n_a} \prod_{b = 1}^{N_c} \sinh \frac{m_a-m_b+2 (k_a-1) \gamma}{2}\right)'} \nonumber \\
&\times \frac{\left(\prod_{a = 1}^{N_a} \prod_{b = 1}^{N_c} \prod_{b = 1}^{n_b} \sinh \frac{\tilde m_a-m_b-2 \mu-2 (k_b-1) \gamma}{2}\right)}{\left(\prod_{a = N_c+1}^{N_f} \prod_{b = 1}^{N_c} \prod_{k_b = 1}^{n_b} \sinh \frac{m_a-m_b-2 k_b \gamma}{2}\right)}
\end{align}
where $'$ denotes that the vanishing factors are omitted. The permutations among $u_i$'s give rise to factor $n!$, which cancels the Weyl group factor $|\mathsf W|$. The first line of \eqref{eq:Aharony res} is simplified to
\begin{align}
\frac{(-1)^{n N_c} e^{\kappa \sum_{a = 1}^{N_c}(n_a m_a+n_a \mu+n_a^2 \gamma)}}{\left(\prod_{a,b = 1}^{N_c} \prod_{k_b = 1}^{n_b} \sinh \frac{m_a-m_b-2 (k_b-n_a-1) \gamma}{2}\right)}.
\end{align}
Combined with the second line of \eqref{eq:Aharony res}, it reproduces the known vortex partition function of 3d $\mathcal N = 2$ $U(N_c)_\kappa$ theory with $(N_f,N_a)$ flavors on $\Omega$-deformed $\mathbb R_\Omega^2 \times S^1$ \cite{Hwang:2012jh}:
\begin{align}
\label{eq:Aharony vort}
\frac{(-1)^{n N_c} e^{\kappa \sum_{a = 1}^{N_c}(n_a m_a+n_a \mu+n_a^2 \gamma)} \left(\prod_{a = 1}^{N_a} \prod_{b = 1}^{N_c} \prod_{k_b = 1}^{n_b} \sinh \frac{\tilde m_a-m_b-2 \mu-2 (k_b-1) \gamma}{2}\right)}{\left(\prod_{a,b = 1}^{N_c} \prod_{k_b = 1}^{n_b} \sinh \frac{m_a-m_b-2 (k_b-n_a-1) \gamma}{2}\right) \left(\prod_{a = N_c+1}^{N_f} \prod_{b = 1}^{N_c} \prod_{k_b = 1}^{n_b} \sinh \frac{m_a-m_b-2 k_b \gamma}{2}\right)}
\end{align}
up to sign, which can be absorbed to the vorticity fugacity.

Next let us examine the $\zeta < 0$ case. Since different $\eta$ is used, different sets of hyperplanes are chosen by the JK-residue rule. Now a set of hyperplanes chosen by the JK-residue rule determine each $u_i$ as follows:
\begin{align}
\label{eq:hyperplanes+}
u_i = \left\{\begin{array}{l}
m_a+\mu-\gamma, \qquad a \in \{N_c+1,\ldots,N_f\} \\
u_j-2 \gamma.
\end{array}\right.
\end{align}
Thus, we evaluate the residue at pole
\begin{align}
u_i = m_a+\mu-(2 k_a-1) \gamma, \qquad \begin{array}{l}
a \in \{N_c+1,\ldots,N_f\}, \\
k_a \in \{1,\ldots,n_a\}, \\
n_a \geq 0, \quad \sum_{a = N_c+1}^{N_f} n_a = n
\end{array}
\end{align}
and obtain the following result:
\begin{align}
& (-1)^{n (N_c-N_f+N_a)} e^{-\kappa \sum_{a = N_c+1}^{N_f} \sum_{k_a = 1}^{n_a} (-n_a m_a-n_a \mu+n_a^2 \gamma)} \times \nonumber \\
& \frac{\left(\prod_{a = 1}^{N_a} \prod_{b = N_c+1}^{N_f} \prod_{k_b = 1}^{n_b} \sinh \frac{-\tilde m_a+m_b+2 \mu-2 k_b \gamma}{2}\right)}{\left(\prod_{a,b = N_c+1}^{N_f} \prod_{k_b = 1}^{n_b} \sinh \frac{-m_a+m_b-2 (k_b-n_a-1) \gamma}{2}\right) \left(\prod_{a = N_c+1}^{N_f} \prod_{k_a = 1}^{n_a} \prod_{b = 1}^{N_c} \sinh \frac{-m_b+m_a-2 k_a \gamma}{2}\right)},
\end{align}
which, up to sign, is the vortex partition function of the dual $U(N_f-N_c)_{-\kappa}$ theory with $(N_a,N_f)$ flavors \cite{Hwang:2012jh,Hwang:2015wna}. Thus, by the explicit computations of the quantum mechanics indices for different $\zeta$, we have shown that the Aharony duality of a 3d $\mathcal N = 2$ SQCD corresponds to the sign flip of the FI parameter in vortex quantum mechanics.
\\

As discussed at the beginning of the section, the shift of the FI parameter from $\zeta > 0 $ to $\zeta < 0$ may accompany a nontrivial jump of the spectrum at $\zeta = 0$, which is called wall-crossing, depending on $N_f$, $N_a$ and $\kappa$. In the context of the JK-residue, such a jump of the index can happen if we have nontrivial residue contributions from asymptotic regions. The existence of an asymptotic pole is a signal of a non-compact Coulomb branch. From \eqref{eq:Aharony 1-loop}, one can find a necessary condition for nontrivial residues at asymptotic regions by taking one $u_i$ very large, $u_i \rightarrow \pm \infty$:
\begin{align}
g^n(u) \sim e^{\left(\kappa \pm \frac{N_a-N_f}{2}\right) u_i}.
\end{align}
This should not vanish in order to have nontrivial residues at asymptotic regions. A necessary condition is thus
\begin{align}
\pm (\kappa \pm \frac{N_a-N_f}{2}) = \pm \kappa-\frac{N_f-N_a}{2} \geq 0.
\end{align}
Since we only allow $|\kappa| \leq \frac{|N_f-N_a|}{2}$, a relevant condition is the following:
\begin{align} \label{eq:cond}
\pm \kappa-\frac{N_f-N_a}{2} = 0.
\end{align}
This is the same condition that the 3d theory has non-compact Coulomb branches \cite{Benini:2011mf}. In the brane picture, this condition implies that there is an infinite parallel direction shared by two NS5-branes, along which D3-branes or D1-branes can move. These moduli of D3-branes and D1-branes are exactly their non-compact Coulomb branches. Therefore, there can be nontrivial wall-crossing of vortex quantum mechanics if the 3d theory has a non-compact Coulomb branch.

Now we should ask if this necessary condition is also sufficient. We show that it is the case by the explicit computation of the wall-crossing spectrum. Let us consider the 1-vortex case first. Recall that the vortex quantum mechanics index is given by \eqref{eq:Aharony ind}. Since the theory is now a rank-1 theory,
\begin{align}
I_{\zeta > 0} = \text{JK-Res}_{\eta = \zeta} \left[g(u) du\right] = \sum_{Q(u*) > 0} \mathrm{Res}_{u = u*} \left[g(u) du\right]
\end{align}
where the residues are summed over the poles whose corresponding charges are positive. Indeed, the JK-residue is independent of the choice of $\eta$. Thus, one can also take $\eta = -\zeta$, in which case asymptotic poles also contribute:
\begin{align}
I_{\zeta > 0} &= \text{JK-Res}_{\eta = -\zeta} \left[g(u) du\right] \nonumber \\
&= -\sum_{Q(u*) < 0} \mathrm{Res}_{u = u*} \left[g(u) du\right]-\mathrm{Res}_{u = \pm \infty} \left[g(u) du\right].
\end{align}
Note that the first term in the last line is nothing but the index with $\zeta < 0$. Thus, as discussed, there is a jump between $I_{\zeta > 0}$ and $I_{\zeta < 0}$ if we have the nonzero asymptotic residue contribution. From \eqref{eq:cond} the asymptotic poles are simple if they exist. One can compute their residues as follows:
\begin{align}
-\mathrm{Res}_{u = \pm \infty} \left[g(u) du\right] &= \lim_{u \rightarrow \infty} g(u)-\lim_{u \rightarrow -\infty} g(u) \\
&= \delta_{2 \kappa,N_f-N_a} \frac{(-1)^{N_a-N_f+N_c} e^{\frac{N_f+N_a}{2} \mu+(N_c-\frac{N_f+N_a}{2}) \gamma}}{\sinh(-\gamma)} \nonumber \\
&\quad -\delta_{2 \kappa,N_a-N_f} \frac{(-1)^{-N_c} e^{-\frac{N_f+N_a}{2} \mu-(N_c-\frac{N_f+N_a}{2}) \gamma}}{\sinh(-\gamma)}. \label{eq:1-wc}
\end{align}
It shows that there is the nontrivial wall-crossing if and only if the condition \eqref{eq:cond} is met. Also we emphasize that \eqref{eq:1-wc} agrees with the 1-particle BPS index of $V_\pm$ \cite{Hwang:2012jh,Hwang:2015wna},\footnote{\eqref{eq:1-wc} is the index of the sector having positive $U(1)_T$ charges where $U(1)_T$ oppositely rotates $V_+$ and $V_-$. The other sector of negative $U(1)_T$ charges is captured by anti-vortices.} which are extra neutral chiral fields on the dual side describing the Coulomb branches of the 3d theory. We will see shortly that the whole wall-crossing factor incorporating multi-vortices is given by the Plethystic exponential of \eqref{eq:1-wc}.

Now let us move on to multi-vortices cases. We first consider the following case:
\begin{align}
\kappa = \frac{N_f-N_a}{2} \neq 0.
\end{align}
There is a pole at $u_i \rightarrow \infty$ and no pole at $u_i \rightarrow -\infty$. Recall that the poles chosen by the JK-residue rule with $\eta = \zeta > 0$ are given by \eqref{eq:Aharony pole}. One can see that those poles are exactly the poles contributing to the contour integral with the unit circle contour. See \cite{Hwang:2015wna} for example. In other words, the JK-residue \eqref{eq:Aharony ind} can be rewritten in the following way:
\begin{align}
\label{eq:unit circle}
I^n_{\zeta > 0} = \frac{1}{|\mathsf W|} \text{JK-Res}_{\eta = \zeta} \left[g^n(u) d^n u\right] = \frac{1}{|\mathsf W|} \oint_{|z_i| = 1} \frac{d^n z}{\prod_{i = 1}^n z_i} g^n(\log z)
\end{align}
where $z_i = e^{u_i}$ and we assume that $\mathrm{Re} (\mu) = \mathrm{Re} (m_a) = 0$ while $\mathrm{Re} (\gamma) < 0$. The contour is taken to be the unit circle traversed counterclockwise. One can check equation \eqref{eq:unit circle} by applying the residue theorem and taking the residues from the inside of the unit circle. On the other hand, one can also evaluate the same integral by taking the residues from the outside of the unit circle. In that case, a contributing pole is determined by a set of hyperplanes:
\begin{align}
\label{eq:hyperplanes0}
z_i &= \left\{\begin{array}{l}
t_a \tau x^{-1}, \\
\infty, \\
z_j x^{-2}
\end{array}\right.
\end{align}
where $t_a = e^{m_a}$, $\tau = e^\mu$ and $x = e^\gamma$.

At the pole, each $z_i$ takes either a finite value or an asymptotic value. One can decompose the hyperplanes \eqref{eq:hyperplanes0} into two sets:
\begin{align}
\{H_1,\ldots,H_m\} \cup \{H_{m+1},\ldots,H_n\}
\end{align}
such that $\{H_1,\ldots,H_k\}$ determines a set of $z_i$'s who take asymptotic values at the pole while $\{H_{k+1},\ldots,H_n\}$ determines the other set of $z_i$'s who take finite values at the pole. Let us define $I$ and $J$, two sets of gauge indices, such that $z_i$ is determined by $\{H_1,\ldots,H_m\}$ if $i \in I$ and is determined by $\{H_{m+1},\ldots,H_n\}$ if $i \in J$. Now we decompose the integrand into three parts:
\begin{align}
& g^n(u) = \nonumber \\
& \frac{e^{\kappa \sum_{i \in I} u_i} \left(\prod_{i \neq j \in I} \sinh \frac{u_i-u_j}{2}\right) \left(\prod_{j \in I} \prod_{a = 1}^{N_a} \sinh \frac{-u_j+\tilde m_a-\mu+\gamma}{2}\right)}{\left(\prod_{i,j \in I} \sinh \frac{u_i-u_j-2 \gamma}{2}\right) \left(\prod_{i \in I} \prod_{b = 1}^{N_c} \sinh \frac{u_i-m_b-\mu-\gamma}{2}\right) \left(\prod_{j \in I} \prod_{a = N_c+1}^{N_f} \sinh \frac{-u_j+m_a+\mu-\gamma}{2}\right)} \nonumber \\
& \times \frac{\left(\prod_{i \in I} \prod_{j \in J} \sinh \frac{u_i-u_j}{2} \sinh \frac{u_j-u_i}{2}\right)}{\left(\prod_{i \in I} \prod_{j \in J} \sinh \frac{u_i-u_j-2 \gamma}{2} \sinh \frac{u_j-u_i-2 \gamma}{2}\right)} \nonumber \\
& \times \frac{e^{\kappa \sum_{i \in J} u_i} \left(\prod_{i \neq j \in J} \sinh \frac{u_i-u_j}{2}\right) \left(\prod_{j \in J} \prod_{a = 1}^{N_a} \sinh \frac{-u_j+\tilde m_a-\mu+\gamma}{2}\right)}{\left(\prod_{i,j \in J} \sinh \frac{u_i-u_j-2 \gamma}{2}\right) \left(\prod_{i \in J} \prod_{b = 1}^{N_c} \sinh \frac{u_i-m_b-\mu-\gamma}{2}\right) \left(\prod_{j \in J} \prod_{a = N_c+1}^{N_f} \sinh \frac{-u_j+m_a+\mu-\gamma}{2}\right)}
\end{align}
where the first line is only determined by $\{H_1,\ldots,H_m\}$, the second line is determined by $\{H_1,\ldots,H_m\}$ and $\{H_{m+1},\ldots,H_n\}$, and the third line is only determined by $\{H_{m+1},\ldots,H_n\}$. One should note that $u_i$ will go to infinity for $i \in I$. Under this limit, the first line becomes
\begin{align}
(-1)^{m (N_a-N_f+N_c)} e^{\frac{m}{2} [(N_f+N_a) \mu+(2 N_c-N_f-N_a) \gamma]} \frac{\left(\prod_{i \neq j \in I} \sinh \frac{u_i-u_j}{2}\right)}{\left(\prod_{i,j \in I} \sinh \frac{u_i-u_j-2 \gamma}{2}\right)}
\end{align}
while the second line becomes 1. We have used the condition $\kappa-\frac{N_f-N_a}{2} = 0$. Thus, the residue can be written in the following simple way:
\begin{align}
I^n_{H_1,\ldots,H_n} &= (-1)^{n+m (N_a-N_f+N_c)} e^{\frac{m}{2} [(N_f+N_a) \mu+(2 N_c-N_f-N_a) \gamma]} \nonumber \\
&\quad \times \mathrm{Res}_{H_1,\ldots,H_m} \left[\frac{\left(\prod_{i \neq j \in I} \sinh \frac{u_i-u_j}{2}\right)}{\left(\prod_{i,j \in I} \sinh \frac{u_i-u_j-2 \gamma}{2}\right)} d^m u\right] \nonumber \\
&\quad \times \mathrm{Res}_{H_{m+1},\ldots,H_n} \left[g^J (u) d^{n-m} u\right]
\end{align}
where
\begin{align}
& g^J (u) = \nonumber \\
& \frac{e^{\kappa \sum_{i \in J} u_i} \left(\prod_{i \neq j \in J} \sinh \frac{u_i-u_j}{2}\right) \left(\prod_{j \in J} \prod_{a = 1}^{N_a} \sinh \frac{-u_j+\tilde m_a-\mu+\gamma}{2}\right)}{\left(\prod_{i,j \in J} \sinh \frac{u_i-u_j-2 \gamma}{2}\right) \left(\prod_{i \in J} \prod_{b = 1}^{N_c} \sinh \frac{u_i-m_b-\mu-\gamma}{2}\right) \left(\prod_{j \in J} \prod_{a = N_c+1}^{N_f} \sinh \frac{-u_j+m_a+\mu-\gamma}{2}\right)}.
\end{align}

The complete index is given by the sum over all possible $\{H_1,\ldots,H_n\}$. Using the permutation symmetries among $z_i$'s, one can fix $I = \{1,\ldots,m\},J = \{m+1,\ldots,n\}$ and multiplies factor ${}_m C_n$. The index is then given by
\begin{align}
\label{eq:ind+asymp}
& \frac{w^n}{n!} \sum_{H_1,\ldots,H_n} I^n_{H_1,\ldots,H_n} \nonumber \\
&= \frac{1}{n!} \sum_{m = 0}^n {}_m C_n \times m! \times (n-m)! \times (-1)^{m (N_a-N_f+N_c)} e^{\frac{m}{2} [(N_f+N_a) \mu+(2 N_c-N_f-N_a) \gamma]} \nonumber \\
&\quad \times \left(\frac{(-w)^m}{m!} \sum_{\cap_{i = 1}^m H_i \in \text{asymp}^+} \mathrm{Res}_{H_1,\ldots,H_m} \left[\frac{\left(\prod_{i \neq j}^m \sinh \frac{u_i-u_j}{2}\right)}{\left(\prod_{i,j = 1}^m \sinh \frac{u_i-u_j-2 \gamma}{2}\right)} d^m u\right]\right) \nonumber \\
&\quad \times \left(\frac{(-w)^{n-m}}{(n-m)!} \sum_{\cap_{i = m+1}^n H_i \in \text{bulk}} \mathrm{Res}_{H_{m+1},\ldots,H_n} [g_{n-m}(u) d^{n-m} u]\right)
\end{align}
where hyperplanes $H_1,\ldots,H_m$ are chosen among
\begin{align}
z_i &= \left\{\begin{array}{l}
\infty, \\
z_j x^{-2}
\end{array}\right.
\end{align}
while hyperplanes $H_{m+1},\ldots,H_n$ are chosen among
\begin{align}
\label{eq:hyperplanes0+}
z_i &= \left\{\begin{array}{l}
t_a \tau x^{-1}, \\
z_j x^{-2}.
\end{array}\right.
\end{align}
Note that the last line of \eqref{eq:ind+asymp} is nothing but the index of $n-m$ vortices with $\zeta < 0$. Negative $\zeta$ is used because the contributing poles are determined by \eqref{eq:hyperplanes0+}, which is equivalent to \eqref{eq:hyperplanes+}.

The remaining thing is to compute the residues in the third line (the second line on the right hand side). One way to compute it is using the following equation:
\begin{align}
\label{eq:asymp ident}
\frac{(-w)^m}{m!} \sum_{\cap_{i = 1}^m H_i \in \text{asymp}^+} \mathrm{Res}_{H_1,\ldots,H_m} \left[\frac{\left(\prod_{i \neq j}^m \sinh \frac{u_i-u_j}{2}\right)}{\left(\prod_{i,j = 1}^m \sinh \frac{u_i-u_j-2 \gamma}{2}\right)} d^m u\right] = \frac{w^m}{m!} \oint_{|z| = 1} \frac{d^m z}{\prod_{i = 1}^m z_i} g'(\log z)
\end{align}
where
\begin{align}
g'(u) = \frac{e^{\frac{1}{2} \sum_{i = 1}^m (u_i-\gamma)} \left(\prod_{i \neq j}^m \sinh \frac{u_i-u_j}{2}\right)}{\left(\prod_{i,j = 1}^m \sinh \frac{u_i-u_j-2 \gamma}{2}\right) \left(\prod_{i = 1}^m \sinh \frac{u_i-\gamma}{2}\right)}.
\end{align}
One can check equation \eqref{eq:asymp ident} by taking the residues outside the unit circle on the right hand side. On the other hand, one can also evaluate the right hand side by taking the residues inside the unit circle. In that case the contributing poles are determined as follows:
\begin{align}
u_i = (2 k-1) \gamma, \qquad k = 1,\ldots,m.
\end{align}
Note that there is no pole at $u_i \rightarrow -\infty$. The integral is thus evaluated as follows:
\begin{align}
\frac{w^m}{m!} \oint_{|z| = 1} \frac{d^m z}{\prod_{i = 1}^m z_i} g'(\log z) &= \frac{w^m e^{\frac{m (m-1)}{2} \gamma} \left(\prod_{k \neq l}^m \sinh \frac{2 (k-l) \gamma}{2}\right)}{\left(\prod_{k,l = 1}^m \sinh \frac{2 (k-l-1) \gamma}{2}\right)' \left(\prod_{k = 2}^m \sinh \frac{2 (k-1) \gamma}{2}\right)} \\
&= \frac{(-w x^{-1})^m}{(x^{-2};x^{-2})_m}.
\end{align}
Substituting this result into \eqref{eq:ind+asymp} and summing over $n \geq 0$, we have
\begin{align}
& \sum_{n = 0}^\infty \frac{w^n}{n!} \sum_{H_1,\ldots,H_n} I^n_{H_1,\ldots,H_n} \nonumber \\
&= \left(\sum_{m = 0}^\infty \frac{((-1)^{N_a-N_f+N_c+1} w \tau^{\frac{N_f+N_a}{2}} x^{N_c-\frac{N_f+N_a}{2}-1})^m}{(x^{-2};x^{-2})_m}\right) \left(\sum_{n = 0}^\infty w^n I^n_{\zeta < 0}\right) \\
&= \mathrm{PE}\left[-\frac{(-1)^{N_a-N_f+N_c+1} w \tau^{\frac{N_f+N_a}{2}} x^{N_c-\frac{N_f+N_a}{2}+1}}{1-x^2}\right] \times \left(\sum_{n = 0}^\infty w^n I^n_{\zeta < 0}\right) \label{eq:wc+}
\end{align}
where we have used $q$-binomial theorem:
\begin{align}
\sum_{n = 0}^\infty \frac{(a;q)_n}{(q;q)_n} z^n = \frac{(a z;q)_\infty}{(z;q)_\infty}.
\end{align}

Now it is clear that the wall-crossing part for each vortex number is organized such that $Z^\text{vort}_{\zeta > 0} \equiv \sum_{n = 0}^\infty w^n I^n_{\zeta > 0}$ factorizes into two parts: $Z^\text{vort}_{\zeta < 0}$ and $Z^\text{wall}$ where the wall-crossing factor $Z^\text{wall}$ is defined as follows:
\begin{align}
\label{eq:wc}
Z^\text{wall} = \mathrm{PE}\left[-\frac{(-1)^{N_a-N_f+N_c+1} w \tau^{\frac{N_f+N_a}{2}} x^{N_c-\frac{N_f+N_a}{2}+1}}{1-x^2}\right].
\end{align}
One should note that this is the same as the BPS index of $V_-$ \cite{Hwang:2012jh,Hwang:2015wna}, which is a neutral chiral field appearing in the dual 3d theory when $\kappa = \frac{N_f-N_a}{2}$:
\begin{align}
\label{eq:V-}
Z_{V_-} = \mathrm{PE} \left[-\frac{\mathsf w \tau^{-A} x^{2-R}}{1-x^2}\right]
\end{align}
where $\mathsf w = (-1)^{N_a-N_f+N_c+1} w$ is the $U(1)_T$ fugacity. $A = -\frac{N_f+N_a}{2}$ and $R = \frac{N_f+N_a}{2}-N_c+1$ are the $U(1)_A$ charge and the $U(1)_R$ charge of $V_-$. Recall that the 3d theory has a Coulomb branch if $\kappa = \frac{N_f-N_a}{2}$. $V_-$ is exactly the operator parameterizing this Coulomb branch. \eqref{eq:wc} shows that the wall-crossing of vortex quantum mechanics captures the information of the Coulomb branch of the 3d theory.

A similar computation can be done for another case:
\begin{align}
\kappa = -\frac{N_f-N_a}{2} \neq 0.
\end{align}
In this case, there is no pole at $u_i \rightarrow \infty$ while there is a pole at $u_i \rightarrow -\infty$. Thus, the roles of $I^n_{\zeta > 0}$ and $I^n_{\zeta < 0}$ are exchanged. The result is as follows:
\begin{align}
\label{eq:wc-}
\sum_{n = 0}^\infty w^n I^n_{\zeta > 0}
&= \mathrm{PE}\left[\frac{(-1)^{N_c+1} w \tau^{-\frac{N_f+N_a}{2}} x^{-N_c+\frac{N_f+N_a}{2}+1}}{1-x^2}\right] \times \left(\sum_{n = 0}^\infty w^n I^n_{\zeta < 0}\right).
\end{align}
Again we observe that the wall-crossing factor is exactly the BPS index of $V_+$ \cite{Hwang:2012jh,Hwang:2015wna}, which is a neutral chiral field appearing in the dual 3d theory when $\kappa = -\frac{N_f-N_a}{2}$:
\begin{align}
\label{eq:V+}
Z_{V_+} = \mathrm{PE} \left[\frac{\mathsf w \tau^A x^R}{1-x^2}\right]
\end{align}
where $\mathsf w = (-1)^{N_c+1} w$, $A = -\frac{N_f+N_a}{2}$ and $R = \frac{N_f+N_a}{2}-N_c+1$.

For the last case: $\kappa = N_f-N_a = 0$, some care is required because there are poles both at $u_i \rightarrow \infty$ and at $u_i \rightarrow -\infty$. Using similar arguments above we show that
\begin{align}
\sum_{n = 0}^\infty \frac{w^n}{|\mathsf W|} \oint_{|z| = 1} \frac{d^n z}{z} g(\log z) &= \mathrm{PE}\left[-\frac{(-1)^{N_c+1} w \tau^{-\frac{N_f+N_a}{2}} x^{-N_c+\frac{N_f+N_a}{2}+1}}{1-x^2}\right] \times \left(\sum_{n = 0}^\infty w^n I^n_{\zeta > 0}\right).
\end{align}
if the residues inside the unit circle are taken while
\begin{align}
\sum_{n = 0}^\infty \frac{w^n}{|\mathsf W|} \oint_{|z| = 1} \frac{d^n z}{z} g(\log z) = \mathrm{PE}\left[-\frac{(-1)^{N_c+1} w \tau^{\frac{N_f+N_a}{2}} x^{N_c-\frac{N_f+N_a}{2}+1}}{1-x^2}\right] \times \left(\sum_{n = 0}^\infty w^n I^n_{\zeta < 0}\right)
\end{align}
if the residues outside the unit circle are taken. Since two results must agree, the two indices with different $\zeta$ satisfy the following identity:
\begin{align}
\label{eq:wc0}
& \sum_{n = 0}^\infty w^n I^n_{\zeta > 0} = \left(\sum_{n = 0}^\infty w^n I^n_{\zeta < 0}\right) \nonumber \\
&\quad \times \mathrm{PE}\left[\frac{(-1)^{N_c+1} w \tau^{-\frac{N_f+N_a}{2}} x^{-N_c+\frac{N_f+N_a}{2}+1}-(-1)^{N_c+1} w \tau^{\frac{N_f+N_a}{2}} x^{N_c-\frac{N_f+N_a}{2}+1}}{1-x^2}\right]
\end{align}
The last factor agrees with the BPS index of $V_+$ and $V_-$.

Combining \eqref{eq:wc+}, \eqref{eq:wc-} and \eqref{eq:wc0}, one can write down a more general identity for any $N_f \geq N_a$ and $|\kappa| \leq \frac{|N_f-N_a|}{2}$:
\begin{align}
\label{eq:ident}
& \sum_{n = 0}^\infty w^n I^n_{\zeta > 0} = \left(\sum_{n = 0}^\infty w^n I^n_{\zeta < 0}\right) \nonumber \\
&\quad \times \mathrm{PE}\left[\frac{\delta_{2 \kappa,N_a-N_f} \tau^{-\frac{N_f+N_a}{2}} x^{-N_c+\frac{N_f+N_a}{2}+1}-\delta_{2 \kappa,N_f-N_a} \tau^{\frac{N_f+N_a}{2}} x^{N_c-\frac{N_f+N_a}{2}+1}}{1-x^2} \mathsf w \right]
\end{align}
where $\mathsf w = (-1)^{\kappa+\frac{N_f-N_a}{2}+N_c+1} w$. We emphasize that the wall-crossing factor is exactly the Plethystic exponential of \eqref{eq:1-wc}, which is the BPS index of $V_\pm$. Indeed, this is a consequence of the equivalence between the Aharony duality and the wall-crossing of vortex quantum mechanics. In the 3d duality perspective, \eqref{eq:ident} is nothing the index equality of a Aharony dual pair on $\mathbb R_\Omega^2 \times S^1$. Since the first line of \eqref{eq:ident} corresponds to the vortex partition functions of the 3d dual pair, the wall-crossing factor should be the contribution of extra chiral fields $V_\pm$ appearing in the dual theory, which turn out to describe the Coulomb branches of the moduli space. This is an indirect way to understand why the wall-crossing factor gives the contribution of the 3d Coulomb branch operators.
\\

In this section, we have shown that vortex quantum mechanics experiences the shift of the 1d FI parameter from $\zeta > 0$ to $\zeta < 0 $, and possibly the nontrivial wall-crossing at $\zeta = 0$, under the Aharony duality. The wall-crossing factor can be identified as the BPS index of the 3d gauge invariant chiral fields describing the Coulomb branches of the moduli space. Furthermore, using this equivalence between the 3d duality and the vortex wall-crossing, we have proven the Aharony duality at the level of vortex partition functions. The vortex partition function is a building block of various supersymmetric partition functions on curved 3-manifolds \cite{Krattenthaler:2011da,Dimofte:2011ju,Pasquetti:2011fj,Beem:2012mb,Hwang:2012jh,Taki:2013opa,Cecotti:2013mba,Fujitsuka:2013fga,Benini:2013yva,Benini:2015noa}. Thus, the identity \eqref{eq:ident} can be used to prove the agreement of various supersymmetric partition functions under the Aharony duality. The analytic proofs of the Aharony duality and its generalizations are worked out for the $S^3_b$ partition function \cite{Willett:2011gp,Benini:2011mf,Amariti:2014lla}, using integral identities of the hyperbolic gamma function found in \cite{Bult:2007}, and for the topologically twisted index on $\Sigma_g \times S^1$ \cite{Closset:2016arn} including the Witten index \cite{Intriligator:2013lca} as a special case. As far as we aware, for other supersymmetric partition functions such as the superconformal index, analytical proofs have been worked out only for the particular gauge rank and the particular number of flavors; e.g., see \cite{Krattenthaler:2011da,Hwang:2012jh}. In appendix \ref{sec:identity} we explain that many identities for 3d supersymmetric partition functions are proven using \eqref{eq:ident}.
\\

\subsection{$\mathcal N = 4$ SQCDs}
\label{sec:N=4 SQCD}

\begin{figure}[tbp]
\centering 
\includegraphics[height=.3\textheight]{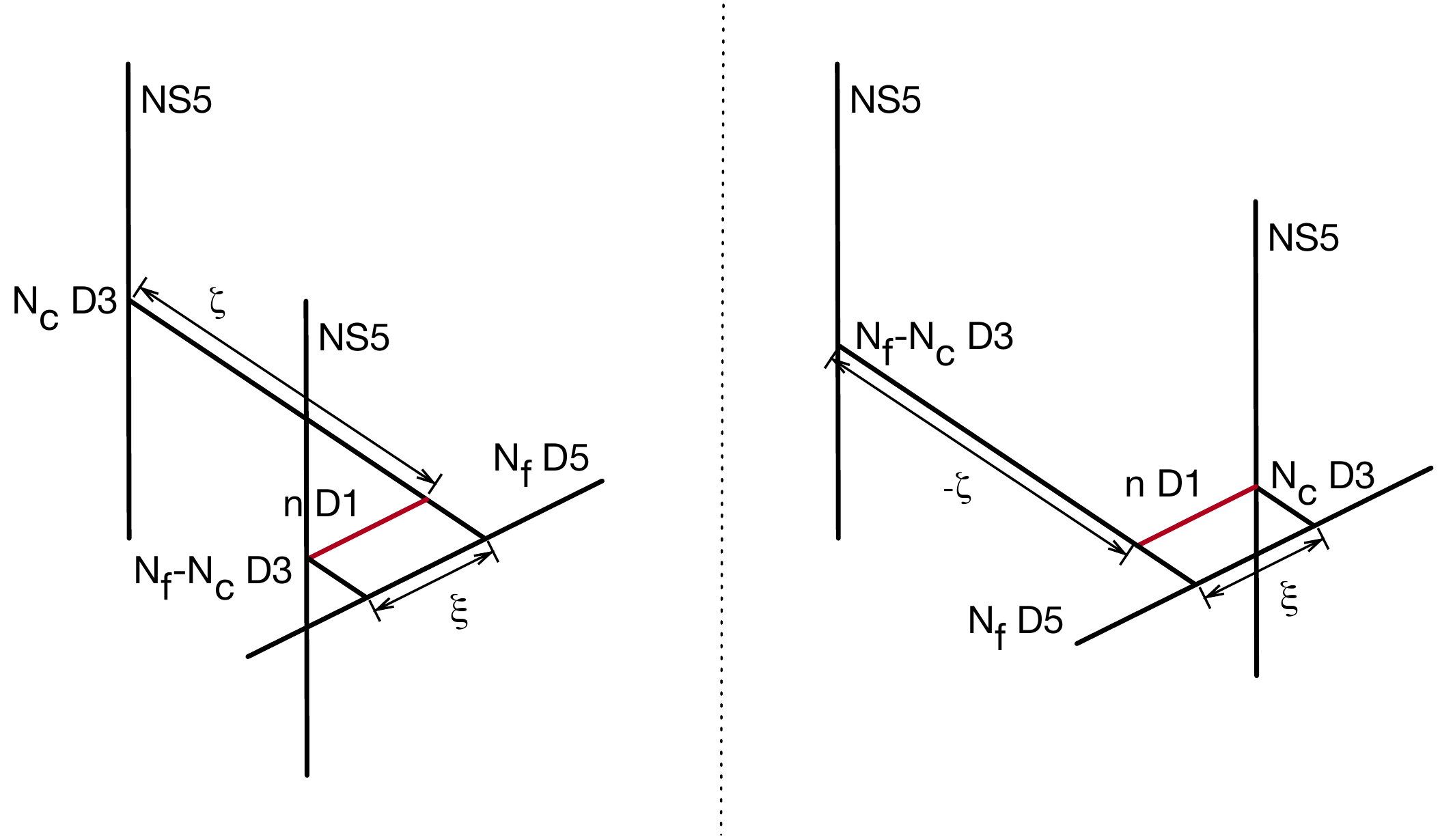}
\caption{\label{fig:N=4 duality} The brane motion representing the Seiberg-like duality of an $\mathcal N = 4$ SQCD. Vortices are realized as D1-branes denoted by the red line in the figure. The spacetime directions occupied by each brane are summarized in table \ref{tab:N=4 branes}. The distance between two NS5-branes along the 6-direction corresponds to 3d FI parameter $\xi$ while that along the 9-direction corresponds to 1d FI parameter $\zeta$.}
\end{figure}

The next example is the $\mathcal N = 4$ $U(N_c)$ gauge theory with $N_f$ hypermultiplets in the fundamental representation. This is a special case of $T_\rho [SU(N)]$ theories with $\rho = [N_c,N_f-N_c]$ and $N = N_f$. As reviewed in section \ref{sec:3d}, this theory has a Seiberg-like dual description, $U(N_f-N_c)$ gauge theory with $N_f$ fundamental hypermultiplets and $2 N_c-N_f$ decoupled free twisted hypermultiplets \cite{Kim:2012uz,Yaakov:2013fza,Gaiotto:2013bwa}. The brane setup in the presence of vortices is given by figure \ref{fig:N=4 duality} where the two NS5-branes are completely parallel.
\begin{table}[tbp]
\centering
\begin{tabular}{|c|cccccccccc|}
\hline
Branes & 0 & 1 & 2 & 3 & 4 & 5 & 6 & 7 & 8 & 9 \\
\hline
NS5 & $\times$ & $\times$ & $\times$ & $\times$ & $\times$ & $\times$ & & & & \\
D5 & $\times$ & $\times$ & $\times$ & & & & $\times$ & $\times$ & $\times$ & \\
D3 & $\times$ & $\times$ & $\times$ & & & & & & & $\times$ \\
D1 & $\times$ & & & & & & $\times$ & & & \\
\hline
\end{tabular}
\caption{\label{tab:N=4 branes} The spacetime directions occupied by the branes in figure \ref{fig:N=4 duality} are marked by $\times$.}
\end{table}
The same argument for the previous example suggests that the Seiberg-like duality between the two 3d $\mathcal N = 4$ theories on D3-branes is equivalent to the wall-crossing of vortex quantum mechanics on D1-branes. In this section, we explicitly realize it by computing the quantum mechanics indices for different 1d FI parameters.
\\

\begin{figure}[tbp]
\centering 
\includegraphics[height=.2\textheight]{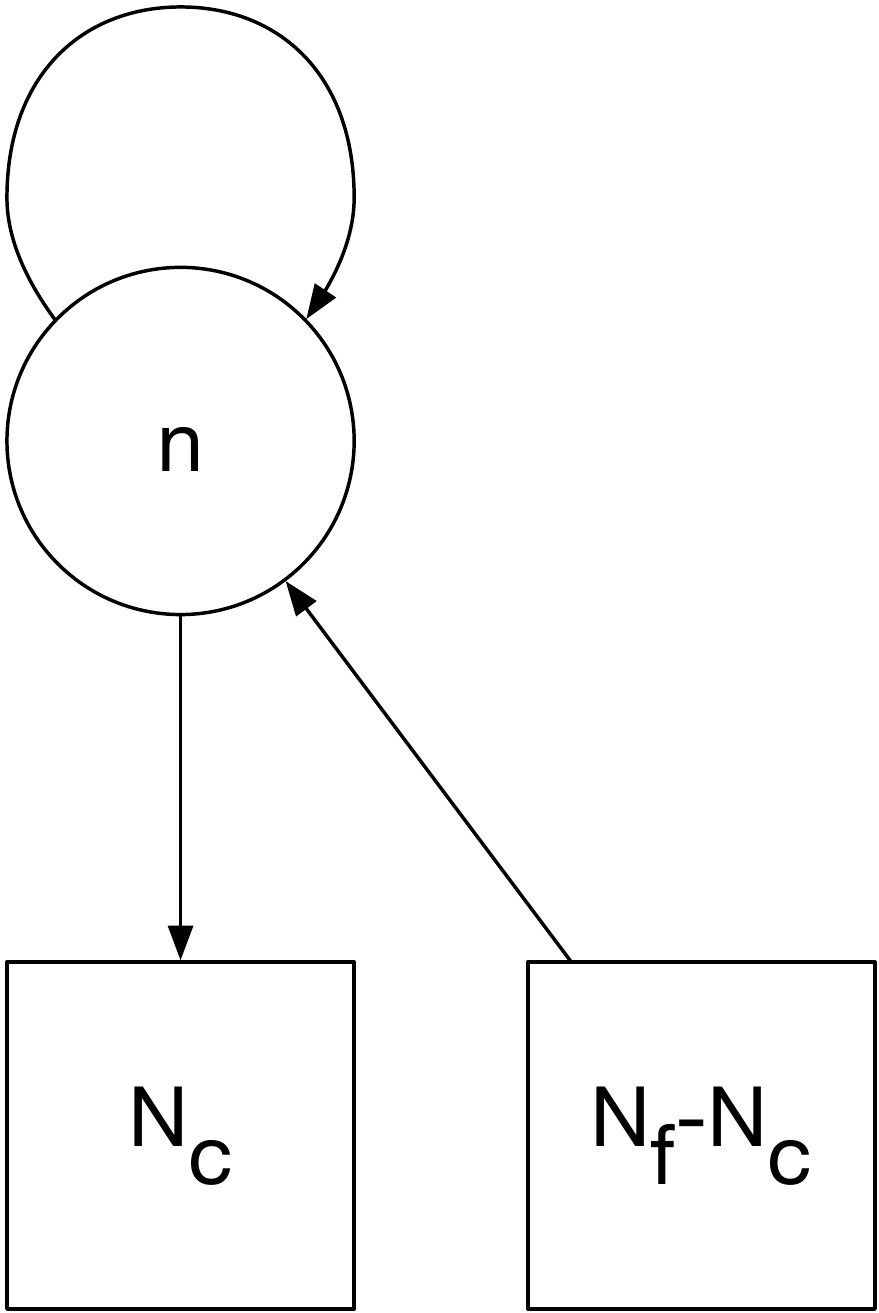}
\caption{\label{fig:N=4 vortex} The quiver diagram representation of vortex quantum mechanics for a 3d $\mathcal N = 4$ SQCD. Each arrow represents the 1d $\mathcal N = 4$ chiral multiplet. This is a truncation of figure \ref{fig:handsaw}.}
\end{figure}

For the 3d $\mathcal N = 4$ $U(N_c)$ theory with $N_f$ flavors, the moduli space of $n$ vortices is described by the $\mathcal N = 4$ gauged quantum mechanics illustrated in figure \ref{fig:N=4 vortex}, which is a truncation of figure \ref{fig:handsaw}.
The refined Witten index of this quantum mechanics is again written as the following JK-residue:
\begin{align}
I^n = \frac{1}{|\mathsf W|} \text{JK-Res}_{\eta = \zeta} \left[g^n(u) d^n u\right]
\end{align}
where $g^n (u)$ is now given by
\begin{align}
\label{eq:N=4 1-loop}
g^n (u) &= \left(\frac{1}{2 \sinh \frac{-2 \mu}{2}}\right)^n \left(\prod_{i \neq j}^n \frac{\sinh \frac{u_i-u_j}{2}}{\sinh \frac{u_i-u_j-2 \mu}{2}}\right) \left(\prod_{i,j = 1}^n \frac{\sinh \frac{u_i-u_j-2 \mu-2 \gamma}{2}}{\sinh \frac{u_i-u_j-2 \gamma}{2}}\right) \nonumber \\
&\quad \times \left(\prod_{i = 1}^n \prod_{b = 1}^{N_c} \frac{\sinh \frac{u_i-m_b-2 \mu-\gamma}{2}}{\sinh \frac{u_i-m_b-\gamma}{2}}\right) \left(\prod_{i = 1}^n \prod_{a = N_c+1}^{N_f} \frac{\sinh \frac{-u_i+m_a-2 \mu-\gamma}{2}}{\sinh \frac{-u_i+m_a-\gamma}{2}}\right).
\end{align}

We first compute the quantum mechanics index for $\zeta > 0$. The JK-residue rule chooses sets of linearly independent hyperplanes each of which determine $u_i$ as follows:
\begin{align}
u_i = \left\{\begin{array}{l}
u_j+2 \mu, \\
u_j+2 \gamma, \\
m_a+\gamma, \qquad a \in \{1,\ldots,N_c\}.
\end{array}\right.
\end{align}
However, a pole intersecting a hyperplane of the first type has the vanishing residue because of zeros of the integrand. Therefore, the contributing poles are written in the following form:
\begin{align}
u_i = m_a+(2 k_a-1) \gamma,\qquad
a \in \{1,\ldots,N_c\}, \quad
k_a \in \{1,\ldots,n_a\}
\end{align}
where $(n_1,\ldots,n_{N_c})$ is a partition of $n$ into $N_c$ nonnegative integers. Again every pair $(a,k_a)$ is assigned to one of $i = 1,\ldots,n$ exactly once. Evaluating the JK-residue, we have the following contribution to the index for a given partition $(n_1,\ldots,n_{N_c})$:
\begin{align}
\label{eq:N=4 res}
& \left(\prod_{a,b = 1}^{N_c} \prod_{k_a = 1}^{n_a} \prod_{k_b = 1}^{n_b} \frac{\sinh \frac{m_a-m_b+2 (k_a-k_b) \gamma}{2}}{\sinh \frac{m_a-m_b-2 \mu+2 (k_a-k_b) \gamma}{2}} \frac{\sinh \frac{m_a-m_b-2 \mu+2 (k_a-k_b-1) \gamma}{2}}{\sinh \frac{m_a-m_b+2 (k_a-k_b-1) \gamma}{2}}\right)' \nonumber \\
& \times \left(\prod_{a = 1}^{N_c} \prod_{k_a = 1}^{n_a} \prod_{b = 1}^{N_c} \frac{\sinh \frac{m_a-m_b-2 \mu+2 (k_a-1) \gamma}{2}}{\sinh \frac{m_a-m_b+2 (k_a-1) \gamma}{2}}\right)' \left(\prod_{b = 1}^{N_c} \prod_{k_b = 1}^{n_b} \prod_{a = N_c+1}^{N_f} \frac{\sinh \frac{-m_b+m_a-2 \mu-2 k_b \gamma}{2}}{\sinh \frac{-m_b+m_a-2 k_b \gamma}{2}}\right).
\end{align}
where $'$ denotes that the vanishing factors are omitted. The Weyl factor $|\mathsf W|$ is canceled by factor $n!$ coming from the permutations among $u_i$'s. \eqref{eq:N=4 res} is further simplified due to the cancelation between the numerator and the denominator. Summing over all possible partitions of $n$, the index with $\zeta > 0$ is given by
\begin{align}
\label{eq:ind+}
& I^n_{\zeta > 0} = \nonumber \\
& \sum_{\substack{n_a \geq 0, \\
\sum n_a = n}} \left(\prod_{a,b = 1}^{N_c} \prod_{k_a = 1}^{n_a} \frac{\sinh \frac{m_a-m_b-2 \mu+2 (k_a-n_b-1) \gamma}{2}}{\sinh \frac{m_a-m_b+2 (k_a-n_b-1) \gamma}{2}}\right) \left(\prod_{b = 1}^{N_c} \prod_{k_b = 1}^{n_b} \prod_{a = N_c+1}^{N_f} \frac{\sinh \frac{-m_b+m_a-2 \mu-2 k_b \gamma}{2}}{\sinh \frac{-m_b+m_a-2 k_b \gamma}{2}}\right),
\end{align}
which reproduces the vortex partition function of the $\mathcal N = 4$ $U(N_c)$ theory with $N_f$ flavors on $\mathbb R_\Omega^2 \times S^1$ \cite{Kim:2012uz}.

On the other hand, for the index with $\zeta < 0$, the JK-residue rule with $\eta = \zeta$ chooses different sets of hyperplanes:
\begin{align}
u_i = \left\{\begin{array}{l}
u_j-2 \mu, \\
u_j-2 \gamma, \\
m_a-\gamma, \qquad a \in \{N_c+1,\ldots,N_f\}.
\end{array}\right.
\end{align}
Since a pole intersecting a hyperplane of the first type has the vanishing residue, a relevant pole is written in the following form:
\begin{align}
u_i = m_a-(2 k_a-1) \gamma, \qquad a \in \{N_c+1,\ldots,N_f\}, \quad k_a \in \{1,\ldots,n_a\}
\end{align}
where $(1,\ldots,n_a)$ is a partition of $n$ into $N_f-N_c$ nonnegative integers. The resulting index of vortex quantum mechanics is given by
\begin{align}
\label{eq:ind-}
& I^n_{\zeta < 0} = \nonumber \\
& \sum_{\substack{n_a \geq 0, \\ \sum n_a = n}} \left(\prod_{a,b = N_c+1}^{N_f} \prod_{k_b = 1}^{n_b} \frac{\sinh \frac{-m_b+m_a-2 \mu+2 (k_b-n_a-1) \gamma}{2}}{\sinh \frac{-m_b+m_a+2 (k_b-n_a-1) \gamma}{2}}\right) \left(\prod_{a = N_c+1}^{N_f} \prod_{k_a = 1}^{n_a} \prod_{b = 1}^{N_c} \frac{\sinh \frac{m_a-m_b-2 \mu-2 k_a \gamma}{2}}{\sinh \frac{m_a-m_b-2 k_a \gamma}{2}}\right).
\end{align}
As expected the index with the negative FI parameter $\zeta < 0$ is the vortex partition function of the dual $U(N_f-N_c)$ theory with $N_f$ flavors. The sign flip of mass $m_a$ is understood because the 3d FI parameter is also flipped under the duality such that the roles of fundamental and anti-fundamental fields are exchanged. This shows that the Seiberg-like duality of a 3d $\mathcal N = 4$ $U(N_c)$ theory with fundamental hypers also corresponds to the sign flip of the FI parameter in its vortex quantum mechanics.
\\

The indices in different FI chambers: $\zeta > 0$ and $\zeta < 0$ do not need to agree due to the non-compact Coulomb branch at $\zeta = 0$. Some states can escape through this non-compact branch. One can trace those escaping states by comparing the indices \eqref{eq:ind+} and \eqref{eq:ind-}. Using the same argument for the previous example we show that two indices \eqref{eq:ind+} and \eqref{eq:ind-} indeed satisfy the following relation:
\begin{align}
\label{eq:N=4 wc}
& \left(\sum_{m = 0}^\infty \frac{w^m}{m!} \sum_{\cap_{i = 1}^m H_i^- \in \text{asymp}^-} \mathrm{Res}_{H_1,\ldots,H_m} \left[g^m_- (u) d^m u\right]\right) \left(\sum_{n = 0}^\infty w^n I^n_{\zeta > 0}\right) \nonumber \\
&= \left(\sum_{m = 0}^\infty \frac{(-w)^m}{m!} \sum_{\cap_{i = 1}^m H_i^+ \in \text{asymp}^+} \mathrm{Res}_{H_1,\ldots,H_m} \left[g^m_+ (u) d^m u\right]\right) \left(\sum_{n = 0}^\infty w^n I^n_{\zeta < 0}\right)
\end{align}
where
\begin{align}
g^m_\pm (u) &= \left(\frac{e^{\pm \frac{1}{2} (2 N_c-N_f) (-2 \mu)}}{2 \sinh \frac{-2 \mu}{2}}\right)^m \left(\prod_{i \neq j}^m \frac{\sinh \frac{u_i-u_j}{2}}{\sinh \frac{u_i-u_j-2 \mu}{2}}\right) \left(\prod_{i,j = 1}^m \frac{\sinh \frac{u_i-u_j-2 \mu-2 \gamma}{2}}{\sinh \frac{u_i-u_j-2 \gamma}{2}}\right).
\end{align}
Hyperplanes $H_1^\pm,\ldots,H_m^\pm$ are chosen among
\begin{align}
u_i &= \left\{\begin{array}{l}
\pm \infty, \\
u_j \mp 2 \mu, \\
u_j \mp 2 \gamma.
\end{array}\right.
\end{align}
Note that $g^m_+ (u)$ and $g^m_- (u)$ are the same when $N_f = 2 N_c$, which is the self-dual case. Therefore, the asymptotic factors in \eqref{eq:N=4 wc} cancel each other such that
\begin{align}
\label{eq:self dual}
I^n_{\zeta > 0} = I^n_{\zeta < 0}
\end{align}
for $N_f = 2 N_c$. \eqref{eq:self dual} implies that there is no spectrum jump at $\zeta = 0$ when $N_f = 2 N_c$. For general $N_f \neq 2 N_c$, one can obtain the nontrivial wall-crossing factor by evaluating the asymptotic residues. However, it turns out that the explicit computations of the asymptotic residues in this example are more complicated than those in the previous example. Instead, it is shown that the explicit form of \eqref{eq:N=4 wc} can be obtained by examining large mass limits of the vortex partition functions \eqref{eq:ind+} and \eqref{eq:ind-} \cite{Hwang:2015wna}:
\begin{align}
\label{eq:N=4 wc 2}
& \sum_{n = 0}^\infty w^n I^n_{\zeta > 0} = \left(\sum_{n = 0}^\infty w^n I^n_{\zeta < 0}\right) \left(\prod_{i = 1}^{2 N_c-N_f} Z_\text{hyper} (x,\tau,w \tau^{2 N_c-N_f-2 i+1})\right)
\end{align}
where $\tau = e^\mu$ and $x = e^\gamma$. $Z_\text{hyper}$ is the contribution of a free twisted hypermultiplet, which is given by
\begin{align}
Z_\text{hyper} (x,\tau,w) = \mathrm{PE} \left[\frac{\tau^{-1} w-\tau w x^2}{1-x^2}\right].
\end{align}
Thus, the wall-crossing factor for the vortex quantum mechanics index corresponds the contribution of the decoupled free twisted hypermultiplets in the dual 3d theory. Indeed, those twisted hypermultiplets describe Coulomb branches of the 3d theory \cite{Gaiotto:2008ak,Kim:2012uz,Yaakov:2013fza,Gaiotto:2013bwa}. Again we observe that the wall-crossing of vortex quantum mechanics captures the information of Coulomb branch of the 3d theory.
\\

We have proven that the vortex quantum mechanics indices in different FI chambers: $\zeta > 0$ and $\zeta < 0$ exactly reproduce the vortex partition functions of a 3d $\mathcal N = 4$ Seiberg-like dual pair. This shows that under the Seiberg-like duality, vortex quantum mechanics for a 3d $\mathcal N = 4$ SQCD experiences the wall-crossing controlled by FI parameter $\zeta$. Furthermore, from the identity \eqref{eq:N=4 wc 2}, we show that the BPS index of the escaping states at the wall $\zeta = 0$ is given by
\begin{align}
Z^\text{wall} &= \prod_{i = 1}^{2 N_c-N_f} Z_\text{hyper} (x,\tau,w \tau^{2 N_c-N_f-2 i+1}) \\
&= \mathrm{PE} \left[\frac{\sinh \left[(2 N-M) \mu\right] \sinh (\mu+\gamma)}{\sinh \mu \sinh \gamma} w\right], \label{eq:wall}
\end{align}
which is also identified as the index of the 3d twisted hypermultiplets describing Coulomb branches of the moduli space.
\\

\subsection{Linear quiver examples}
\subsubsection{$T_\rho [SU(N)]$}

We have seen that for a 3d SQCD, the Seiberg-like duality is equivalent to the wall-crossing of vortex quantum mechanics controlled by the 1d FI parameter $\zeta$. In this section, we would like to ask whether this phenomenon can be generalized to more complicated cases such as linear quiver gauge theories we examined in section \ref{sec:vortex QM}. We will see that even in such cases, the equivalence between the 3d Seiberg-like duality and the wall-crossing of vortex quantum mechanics is observed.
\\

\begin{figure}[tbp]
\centering 
\includegraphics[height=.25\textheight]{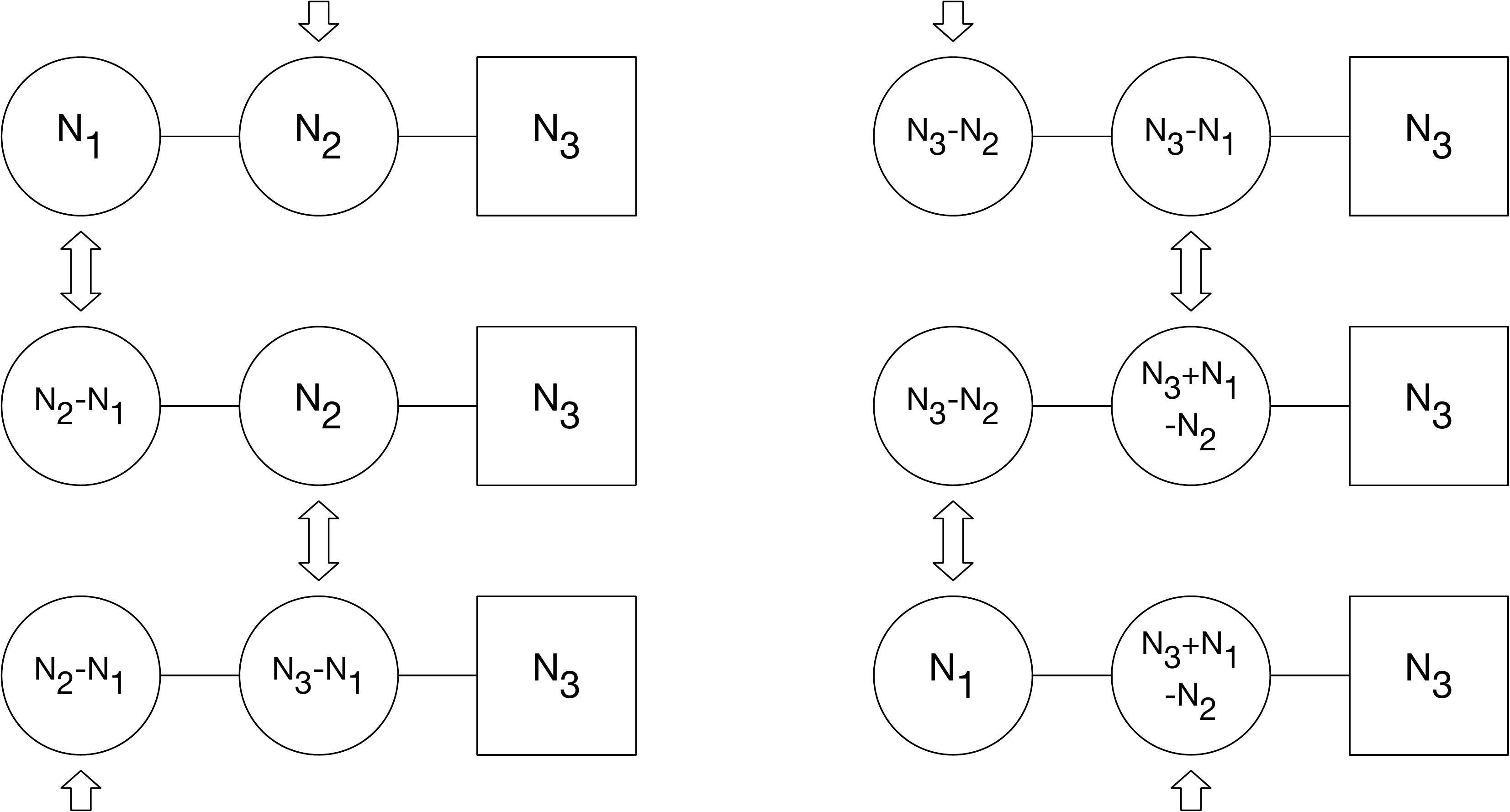}
\caption{\label{fig:T[SU(N)]} The Seiberg-like duality chain of $T_{[N_1,N_2-N_1,N_3-N_2]} [SU(N_3)]$.}
\end{figure}

\begin{figure}[tbp]
\centering 
\includegraphics[height=.5\textheight]{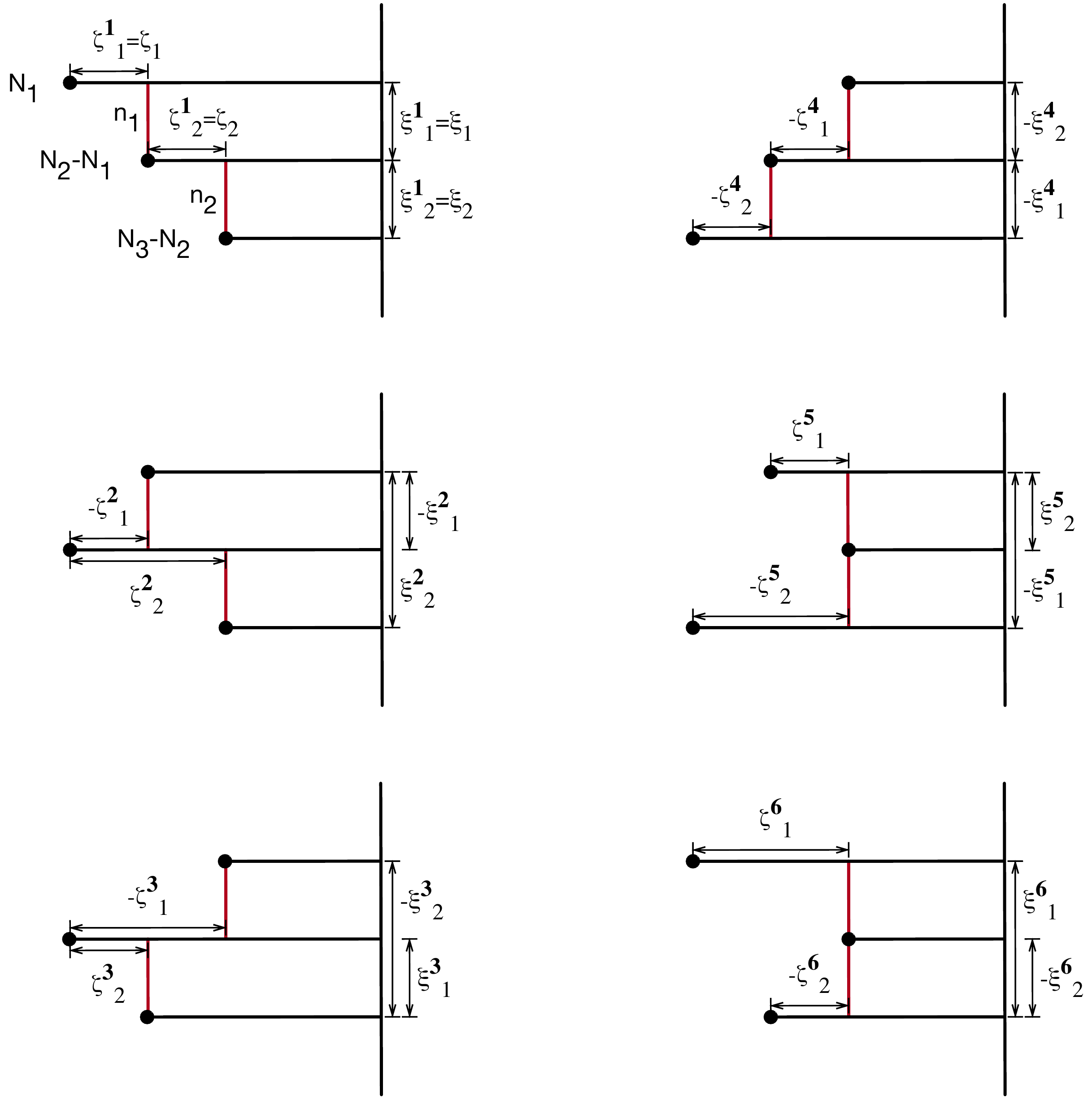}
\caption{\label{fig:T[SU(N)] duality} The brane setups representing the duality chain in figure \ref{fig:T[SU(N)]} are shown. NS5-branes, D5-branes, D3-branes, D1-branes are denoted by black dots, black vertical lines, black horizontal lines and red vertical lines respectively. The vertical distances between NS5-branes correspond to 3d FI parameters $\xi_l$ while the horizontal distances between NS5-branes correspond to 1d FI parameters $\zeta_l$.}
\end{figure}

First let us consider $T_\rho [SU(N)]$ theories. The vortex partition functions of $T_\rho [SU(N)]$ theories are given by \eqref{eq:T[SU(N)] result}. We should remind you that this result is for positive 3d FI parameters $\xi_1,\ldots,\xi_{L} > 0$. In order to examine the Seiberg-like dualities of $T_\rho [SU(N)]$ theories, however, we have to relax this positive FI condition because the Seiberg-like dualities incorporate nontrivial FI mappings. For concreteness, let us consider $T_{[N_1,N_2-N_1,N_3-N_2]} [SU(N_3)]$ having two gauge nodes. We have a duality chain including this theory as shown in figure \ref{fig:T[SU(N)]}.
The duality chain contains all possible ranges of the FI parameters. If we assume $\xi^{\mathbf 1}_1 = \xi_1 > 0$ and $\xi^{\mathbf 1}_2 = \xi_2 > 0$, each theory in the duality chain has the FI parameters in the following ranges:
\begin{align}
\label{eq:FI map}
\begin{array}{lccclcc}
\xi^{\mathbf 1}_1 = \xi_1  &>& 0, &\qquad& \xi^{\mathbf 1}_2 = \xi_2  &>& 0, \\
\xi^{\mathbf 2}_1 = -\xi_1  &<& 0, &\qquad& \xi^{\mathbf 2}_2 = \xi_1+\xi_2  &>& |\xi^{\mathbf 2}_1|, \\
\xi^{\mathbf 3}_1 = \xi_2  &>& 0, &\qquad& \xi^{\mathbf 3}_2 = -\xi_1-\xi_2  &<& -|\xi^{\mathbf 3}_1|, \\
\xi^{\mathbf 4}_1 = -\xi_2  &<& 0, &\qquad& \xi^{\mathbf 4}_2 = -\xi_1  &<& 0, \\
\xi^{\mathbf 5}_1 = -\xi_1-\xi_2  &<& -|\xi^{\mathbf 5}_2|, &\qquad& \xi^{\mathbf 5}_2 = \xi_1  &>& 0, \\
\xi^{\mathbf 6}_1 = \xi_1+\xi_2  &>& |\xi^{\mathbf 6}_2|, &\qquad& \xi^{\mathbf 6}_2 = -\xi_2  &<& 0 \\
\end{array}
\end{align}
where $\xi^{\mathbf k}_l$ is an FI parameter of the $\mathbf k$-th theory in the duality chain. These FI mappings under the Seiberg-like dualities can be read off from the brane setup, figure \ref{fig:T[SU(N)] duality}.
To the best of our knowledge, the vortex partition functions of $T_\rho [SU(N)]$ with general FI parameter ranges have not been investigated in the literatures. For this reason, so far it is not manifest to study the Seiberg-like duality of $T_\rho [SU(N)]$ using the vortex partition function.

\begin{figure}[tbp]
\centering 
\includegraphics[height=.2\textheight]{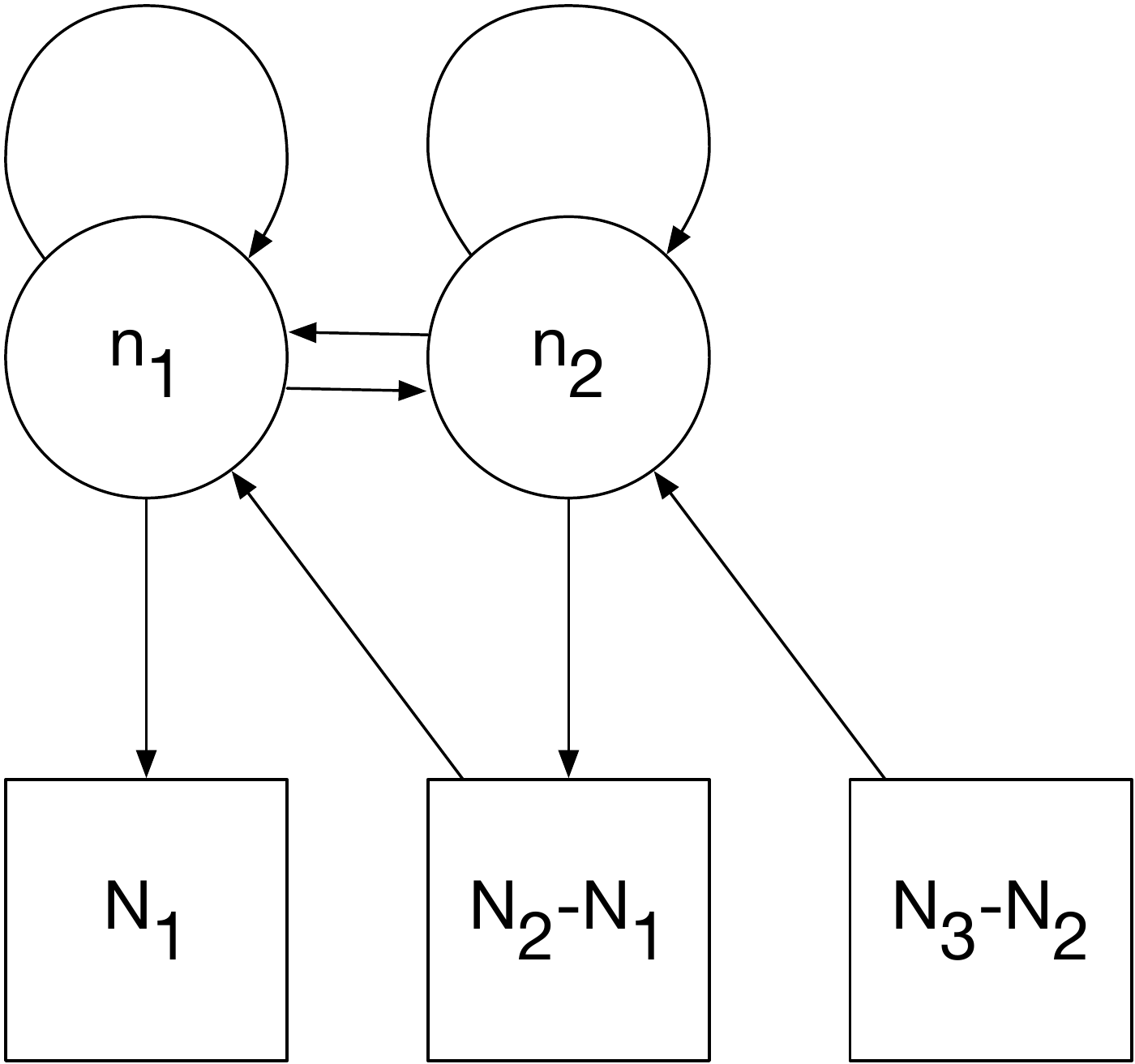}
\caption{\label{fig:T[SU(N)] vortex} Vortex quantum mechanics for $T_{[N_1,N_2-N_1,N_3-N_2]} [SU(N_3)]$.}
\end{figure}

On the other hand, we have seen that two vortex quantum mechanics for a 3d Seiberg-like dual pair are related by a change of 1d FI parameters, at least for SQCDs. In other words, the two vortex quantum mechanics should be the same except the FI parameters. We claim that this relation still holds for more complicated theories such as $T_\rho [SU(N)]$ we are now considering. For example, the first theory in the duality chain has vortex quantum mechanics described in figure \ref{fig:T[SU(N)] vortex}.
The vortex quantum mechanics has positive FI parameters, $\zeta^{\mathbf 1}_1 > 0$ and $\zeta^{\mathbf 1}_2 > 0$. The second theory in the duality chain then should have the same vortex quantum mechanics except FI parameters, which are now in different ranges: $\zeta^{\mathbf 2}_1 < 0$ and $\zeta^{\mathbf 2}_2 > |\zeta^{\mathbf 2}_1|$. Indeed, we claim that all those six theories in the 3d duality chain have the same vortex quantum mechanics with different 1d FI parameters in the following ranges:
\begin{align}
\label{eq:T[SU(N)] FI}
\begin{array}{ccccccc}
\zeta^{\mathbf 1}_1 \quad &>& 0, &\qquad& \zeta^{\mathbf 1}_2 \quad &>& 0, \\
\zeta^{\mathbf 2}_1 \quad &<& 0, &\qquad& \zeta^{\mathbf 2}_2 \quad &>& |\zeta^{\mathbf 2}_1|, \\
\zeta^{\mathbf 3}_1 \quad &<& -|\zeta^{\mathbf 3}_2|, &\qquad& \zeta^{\mathbf 3}_2 \quad &>& 0, \\
\zeta^{\mathbf 4}_1 \quad &<& 0, &\qquad& \zeta^{\mathbf 4}_2 \quad &<& 0, \\
\zeta^{\mathbf 5}_1 \quad &>& 0, &\qquad& \zeta^{\mathbf 5}_2 \quad &<& -|\zeta^{\mathbf 5}_1|, \\
\zeta^{\mathbf 6}_1 \quad &>& |\zeta^{\mathbf 6}_2|, &\qquad& \zeta^{\mathbf 6}_2 \quad &<& 0
\end{array}
\end{align}
where $\zeta^{\mathbf k}_l$ is an FI parameter of vortex quantum mechanics for the $\mathbf k$-th theory in the duality chain. An interesting thing is that the number of phases of the 3d duality chain and that of 1d FI parameters are the same. This is a clue that the 3d Seiberg-like duality and the wall-crossing of vortex quantum mechanics are related.
\\

One should note that quantum mechanics in figure \ref{fig:T[SU(N)] vortex} with the FI parameters \eqref{eq:T[SU(N)] FI} is exactly the world-volume theory of D1-branes in figure \ref{fig:T[SU(N)] duality}. The Seiberg-like duality of $T_\rho [SU(N)]$ and the wall-crossing of its vortex quantum mechanics are inferred from the same brane motion. Thus, we expect that the quantum mechanics with the FI parameters \eqref{eq:T[SU(N)] FI} correctly describes the vortex moduli spaces of the 3d theories in the duality chain.

Furthermore, we provide additional evidence by explicitly computing the quantum mechanics indices for different FI parameters. In the previous section, we have seen that the vortex partition functions of an $\mathcal N = 4$ SQCD dual pair agree up to the contribution of decoupled twisted hypermultiplets. The number of the decoupled twisted hypermultiplets is determined by the rank difference between the gauge groups of the dual pair. This is a consequence of the fact that a dual pair must have the same number of Coulomb branches, which is determined by the gauge group rank \cite{Gaiotto:2008ak,Kim:2012uz,Yaakov:2013fza,Gaiotto:2013bwa}. This is still true for $T_\rho [SU(N)]$ theories. Indeed, we have checked that the following relations hold among the six quantum mechanics indices in different FI chambers:\footnote{This is numerically checked for various $N_1,N_2,N_3$ up to $n_1,n_2 \leq 2$.}
\begin{align}
\begin{aligned}
\label{eq:T[SU(N)] wc}
Z_{\mathbf 1}/Z_{\mathbf 2} &= Z^\text{wall}_{N_1,N_2} (w_1), \\
Z_{\mathbf 2}/Z_{\mathbf 3} &= Z^\text{wall}_{N_2,N_3+N_2-N_1} (w_1 w_2), \\
Z_{\mathbf 3}/Z_{\mathbf 4} &= Z^\text{wall}_{N_2-N_1,N_3-N_1} (w_2), \\
Z_{\mathbf 4}/Z_{\mathbf 5} &= Z^\text{wall}_{N_3-N_1,2 N_3-N_2} (w_1), \\
Z_{\mathbf 5}/Z_{\mathbf 6} &= Z^\text{wall}_{N_3-N_2,N_3-N_2+N_1} (w_1 w_2), \\
Z_{\mathbf 6}/Z_{\mathbf 1} &= Z^\text{wall}_{N_3-N_2+N_1,N_1+N_3} (w_2)
\end{aligned}
\end{align}
where $Z_{\mathbf k}$ on the left hand side is the generating function of the vortex indices in the $\mathbf k$-th FI chamber:
\begin{align}
Z_{\mathbf k} = \sum_{n_1 = 0}^\infty \sum_{n_2 = 0}^\infty w_1^{n_1} w_2^{n_2} I_{\mathbf k}^{n_1,n_2}
\end{align}
while $Z^\text{wall}_{N,M}$ on the right hand side is the wall-crossing factor:
\begin{align}
Z^\text{wall}_{N,M} (w) = \mathrm{PE} \left[\frac{\sinh \left[(2 N-M) \mu\right] \sinh (\mu+\gamma)}{\sinh \mu \sinh \gamma} w\right].
\end{align}
Here we allow negative $2 N-M$ as well such that
\begin{align}
Z^\text{wall}_{N,M} (w) = Z^\text{wall}_{M-N,M} (w)^{-1}
\end{align}
for $2 N-M < 0$. \eqref{eq:wall} tells us that $Z^\text{wall}_{N,M}$ with $2 N \geq M$ is exactly the contribution of $2 N-M$ decoupled twisted hypermultiplets. Thus, the wall-crossing factors in \eqref{eq:T[SU(N)] wc} reflect the correct number of decoupled twisted hypermultiplets at each duality step. Although we have examined a two gauge node example, this behavior of the quantum mechanics index is expected for the other $T_\rho [SU(N)]$ theories as well. Therefore, we expect that the Seiberg-like duality of general $T_\rho [SU(N)]$ is equivalent to the wall-crossing of its vortex quantum mechanics.
\\

\subsubsection{$\mathcal N = 2$ linear quiver examples}

Next let us consider $\mathcal N = 2$ linear quiver examples, which we have examined in section \ref{sec:N=2}. We illustrate some examples that exhibit the equivalence of the 3d Seiberg-like duality and the wall-crossing of vortex quantum mechanics. The relation between the Seiberg-like duality and the vortex wall-crossing for general $\mathcal N = 2$ theories will be worth studying, which we relegate to future work.

\begin{figure}[tbp]
\centering 
\includegraphics[height=.25\textheight]{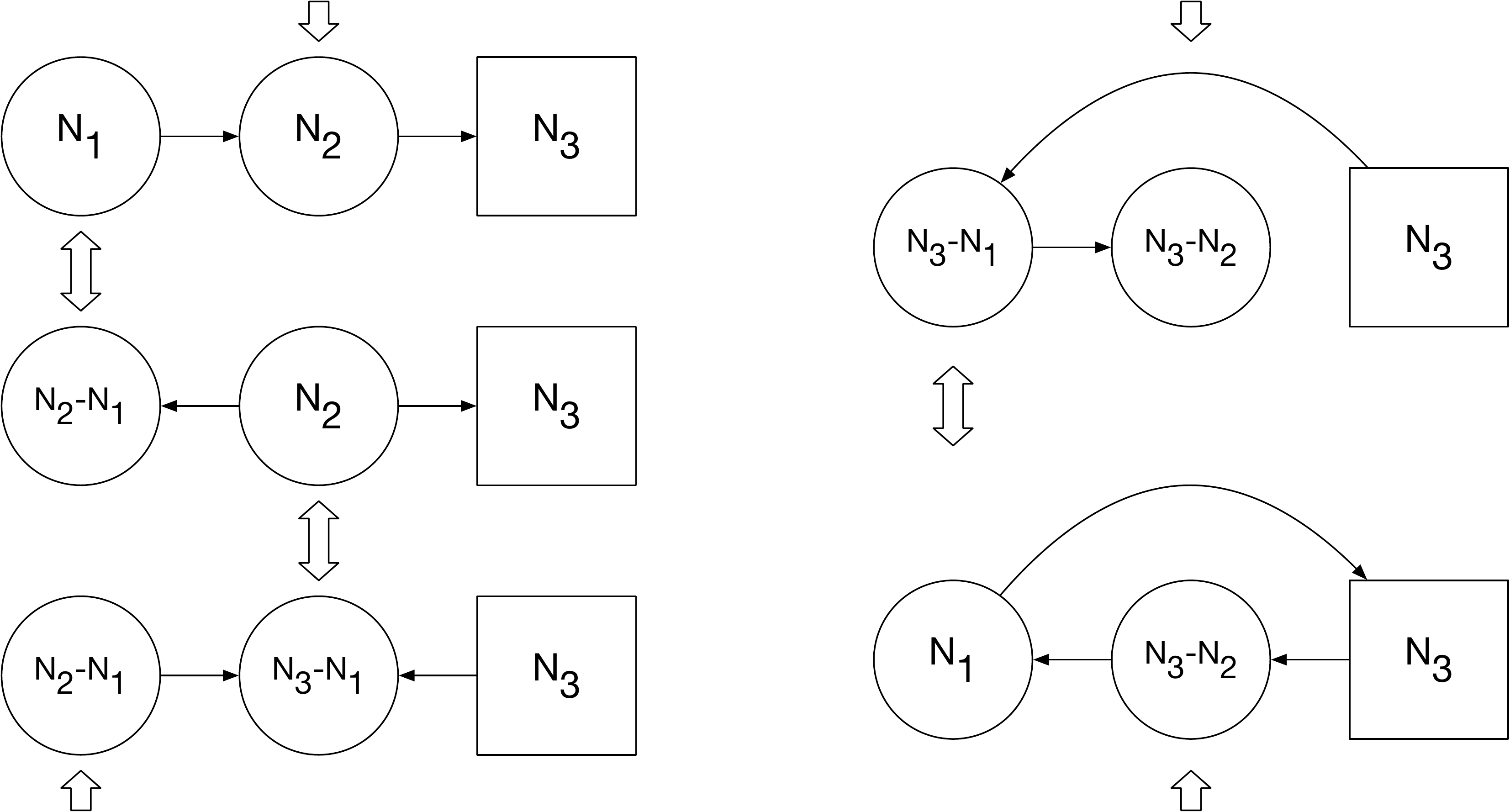}
\caption{\label{fig:N=2} A typical duality chain of an $\mathcal N = 2$ linear quiver theory we are considering. Unlike $T_\rho [SU(N)]$, an arrow has a direction which can be flipped under the duality. In addition, a new arrow can appear after the duality action.}
\end{figure}

The examples we are considering in this section have a duality chain illustrated in figure \ref{fig:N=2}.
The corresponding vortex quantum mechanics is given by figure \ref{fig:def vortex} with $L = 2$.
One should note that, unlike the previous $T_\rho [SU(N)]$ example, vortex quantum mechanics now has only five FI chambers instead of six:
\begin{align}
\label{eq:N=2 FI}
\begin{array}{ccccccc}
\zeta^{\mathbf 1}_1 \quad &>& 0, &\qquad& \zeta^{\mathbf 1}_2 \quad &>& 0, \\
\zeta^{\mathbf 2}_1 \quad &<& 0, &\qquad& \zeta^{\mathbf 2}_2 \quad &>& |\zeta^{\mathbf 2}_1|, \\
\zeta^{\mathbf 3}_1 \quad &<& -|\zeta^{\mathbf 3}_2|, &\qquad& \zeta^{\mathbf 3}_2 \quad &>& 0, \\
\zeta^{\mathbf 4}_1 \quad &<& 0, &\qquad& \zeta^{\mathbf 4}_2 \quad &<& 0, \\
\zeta^{\mathbf 5}_1 \quad &>& 0, &\qquad& \zeta^{\mathbf 5}_2 \quad &<& 0.
\end{array}
\end{align}
The fifth chamber was divided into two chambers: $|\zeta^{\mathbf 5}_1| < |\zeta^{\mathbf 5}_2|$ and $|\zeta^{\mathbf 5}_1| > |\zeta^{\mathbf 5}_2|$ for $T_\rho [SU(N)]$ while it is not for the current example. This is because there is only one bi-fundamental chiral multiplet between adjacent gauge nodes of vortex quantum mechanics. Indeed, the duality chain in figure \ref{fig:N=2} also includes only five theories instead of six. This is the first clue that the 3d Seiberg-like duality relates to the wall-crossing of vortex quantum mechanics for $\mathcal N = 2$ linear quiver theories as well.
\\

\begin{figure}[tbp]
\centering 
\includegraphics[height=.2\textheight]{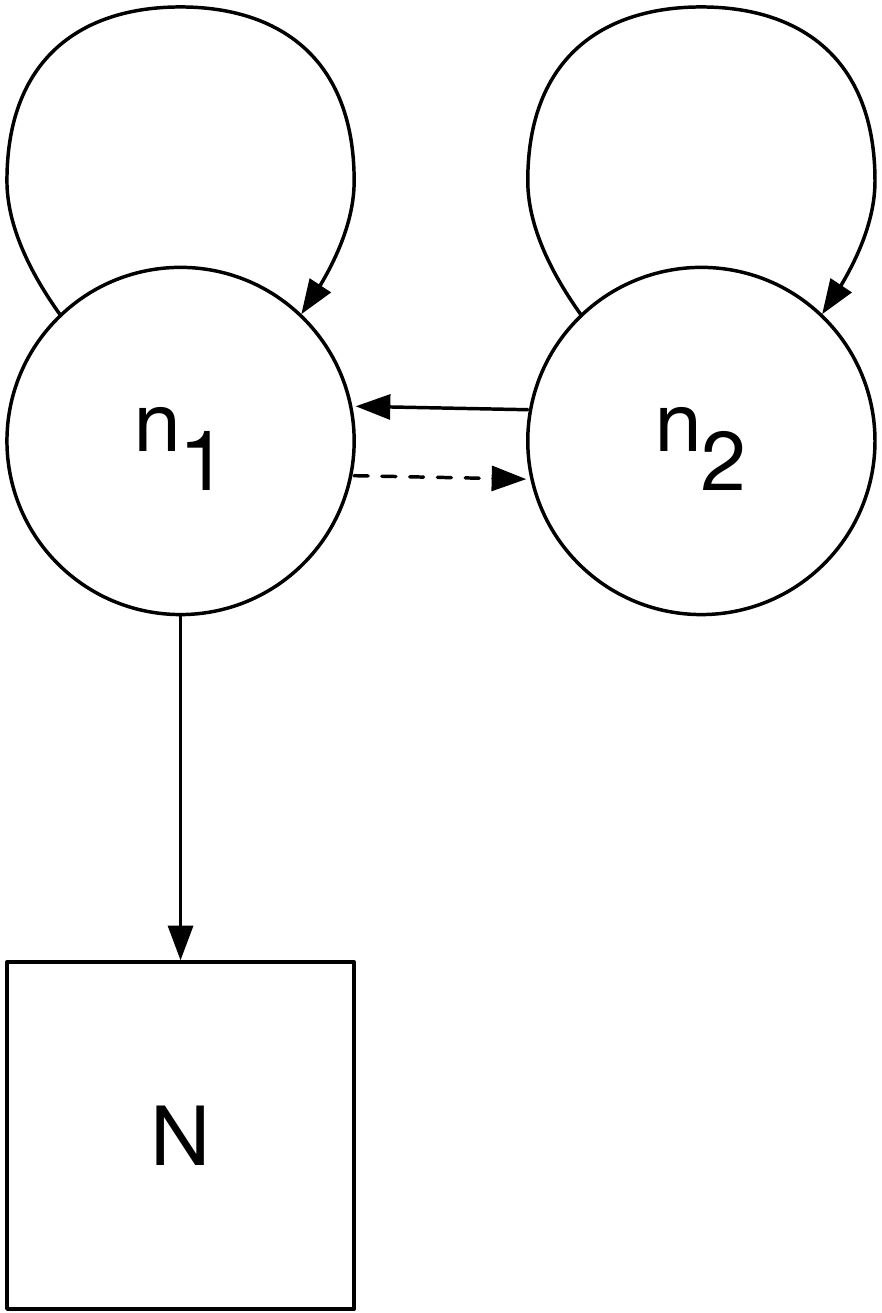}
\caption{\label{fig:NNN vortex} Vortex quantum mechanics for a $(N,N,N)$-type quiver theory is illustrated. If the parent 3d theory has CS/BF interactions, the quantum mechanics also has the corresponding Wilson lines shown in \eqref{eq:Wilson lines}.}
\end{figure}

The simplest example is an $(N,N,N)$-type quiver theory. The theory has gauge group $U(N) \times U(N)$ and flavor group $U(N)$. We introduce the CS interaction for each gauge node as well as the BF interaction between the $U(1)$ parts of them. The level of those CS and BF interactions are chosen as follows:
\begin{gather}
\begin{gathered}
\label{eq:NNN CS}
\kappa^{(1)} = \frac{N}{2}, \qquad \kappa^{(2)} = 0, \\
\Delta \kappa_{U(1)}^{(1)} = 0, \qquad \Delta \kappa_{U(1)}^{(2)} = 1, \\
\kappa_{U(1)}^{(1,2)} = -\frac{1}{2}
\end{gathered}
\end{gather}
where $\kappa^{(l)}$ is the $U(N)$ CS level while $\Delta \kappa_{U(1)}^{(l)}$ is the additional CS level shift for the $U(1)$ part of $U(N)$.\footnote{In other words, the Lagrangian terms are given by 
\begin{align}
\frac{\kappa^{(l)}}{4 \pi} \mathrm{Tr} \left(A^{(l)} d A^{(l)}-\frac{2 i}{3} A^{(l)}{}^3\right)+\frac{\Delta \kappa_{U(1)}^{(l)}}{4 \pi} \mathrm{Tr} A^{(l)} d \mathrm{Tr} A^{(l)}.
\end{align}
} $k_{U(1)}^{(l,l+1)}$ is the BF level. Those ranks of the nodes and CS/BF levels are chosen such that the theory has simple Seiberg-like duals as we explain shortly. Vortex quantum mechanics for this theory is described in figure \ref{fig:NNN vortex} with the Wilson lines \eqref{eq:Wilson lines}.

\begin{figure}[tbp]
\centering 
\includegraphics[height=.25\textheight]{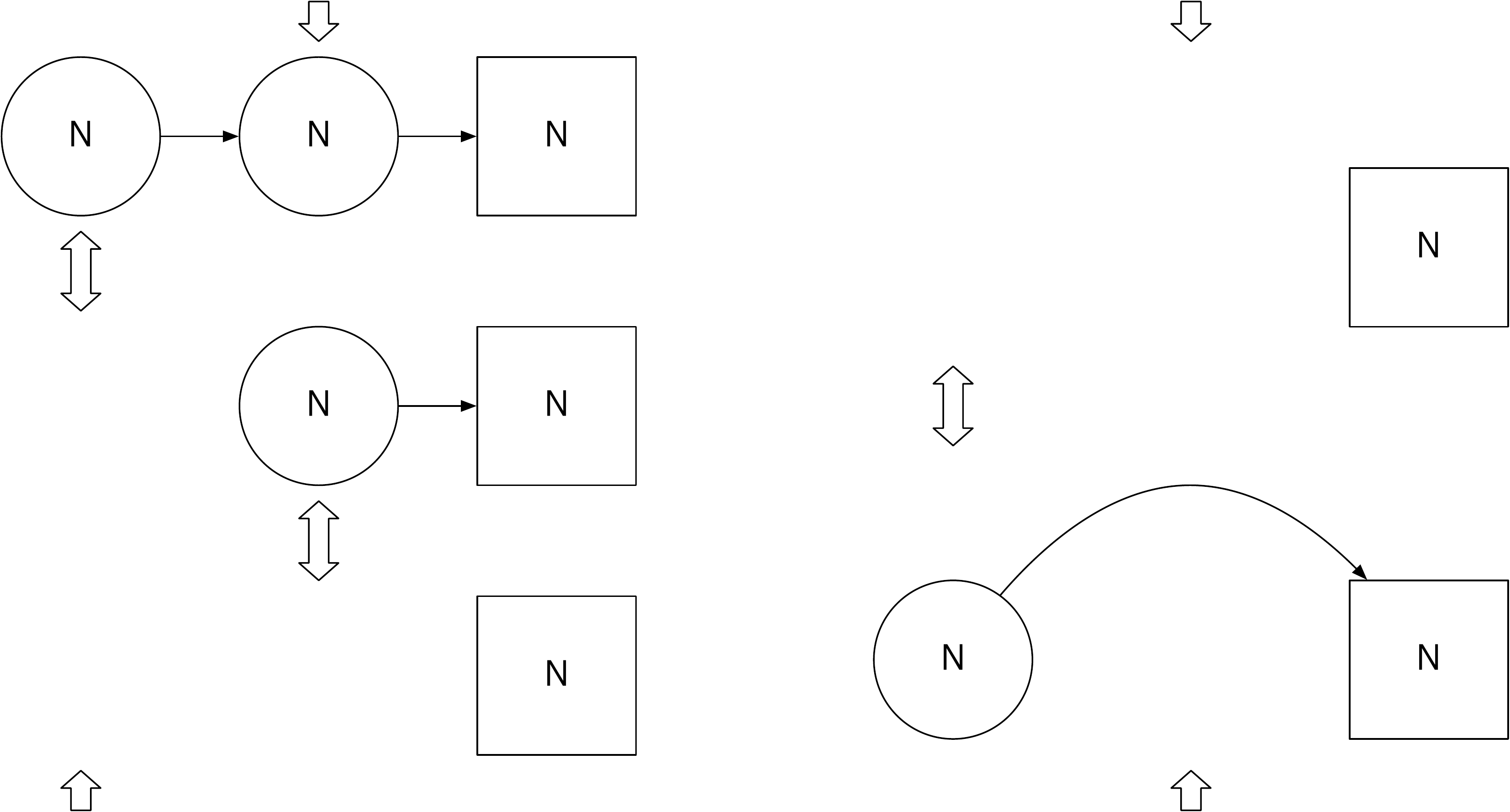}
\caption{\label{fig:NNN} The Seiberg-like duality chain of a $(N,N,N)$-type quiver theory is represented. This is a special case of figure \ref{fig:N=2}. Since every node has the same rank, a node to which the duality action is applied disappears after the duality action. Each theory may have nontrivial CS/BF interactions, which is explained in the paragraph before \eqref{eq:NNN FI}.}
\end{figure}

We illustrate the duality chain of this theory in figure \ref{fig:NNN}, which is a special case of figure \ref{fig:N=2}.
The duality chain is basically obtained by the rule examined in \cite{Aharony:1997gp,Benini:2011mf} with some additional ingredients regarding $U(1)$ factors. The first dual theory, i.e., the second theory in the duality chain, can be obtained by taking the duality action on the leftmost node. Following the duality rule in \cite{Aharony:1997gp,Benini:2011mf}, the dual gauge rank is given by $N-N = 0$; i.e, the first node should vanish. Since the number of fundamental chiral multiplets charged under the first gauge node is equal to the twice of the CS level of that node, we expect that an extra decoupled chiral multiplet appears in the dual theory as it happens in a single gauge node case. This extra chiral multiplet corresponds to a gauge invariant monopole operator in the original theory, which describes a Coulomb branch of the moduli space. Furthermore, the duality action also has a nontrivial effect on the CS level of the second node. In \cite{Benini:2011mf}, it is argued that the CS level of the second node is shifted by the amount of the CS level of the first node and becomes $\frac{N}{2}$.\footnote{We are considering the $U(N)$ CS level.}

On the other hand, the duality effect related to the BF interaction has not been discussed in the literatures. Here we trace this effect of the BF interaction by examining the Abelian example: the $(1,1,1)$ theory and its dual theory. The theory has the CS and BF interactions of the levels in \eqref{eq:NNN CS}. Since there is no distinction between the $U(N)$ CS level and the $U(1)$ CS level for $N = 1$, each gauge node just has the CS interaction of level $\kappa_l \equiv \kappa^{(l)}+\Delta \kappa_{U(1)}^{(l)}$. Without the BF interaction, the rule in \cite{Benini:2011mf} tells us that the dual theory has CS level $\kappa_1+\kappa_2 = \frac{3}{2}$. However, the level half BF interaction is not avoidable due to the regularization of the fermion in the bi-fundamental chiral multiplet. Indeed, we will see that if the BF interaction of level $-\frac{1}{2}$ is included, the dual theory has CS level $\frac{1}{2}$ rather than $\frac{3}{2}$.
\\

In order to see this effect, let us first analyze the vacuum moduli space of the $(1,1,1)$ theory. In section \ref{sec:3d}, we have reviewed that a 3d $\mathcal N = 2$ $U(1)$ theory has three types of vacua: Coulomb, Higgs and topological vacuum, For general gauge groups, vacua of mixed types are also available. The vacuum equations are given by
\begin{gather}
\sum_i 2 \pi n^{(l)}_i |Q_i|^2-F^{(l)} (\sigma) = 0, \\
m_i (\sigma)^2 |Q_i|^2 = 0
\end{gather}
where $l$ labels each $U(1)$ gauge node and $i$ labels each charged chiral multiplet. $F^{(l)} (\sigma) \equiv \xi_\text{eff}+\kappa_\text{eff} \sigma$ is given by equation \eqref{eq:F}. For the $(1,1,1)$ theory,
\begin{align}
F^{(1)} (\sigma) &= \xi_1+\frac{1}{2} \sigma_1-\frac{1}{2} \sigma_2+\frac{1}{2} |\sigma_1-\sigma_2|, \\
F^{(2)} (\sigma) &= \xi_2+\sigma_2-\frac{1}{2} \sigma_1+\frac{1}{2} |\sigma_2|-\frac{1}{2} |\sigma_1-\sigma_2|
\end{align}
where the contribution of the BF interaction is taken into account. For $\xi_l > 0$, $F^{(1)} (\sigma)$ is always positive. Therefore, $m_{12} (\sigma) = \sigma_1-\sigma_2$, mass of bi-fundamental $Q_{12}$, should vanish so that $Q_{12}$ can have the nonzero vacuum expectation value. Then $F^{(2)} (\sigma)$ is also positive such that the theory only has a Higgs vacuum at $\sigma_1 = \sigma_2 = 0$. $Q_{12}$ and $Q_{23}$ are determined by
\begin{gather}
\begin{gathered}
|Q_{12}|^2-\frac{\xi_1}{2 \pi} = 0, \\
|Q_{23}|^2-|Q_{12}|^2-\frac{\xi_2}{2 \pi} = 0.
\end{gathered}
\end{gather}
Thus, the vacuum moduli space of the theory is $\mathbb CP^0$. If either $\xi_1$ or $\xi_1+\xi_2$ vanishes, new Coulomb vacua appear. They are parameterized by $\sigma_1' \equiv \sigma_1-\sigma_2 \leq 0$ or $\sigma_2 \leq 0$ respectively. This is consistent with the duality chain in figure \ref{fig:NNN}. In the third theory, for example, those two Coulomb branches are described by two chiral multiplets of masses $\xi_1$ and $\xi_1+\xi_2$.

Now let us ask what the correct CS level of the dual theory is. The answer is more clear if we allow a generic value for the second CS level of the original theory. For later use, we also attach more flavors; i.e., we increase the rank of the last flavor node. With CS level $\kappa$ of the second node and rank $N$ of the last node, $F^{(l)} (\sigma)$'s for the original theory are rewritten in the following way:
\begin{align}
F^{(1)} (\sigma) &= \xi_1+\frac{1}{2} \sigma_1-\frac{1}{2} \sigma_2+\frac{1}{2} |\sigma_1-\sigma_2|, \\
F^{(2)} (\sigma) &= \xi_2+\kappa \sigma_2-\frac{1}{2} \sigma_1+\frac{N}{2} |\sigma_2|-\frac{1}{2} |\sigma_1-\sigma_2|.
\end{align}
Assuming $\xi_l > 0$, $F^{(1)} (\sigma)$ is always positive as before. Thus, the first gauge node only allows a Higgs vacuum solution. We ask whether there exists a Coulomb or topological vacuum solution for the second node, which satisfies $F^{(2)} (\sigma) = 0$. Since the first gauge node has a Higgs vacuum solution, $\sigma_1-\sigma_2$ should vanish. $F^{(2)} (\sigma)$ is then written as follows:
\begin{align}
F^{(2)} (\sigma) &= \xi_2+\kappa \sigma_2-\frac{1}{2} \sigma_2+\frac{N}{2} |\sigma_2|
\end{align}
such that $F^{(2)} (\sigma) = 0$ has a topological vacuum solution:
\begin{align}
\label{eq:top}
\sigma_2 = \left\{\begin{array}{cc}
-\frac{\xi_2}{\kappa+\frac{N-1}{2}}, \qquad & \kappa < -\frac{N-1}{2}, \\
-\frac{\xi_2}{\kappa-\frac{N+1}{2}}, \qquad & \kappa > \frac{N+1}{2}.
\end{array}\right.
\end{align}
Therefore, the theory has a Higgs-topological vacuum if $\kappa < -\frac{N-1}{2}$ or $\kappa > \frac{N+1}{2}$. As noted in section \ref{sec:3d} the effective theory at this classical vacuum is the $\mathcal N = 2$ $U(1)_{\kappa_\text{eff}}$ CS theory where $\kappa_\text{eff} = \kappa+\frac{N-1}{2}$ for $\kappa < -\frac{N-1}{2}$ or $\kappa_\text{eff} = \kappa-\frac{N+1}{2}$ for $\kappa > \frac{N+1}{2}$. Thus, the actual number of quantum topological vacua is $|\kappa_\text{eff}|$. In addition, there is Higgs-Higgs vacua at $\sigma_1 = \sigma_2 = 0$ regardless of the value of $\kappa$, which are determined by
\begin{gather}
\begin{gathered}
\label{eq:Higgs vacuum}
|Q_{12}|^2-\frac{\xi_1}{2 \pi} = 0, \\
\sum_{a = 1}^N |Q_{23,a}|^2-|Q_{12}|^2-\frac{\xi_2}{2 \pi} = 0.
\end{gathered}
\end{gather}
This defines the Higgs branch of the moduli space, which is given by $\mathbb CP^{N-1}$. This splits into $N$ separate vacua if we turn on small real masses for $Q_{23,a}$'s.\footnote{If masses of $Q_{23,a}$ are much smaller than $\xi_2$, the topological vacuum solution \eqref{eq:top} doesn't change.} As a result, the number of vacua, or the Witten index, of the theory is given by
\begin{align}
I = \left\{\begin{array}{cc}
-\kappa+\frac{1}{2} N+\frac{1}{2}, \qquad & \kappa < -\frac{N-1}{2}, \\
N, \qquad & -\frac{N-1}{2} \leq \kappa \leq \frac{N+1}{2}, \\
\kappa+\frac{1}{2} N-\frac{1}{2}, \qquad & \frac{N+1}{2} < \kappa.
\end{array}\right.
\end{align}
When an FI parameter is turned off, we also have Coulomb vacua. If $\xi_1 = 0$, there is a Coulomb branch of the moduli space parameterized by $\sigma_1' = \sigma_1-\sigma_2 \leq 0$. If $\xi_1+\xi_2 = 0$ and $\kappa = \frac{N+1}{2}$, there is another Coulomb branch parameterized by $\sigma_2 \leq 0$; if we choose $\kappa = -\frac{N-1}{2}$, the second Coulomb branch is parameterized by positive $\sigma_2 \geq 0$.

Taking the duality action on the first node, the dual theory is given by the $U(1)_{\kappa'}$ theory with $N$ fundamental chiral multiplets. We want to determine CS level $\kappa'$ that gives the same Witten index as the original theory. First note that $F(\sigma)$ for the dual theory is given by
\begin{align}
F(\sigma) = \xi_2+\kappa' \sigma+\frac{N}{2} |\sigma|.
\end{align}
Depending on the value of $\kappa'$, $F(\sigma) = 0$ allows the following solution:
\begin{align}
\sigma = \left\{\begin{array}{cc}
-\frac{\xi_2}{\kappa'+\frac{N}{2}}, \qquad & \kappa' < -\frac{N}{2}, \\
-\frac{\xi_2}{\kappa'-\frac{N}{2}}, \qquad & \kappa' > \frac{N}{2},
\end{array}\right.
\end{align}
which has topological multiplicity $-\kappa'-\frac{N}{2}$ or $\kappa'-\frac{N}{2}$ respectively. Taking Higgs vacua into account, the Witten index of the dual theory is given by
\begin{align}
I = \left\{\begin{array}{cc}
-\kappa'+\frac{1}{2} N, \qquad & \kappa' < -\frac{N}{2}, \\
N, \qquad & -\frac{N}{2} \leq \kappa' \leq \frac{N}{2}, \\
\kappa'+\frac{1}{2} N, \qquad & \frac{N}{2} < \kappa'.
\end{array}\right.
\end{align}
Thus, the dual theory must have CS level $\kappa' = \kappa-\frac{1}{2}$ in order to have the same Witten index as the original theory. This is different from the duality rule in \cite{Benini:2011mf}, $\kappa' = \kappa+\frac{1}{2}$, where the shift by $\frac{1}{2}$ is the result of the CS interaction of the first node. Here we see that there is an additional shift by $-1$, which should be the result of the BF interaction of level $-\frac{1}{2}$.

In general we expect that our level half BF interaction yields another CS level shift for the second gauge node, but only for the $U(1)$ part because the BF interaction is only relevant for the $U(1)$ parts of the gauge nodes. It turns out that this expectation is consistent with what we observe from the vortex partition function analysis. Thus, adopting this additional shift due to the presence of the BF interaction, the CS level for the $U(1)$ part of the second node is shifted by $-1$ after the duality action. This is why we have engineered the original $(N,N,N)$ theory to have the CS/BF levels as in \eqref{eq:NNN CS}, especially $\Delta \kappa_{U(1)}^{(2)} = 1$. The second theory in the duality chain is then the $U(N)_{\frac{N}{2}}$ theory without the additional CS level shift for the $U(1)$ part. The theory also contains $N$ fundamental chiral multiplets and one decoupled chiral multiplet. The gauge theory sector is a familiar SQCD. We know that this theory is dual to a single chiral multiplet. Thus, the next dual theory, the third theory in the duality chain, is a theory of two chiral multiplets. These two chiral multiplets describe the two Coulomb branches of the moduli space, which we have observed in the vacuum analysis. The fifth and fourth theories can be obtained by sequentially taking the duality actions on the second node and the first node of the original theory. They are also given by the $U(N)_{\frac{N}{2}}$ theory and the two-chiral theory. Those five theories in the duality chain have the following FI parameters:
\begin{align}
\label{eq:NNN FI}
\begin{array}{lll}
\xi^{\mathbf 1}_1 = \xi_1, &\qquad& \xi^{\mathbf 1}_2 = \xi_2, \\
\xi^{\mathbf 2}_1 = \varnothing, &\qquad& \xi^{\mathbf 2}_2 = \xi_1+\xi_2, \\
\xi^{\mathbf 3}_1 = \varnothing, &\qquad& \xi^{\mathbf 3}_2 = \varnothing, \\
\xi^{\mathbf 4}_1 = \varnothing, &\qquad& \xi^{\mathbf 4}_2 = \varnothing, \\
\xi^{\mathbf 5}_1 = \xi_1, &\qquad& \xi^{\mathbf 5}_2 = \varnothing
\end{array}
\end{align}
where $\varnothing$ means that the corresponding gauge node vanishes. We assume $\xi_1,\xi_2 > 0$.

We also comment that if the 3d gauge group is Abelian, our construction of vortex quantum mechanics has ambiguity for the Wilson line part, which reflects the 3d CS action. For the Abelian 3d gauge group, there is no distinction between $\kappa^{(l)}$ and $\Delta \kappa_{U(1)}^{(l)}$. However, their counterparts in vortex quantum mechanics are distinguished. One is realized as a gauge Wilson line while the other is realized as a global Wilson line. Even though both constructions of vortex quantum mechanics give the same vortex partition function, their wall-crossing behaviors are completely different. We propose that this ambiguity can be resolved by demanding the wall-crossing behavior compatible with the 3d Seiberg-like duality. For the Abelian $(1,1,1)$ theory, we follow the prescription that the assignment of CS levels for non-Abelian gauge groups \eqref{eq:NNN CS} still holds for the Abelian gauge group. Then, as we will see, the proposed construction of vortex quantum mechanics for the $(1,1,1)$ theory shows claimed wall-crossing behavior compatible with the 3d Seiberg-like duality.
\\

Now let us move on to the vortex partition functions. Recall that vortex quantum mechanics for the $(N,N,N)$ theory is given by figure \ref{fig:NNN vortex}. We propose that the other theories in the duality chain share the same vortex quantum mechanics with different 1d FI parameters shown in \eqref{eq:N=2 FI}. Using the JK-residue formula \eqref{eq:JK}, one can obtain the index of vortex quantum mechanics in each FI chamber. The results are organized as follows:
\begin{align}
\begin{aligned}
\label{eq:ind NNN}
Z_{\mathbf 1} &= \sum_{n_1}^\infty \sum_{n_2 = 0}^\infty (-w_1)^{n_1} w_2^{n_2} I^{n_1,n_2}_{\mathbf 1} = \mathrm{PE} \left[-\frac{x^{\frac{N}{2}+1} w_1}{1-x^2}-\frac{x^{\frac{N}{2}+1} \tau^{\frac{N}{2}} w_1 w_2}{1-x^2}\right], \\
Z_{\mathbf 2} &= \sum_{n_1}^\infty \sum_{n_2 = 0}^\infty (-w_1)^{n_1} w_2^{n_2} I^{n_1,n_2}_{\mathbf 2} = \mathrm{PE} \left[-\frac{x^{\frac{N}{2}
+1} \tau^{\frac{N}{2}} w_1 w_2}{1-x^2}\right], \\
Z_{\mathbf 3} &= \sum_{n_1}^\infty \sum_{n_2 = 0}^\infty (-w_1)^{n_1} w_2^{n_2} I^{n_1,n_2}_{\mathbf 3} = 1, \\
Z_{\mathbf 4} &= \sum_{n_1}^\infty \sum_{n_2 = 0}^\infty (-w_1)^{n_1} w_2^{n_2} I^{n_1,n_2}_{\mathbf 4} = 1, \\
Z_{\mathbf 5} &= \sum_{n_1}^\infty \sum_{n_2 = 0}^\infty (-w_1)^{n_1} w_2^{n_2} I^{n_1,n_2}_{\mathbf 1} = \mathrm{PE} \left[-\frac{x^{\frac{N}{2}+1} w_1}{1-x^2}\right]
\end{aligned}
\end{align}
where we have introduced the additional negative sign to vorticity fugacity $w_1$ to simplify the expressions. We have numerically checked \eqref{eq:ind NNN} up to $n_1,n_2 \leq 3$ for $N =1,2,3$.

Since the theories in the duality chain except the first one have a single or no gauge node, we already know their vortex partition functions. The quantum mechanics indices \eqref{eq:ind NNN} in different FI chambers correctly reproduce them. Furthermore, the wall-crossing factors, which are given by the ratios of two indices in adjacent FI chambers, are exactly the contributions of the gauge singlet chiral multiplets appearing after the duality actions.\footnote{See equation \eqref{eq:V-}.} Thus, the $(N,N,N)$ theory is the first example showing that the proposed equivalence between the 3d Seiberg-like duality and the wall-crossing of vortex quantum mechanics can be generalized to an $\mathcal N = 2$ linear quiver gauge theory.
\\

Our next example is a $(1,1,N)$-type quiver theory. The $(N,N,N)$ theory we have considered so far flows to a free theory of two chiral multiplets in the IR limit. The $(1,1,N)$ theory, on the other hand, flows to an interacting IR fixed point. This example shows that the equivalence between the 3d Seiberg-like duality and the vortex wall-crossing can be generalized to an $\mathcal N = 2$ linear quiver theory flowing to an interacting IR fixed point.

\begin{figure}[tbp]
\centering 
\includegraphics[height=.25\textheight]{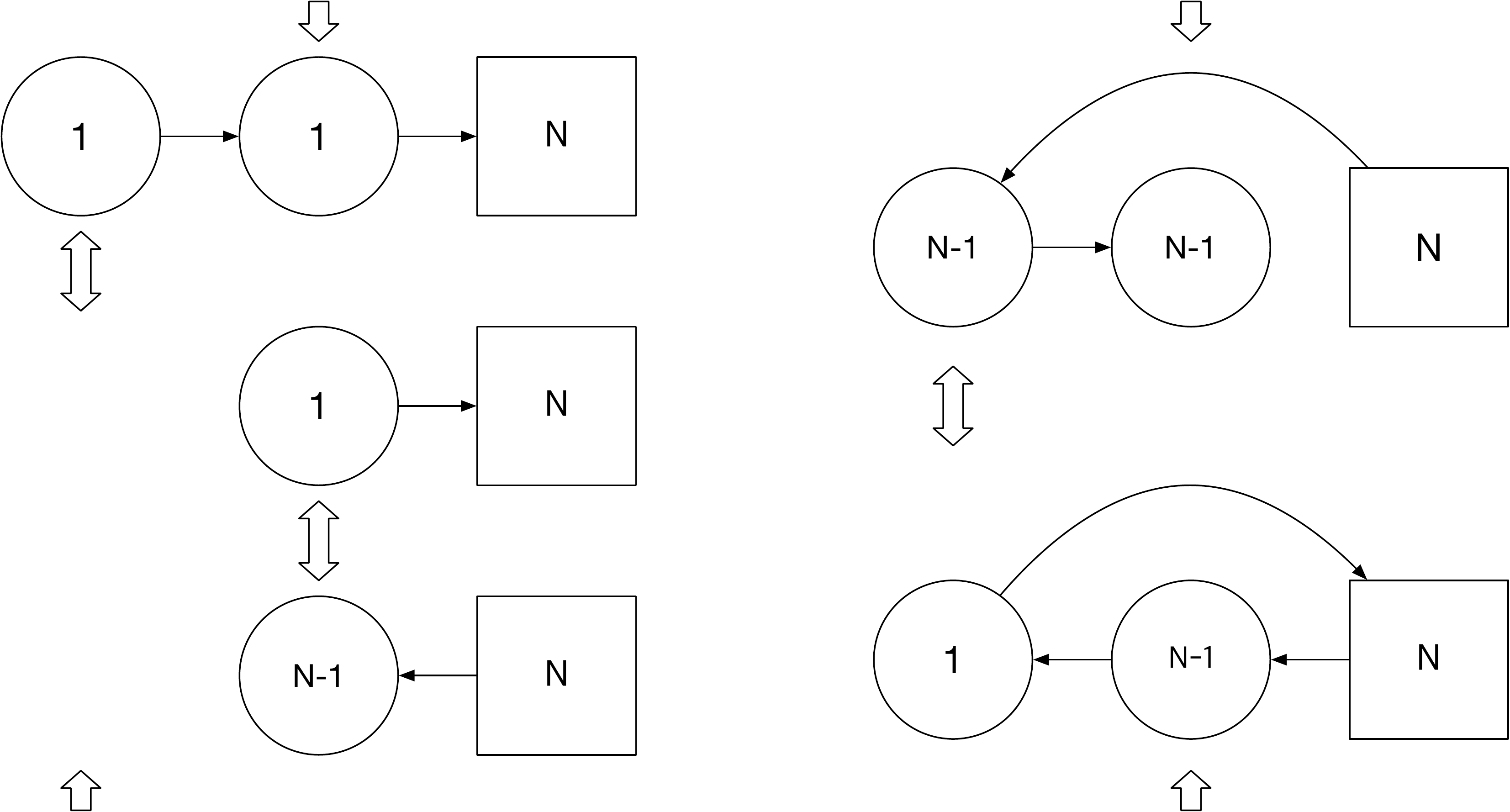}
\caption{\label{fig:11N} The Seiberg-like duality chain of a $(1,1,N)$-type quiver theory is represented. Note that in the third duality action, the roles of the first node and the second node are exchanged.}
\end{figure}

The duality chain containing the $(1,1,N)$ theory is shown in figure \ref{fig:11N}.
The CS/BF levels of the first theory are chosen as follows:
\begin{gather}
\begin{gathered}
\label{eq:11N CS}
\kappa^{(1)} = \frac{1}{2}, \qquad \kappa^{(2)} = \frac{N-1}{2}, \\
\Delta \kappa_{U(1)}^{(1)} = 0, \qquad \Delta \kappa_{U(1)}^{(2)} = 1, \\
\kappa_{U(1)}^{(1,2)} = -\frac{1}{2}
\end{gathered}
\end{gather}
so that the second theory is given by the $U(1)_{\frac{N}{2}}$ theory with $N$ fundamental chiral multiplets and one decoupled chiral multiplet $V_{1-}$. Again this is a familiar SQCD, but now it flows to an interacting IR fixed point. Its Seiberg-like (or Aharony) dual theory is given by the $U(N-1)_{-\frac{N}{2}}$ theory with $N$ anti-fundamental chiral multiplets and two gauge singlet chiral multiplets $V_{1-}$ and $V_{2-}$. $V_{1-}$ is from the first duality action and $V_{2-}$ is from the second duality action. $V_{2-}$ couples to the gauge theory via the following superpotential \cite{Aharony:1997gp,Benini:2011mf}:
\begin{align}
\label{eq:sup}
W = V_{2-} v_+
\end{align}
where $v_+$ is the gauge invariant monopole operator of the $U(N-1)_{-\frac{N}{2}}$ theory. Both $V_{1-}$ and $V_{2-}$ describe the Coulomb branches of the moduli space while $v_+$ is lifted by the superpotential \eqref{eq:sup}. The fifth and fourth theories are obtained by taking the duality actions on the second node and the first node of the first theory sequentially. The FI parameters of those five theories in the duality chain are as follows:
\begin{align}
\label{eq:11N FI}
\begin{array}{lll}
\xi^{\mathbf 1}_1 = \xi_1, &\qquad& \xi^{\mathbf 1}_2 = \xi_2, \\
\xi^{\mathbf 2}_1 = \varnothing, &\qquad& \xi^{\mathbf 2}_2 = \xi_1+\xi_2, \\
\xi^{\mathbf 3}_1 = \varnothing, &\qquad& \xi^{\mathbf 3}_2 = -\xi_1-\xi_2, \\
\xi^{\mathbf 4}_1 = -\xi_1, &\qquad& \xi^{\mathbf 4}_2 = -\xi_2, \\
\xi^{\mathbf 5}_1 = \xi_1, &\qquad& \xi^{\mathbf 5}_2 = -\xi_2.
\end{array}
\end{align}
\\

\begin{figure}[tbp]
\centering 
\includegraphics[height=.2\textheight]{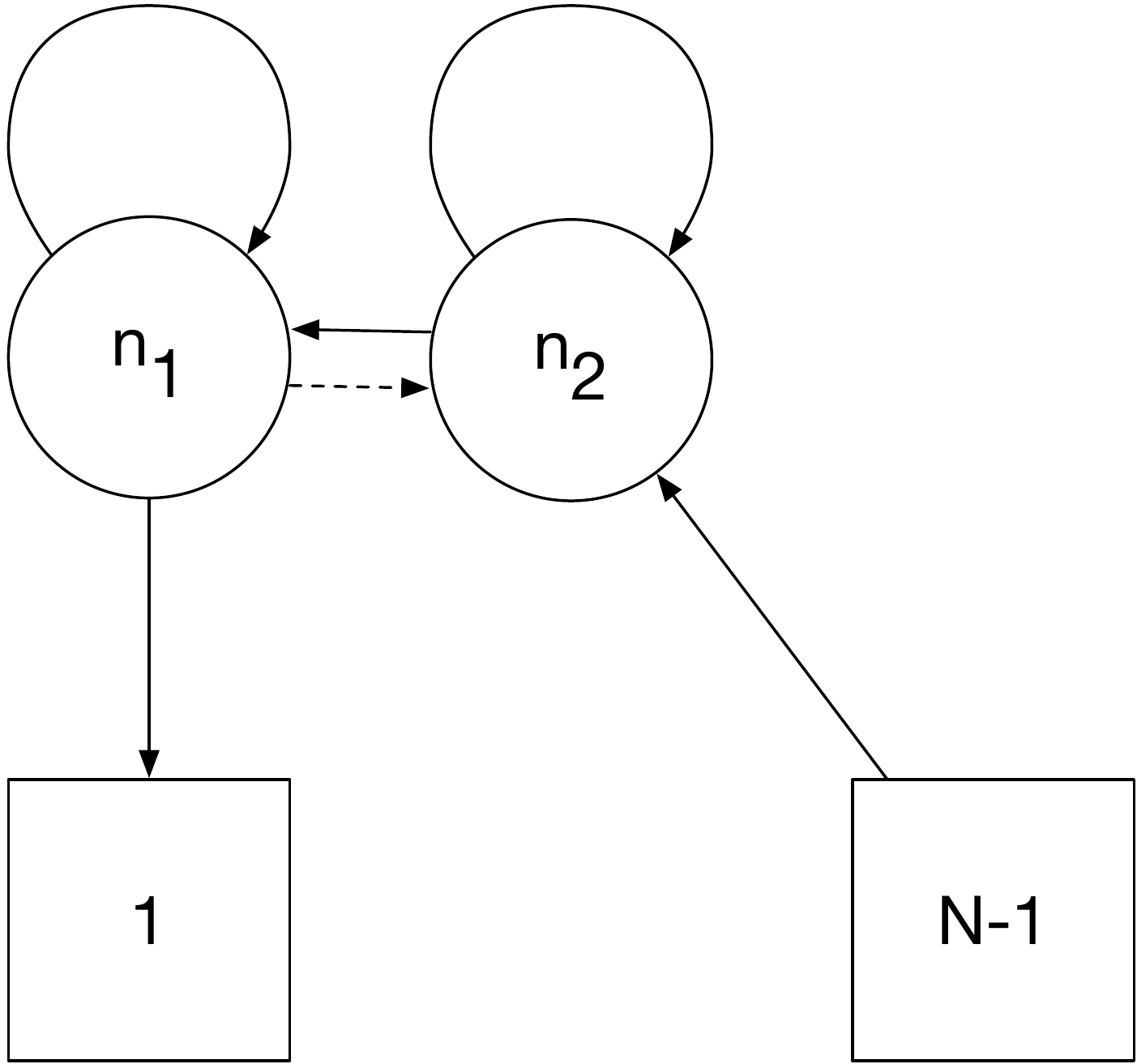}
\caption{\label{fig:11N vortex} Vortex quantum mechanics for a $(1,1,N)$-type quiver theory is illustrated. If the parent 3d theory has CS/BF interactions, the quantum mechanics also has the corresponding Wilson lines shown in \eqref{eq:Wilson lines}.}
\end{figure}

From the result of section \ref{sec:N=2}, vortex quantum mechanics for the $(1,1,N)$ theory is represented by the quiver diagram shown in figure \ref{fig:11N vortex}.
The index in each FI chamber can be computed using the JK-residue formula \eqref{eq:JK}. One should note that, except the fifth theory, the theories in the duality chain are equivalent to linear quiver gauge theories with positive FI parameters up to charge conjugation. We already know the vortex partition functions of those types of theories from \eqref{eq:N=2 result} using charge conjugation. Indeed, we have confirmed that the index in each FI chamber of quantum mechanics in figure \ref{fig:11N vortex} correctly reproduces the vortex partition function from \eqref{eq:N=2 result}\footnote{We need some care for the fourth theory. The fourth theory has the following gauge CS/BF interactions:
\begin{gather}
\begin{gathered}
\kappa^{(1)} = -\frac{1}{2}, \qquad \kappa^{(2)} = -\frac{N-1}{2}, \\
\Delta \kappa_{U(1)}^{(1)} = 0, \qquad \Delta \kappa_{U(1)}^{(2)} = 1, \\
\kappa_{U(1)}^{(1,2)} = -\frac{1}{2}.
\end{gathered}
\end{gather}
Moreover, the theory includes the BF interactions of levels $\pm \frac{N}{2}$ between global $U(1)_A$ and each gauge $U(1)$. Their contributions to the vortex partition function are realized as shifts of fugacities $w_1 \rightarrow w_1 \tau^{-\frac{N}{2}}$ and $w_2 \rightarrow w_2  \tau^{\frac{N}{2}}$, which should be taken into account when we compute the vortex partition function from \eqref{eq:N=2 result} using charge conjugation.} except the fifth theory, for which we don't have other independent computations of the vortex partition function.

Since those four vortex partition functions are now more complicated, we don't explicitly write them down here. Nevertheless, the wall-crossing factors given by the ratios of two indices in adjacent FI chambers are still written in terms of simple Plethystic exponentials:
\begin{align}
\begin{aligned}
Z^\text{wall}_1 &= Z_{\mathbf 1}/Z_{\mathbf 2} = \mathrm{PE} \left[-\frac{x^{\frac{3}{2}} w_1}{1-x^2}\right], \\
Z^\text{wall}_2 &= Z_{\mathbf 2}/Z_{\mathbf 3} = \mathrm{PE} \left[-\frac{x^{2-\frac{N}{2}} \tau^{\frac{N}{2}} w_1 w_2}{1-x^2}\right], \\
Z^\text{wall}_3 &= Z_{\mathbf 3}/Z_{\mathbf 4} = \mathrm{PE} \left[\frac{x^{\frac{5}{2}-\frac{N}{2}} \tau^{\frac{N}{2}} w_2}{1-x^2}\right]
\end{aligned}
\end{align}
where we have used vorticity fugacities $-w_1$ and $-w_2$ with additional negative signs for simplification. Each wall-crossing factor should reproduce the contribution of gauge singlet chiral multiplet appearing after each duality action. Those chiral multiplets are summarized in table \ref{tab:monopoles} with their global symmetry charges.
\begin{table}[tbp]
\centering
\begin{tabular}{|c|cccc|}
\hline
 & $U(1)_R$ & $U(1)_A$ & $U(1)_T^{(1)}$ & $U(1)_T^{(2)}$ \\
\hline
$V_{1-}$ & $\frac{1}{2}$ & $0$ & $-1$ & $0$ \\
$V_{2-}$ & $\frac{N}{2}$ & $\frac{N}{2}$ & $-1$ & $-1$ \\
$V_{3+}$ & $\frac{5}{2}-\frac{N}{2}$ & $\frac{N}{2}$ & $0$ & $1$ \\
\hline
$v_{1-}^{\mathbf 1}$ & $\frac{1}{2}$ & $0$ & $-1$ & $0$ \\
$v_{2-}^{\mathbf {1,2}}$ & $\frac{N}{2}$ & $\frac{N}{2}$ & $-1$ & $-1$ \\
$v_{2+}^{\mathbf {3,4}}$ & $2-\frac{N}{2}$ & $-\frac{N}{2}$ & $1$ & $1$ \\
$v_{3-}^{\mathbf 4}$ & $\frac{N}{2}-\frac{1}{2}$ & $-\frac{N}{2}$ & $0$ & $-1$ \\
\hline
\end{tabular}
\caption{\label{tab:monopoles} The global symmetry charges of gauge singlet chiral multiplets $V_{1-},V_{2-},V_{3+}$ and those of the gauge invariant monopole operators are summarized. The $U(1)_R$ charges are UV values assigning the vanishing charges to the bi-fundamental chiral multiplets. The IR superconformal values are given by combinations of the UV R-charge and other global $U(1)$ charges, which can be determined by $Z$-extremization \cite{Jafferis:2010un}.}
\end{table}

As seen in the vacuum analysis, the $(1,1,N)$ theory with the CS/BF levels \eqref{eq:11N CS}, and its dual theories, have two Coulomb branches for vanishing FI parameters. Those Coulomb branches are parameterized by two chiral operators among extra chiral fields $V_{1,2-}$ and the gauge invariant monopole operators of the theory. The others should be massive. Thus, we expect the following superpotential for each theory:
\begin{align}
\begin{aligned}
\label{eq:sups}
W_{\mathbf 1} &= 0, \\
W_{\mathbf 2} &= 0, \\
W_{\mathbf 3} &= V_{2-} v_{2+}^{\mathbf 3}, \\
W_{\mathbf 4} &= V_{2-} v_{2+}^{\mathbf 4}+V_{3+} v_{3-}^{\mathbf 4}
\end{aligned}
\end{align}
where $v_{i \pm}^{\mathbf k}$'s are gauge invariant monopole operators of the $\mathbf k$-th theory whose global charges are summarized in table \ref{tab:monopoles}. The Coulomb branches are parameterized by monopole operators $v_{1,2-}^{\mathbf 1}$ in the first theory, parameterized by chiral field $V_{1-}$ and monopole operator $v_{2-}^{\mathbf 2}$ in the second theory, and parameterized by chiral fields $V_{1,2-}$ in the third theory and in the fourth theory. For the third and fourth theories, the gauge invariant monopole operators become massive due to the superpotentials in \eqref{eq:sups}. In addition, monopole operator $v_{3-}^{\mathbf 4}$ is supposed to be dualized to a single chiral field. Thus, the last term in $W_{\mathbf 4}$ induces the mass terms for both $V_{3+}$ and (dualized) $v_{3-}^{\mathbf 4}$ such that both of them are integrated out in the IR limit. This explains how we have the correct number of the chiral operators parameterizing the Coulomb branches. We also comment that the superpotentials in \eqref{eq:sups} should be understood as the effective descriptions of the interactions between the monopole operators and the extra chiral fields. For a single gauge node case, i.e, the Aharony duality, this kind of superpotential can be derived from the Giveon-Kutasov duality \cite{Giveon:2008zn}, whose superpotential doesn't explicitly depend on nonperturbative monopole operators \cite{Intriligator:2013lca}.
\\

The fifth theory in the duality chain, on the other hand, is not a type of a linear quiver theory we have examined in section \ref{sec:N=2}. The fifth theory has a loop in its quiver diagram representation (see figure \ref{fig:11N}) and has FI parameters of opposite signs (see \eqref{eq:11N FI}). For this reason, we cannot independently compute the vortex partition function of the fifth theory using \eqref{eq:N=2 result} to confirm whether the quantum mechanics index in the fifth FI chamber, $Z_{\mathbf 5}$, gives the correct vortex partition function or not. We still, however, observe that $Z_{\mathbf 5}$ shows desired wall-crossing behavior: its ratios to the adjacent indices are expressed by simple Plethystic exponentials:
\begin{align}
\begin{aligned}
\label{eq:wall11N}
Z^\text{wall}_4 &= Z_{\mathbf 4}/Z_{\mathbf 5} = \mathrm{PE} \left[\frac{x^{\frac{3}{2}} w_1}{1-x^2}\right], \\
Z^\text{wall}_5 &= Z_{\mathbf 5}/Z_{\mathbf 1} = \mathrm{PE} \left[-\frac{x^{\frac{5}{2}-\frac{N}{2}} \tau^{\frac{N}{2}} w_2}{1-x^2}+\frac{x^{2-\frac{N}{2}} \tau^{\frac{N}{2}} w_1 w_2}{1-x^2}\right],
\end{aligned}
\end{align}
which can be identified as the contributions of chiral multiplets. If $Z_{\mathbf 5}$ indeed reproduces the vortex partition function of the fifth theory, \eqref{eq:wall11N} tells us that the fifth theory contains extra chiral multiplets $V_{2-}$ and $V_{3+}$. Considering the two Coulomb branches of the moduli space, $V_{3+}$ should be massive while $V_{2-}$ should remain massless so as to parameterize a Coulomb branch. The other Coulomb branch should be parameterized by a gauge invariant monopole operator of the theory. We admit that it is not clear if our $Z_{\mathbf 5}$ does give the vortex partition function of the fifth theory. Nevertheless, our wall-crossing approach could be a hint for understanding the vortex dynamics and the Aharony duality of this theory.
\\

So far we have seen that the equivalence between the 3d Seiberg-like duality and the wall-crossing of vortex quantum mechanics is observed in $\mathcal N = 2$ linear quiver examples as well. Here our analysis is restricted to some examples for which we can independently compute the vortex partition functions of dual theories so that we can test the proposal. It will be an interesting problem to test the proposal for general 3d $\mathcal N = 2$ theories.

In particular it is a very nontrivial observation that the wall-crossing spectra of an infinite number of vortex quiver theories are organized as a simple Plethystic exponential, which is identified as the 3d determinant of extra gauge singlet chiral multiplets appearing in the dual theory. One can demand this wall-crossing behavior as a constraint beyond the vortex partition function, which is relatively easy to compute but doesn't constrain the vortex quantum mechanics much. Indeed, we have used the wall-crossing behavior to fine-tune vortex quantum mechanics for 3d Abelian theories that have ambiguity not distinguished by the vortex partition function. Although our analysis is not a proof of the equivalence between the 3d Seiberg-like duality and the vortex wall-crossing, the observations we have made strongly suggest that they have a close relation to each other.
\\

\subsection{Fundamental vortices and particle-vortex duality}

We have examined two types of 3d linear quiver gauge theories preserving either $\mathcal N = 4$ or $\mathcal N = 2$ supersymmetry. In this section, we would like to address the role of \emph{fundamental vortices}, which is explained shortly, in the Seiberg-like dualities of those 3d theories. Let us go back to the $T_\rho [SU(N)]$ example. We considered the 3d duality chain shown in figure \ref{fig:T[SU(N)]}, whose corresponding 1d vortex quiver is given by figure \ref{fig:T[SU(N)] vortex}. While the vortex quiver and the ranks of the nodes in figure \ref{fig:T[SU(N)] vortex} remain the same, the corresponding vortex numbers, or the Chern numbers, defined by $q^{(l)} = \frac{1}{2 \pi} \mathrm{Tr} \int F^{(l)}$ shift nontrivially under the Seiberg-like duality. The Chern numbers $(q^{(1)},q^{(2)}) = (n_1,n_2)$ in the first theory, for example, are mapped to $(n_2-n_1,n_2)$ in the second theory, as inferred from the FI parameter map \eqref{eq:FI map} and BPS vortex mass $\sum_l q^{(l)} \xi_l^{\mathbf k}$. As such, the basic building blocks for the vortex sector are not going to be dictated by ``unit'' Chern numbers.

On the other hand, every BPS vortex has positive mass, $\sum_l q^{(l)} \xi_l^{\mathbf k} > 0$, whose collection is spanned by vortices with some minimal masses, e.g., $\xi_1$ and $\xi_2$ prior to the duality map for the current example. The vortices with
such minimal masses are what we refer as the fundamental vortices, and can be regarded as basic building blocks for topological sectors of the theory. The other BPS vortices are composite states of the fundamental vortices and their masses are given by linear combinations of the minimal masses with nonnegative integer coefficients.

For the first theory in the duality chain, one can easily identify the fundamental vortices as those with vortex numbers (1,0) and (0,1). As we mentioned, the vortex numbers (1,0) and (0,1) are mapped to vortex numbers (-1,1) and (1,1) in the second theory, yet the corresponding BPS masses $\xi_1$ and $\xi_2$ are unchanged. Thus, the fundamental vortices are mapped to the fundamental vortices under the duality, and they are described by the same 1d quiver on either sides of the dual pair. The ranks of the 1d vortex quiver now has natural interpretations as the numbers of fundamental vortices rather than the vortex numbers, which are not invariant under the Seiberg-like duality.

Note that vortices with the ``unit'' Chern numbers, such as (1,0) and (0,1), and the fundamental vortices coincide for the first theory while this is not generally true for other theories in the duality chain. Again for the current example, the fundamental vortices in the second theory have Chern numbers (-1,0) and (1,1). The (0,1) vortex in the second theory, which has the ``unit'' Chern
number, is indeed a composite state of the above two fundamental vortices. In our discussion, nevertheless, one always finds a canonical theory in a given duality chain where the fundamental vortices carry ``unit'' Chern numbers. In such a canonical theory,
the vortex quiver is thus naturally defined.

In the example we discussed, the canonical theory is always the first theory in the duality chain where we have chosen positive 3d
FI parameters $\xi_1,\ldots,\xi_L > 0$. If we consider general 3d FI parameters, on the other hand, the canonical theory would be different. Indeed, inferring from the equivalence between the 3d Seiberg-like duality and the wall-crossing of vortex quantum mechanics, one can define vortex quantum mechanics for a $T_\rho [SU(N)]$ theory having general FI parameters. For example, vortex quantum mechanics for $T_{[M_1,M_2-M_1,M_3-M_2]} [SU(M_3)]$ with general $\xi_1$, $\xi_2$ is given by figure \ref{fig:T[SU(N)] vortex} with parameters in table \ref{tab:general FI}, which can be straightforwardly generalized to the other $T_\rho [SU(N)]$ theories.
\begin{table}[tbp]
\centering
\begin{tabular}{|l|l||l|l|l|l|l|}
\hline
$\xi_1$ & $\xi_2$ & $N_1$ & $N_2-N_1$ & $N_3-N_2$ & $\zeta_1$ & $\zeta_2$ \\
\hline
$> 0$ & $> 0$ & $M_1$ & $M_2-M_1$ & $M_3-M_2$ & $> 0$ & $> 0$ \\
$< 0$ & $> |\xi_1|$ & $M_2-M_1$ & $M_1$ & $M_3-M_2$ & $< 0$ & $> |\zeta_1|$ \\
$> 0$ & $< -|\xi_1|$ & $M_2-M_1$ & $M_3-M_2$ & $M_1$ & $< -|\zeta_2|$ & $> 0$ \\
$< 0$ & $< 0$ & $M_3-M_2$ & $M_2-M_1$ & $M_1$ & $< 0$ & $< 0$ \\
$< -|\xi_2|$ & $> 0$ & $M_3-M_2$ & $M_1$ & $M_2-M_1$ & $> 0$ & $< -|\zeta_1|$ \\
$> |\xi_2|$ & $< 0$ & $M_1$ & $M_3-M_2$ & $M_2-M_1$ & $> |\zeta_2|$ & $< 0$ \\
\hline
\end{tabular}
\caption{\label{tab:general FI} Parameters of vortex quantum mechanics for $T_{[M_1,M_2-M_1,M_3-M_2]} [SU(M_3)]$ theory having FI parameters $\xi_1,\xi_2$. For the 3d FI parameters $\xi_1,\xi_2$ in the above ranges, vortex quantum mechanics is given by the quiver diagram in figure \ref{fig:T[SU(N)] vortex} with the above flavor ranks $N_1,N_2-N_1,N_3-N_2$ and 1d FI parameters $\zeta_1,\zeta_2$.}
\end{table}
When $\xi_1 < 0$ and $\xi_2 > |\xi_1|$, the vortex quiver of ranks $(n_1,n_2)$ corresponds to vortex numbers $(n_2-n_1,n_2)$; the (first) theory is thus not the canonical one. Instead, the canonical theory is given by the second theory in the duality chain, which is obtained by taking the duality action on the leftmost node. In that case, both 3d FI parameters become positive due to the effect of the duality, and the vortex numbers are equal to the ranks of the vortex quiver. For the other ranges of $\xi_1$, $\xi_2$, the canonical theory can be found in a similar manner: it is given by the theory in a duality chain having all positive FI parameters. This is generally true, at least for the $\mathcal N = 4, 2$ linear quiver gauge theories we examined.
\\

So far we have only discussed generic cases that the fundamental vortices are mapped to the fundamental vortices under the duality. However, there are some obvious exceptions. For example, let us consider the $T_{[N,0]} [SU(N)]$ theory, i.e., the $U(N)$ theory with $N$ hypermultiplets. The theory has BPS vortices, whose spectrum is captured by the refined Witten index of vortex quantum mechanics we considered. The fundamental vortex carries vortex number 1 and has mass $\xi$ assuming the positive FI parameter $\xi > 0$. However, if we consider its Seiberg-like duality, the dual theory is a non-gauge theory of $N$ twisted hypermultiplets. Since no gauge symmetry exists on the dual side, there appears no obvious way to talk about vortices, fundamental or not. Does this mean
that the notion of fundamental vortices cannot be sustained generally?

A resolution comes, paradoxically, from realizing that such exceptional cases should be considered as part of more general phenomena. Whenever a wall-crossing occurs, the collection of 1d quantum states that contribute to the vortex partition function changes. Even when we can talk about semiclassical fundamental vortices on both sides, the 1d wall-crossing means that there are no 1-1 map between quantum BPS states in the topological sector. Despite this, however, the common vortex quiver theory serves as an intermediate and very useful device that computes the vortex partition functions on both side of duality. The extreme cases we just mentioned are merely special limits where the absence of the contributing quantum states become glaringly obvious due to the lack of semiclassical vortices.

This naturally leads to the main question of this note: when there is 1d wall-crossing, what happens to those quantum states that account for the difference? Since 3d partition function remains invariant and since wall-crossing means appearance of and disappearance of supersymmetric quantum states made up of fundamental vortices, the latter discrepancy must be canceled by something else. As is obvious with Aharony dual pairs, this restoration is achieved by extra neutral chiral multiplets or hypermultiplets, contributing perturbative states. Since the two theories connected by the duality are supposed to be equivalent at full quantum level, we can say that some part of quantum states built from fundamental vortices are replaced by elementary excitations on the dual side, and vice versa. In other words, the wall-crossing of the quiver theory for the fundamental vortices is a 1d manifestation of a 3d {\it particle-vortex duality}.  The exceptional cases above, where vortex quantum mechanics wall-crosses into a trivial chamber, signals that the quantum vortices in the 3d theory are completely replaced by perturbative particle-like states in the dual description.

Another type of limiting cases occur when there is no wall-crossing while at least one side is devoid of the 3d gauge group. Let us consider an $\mathcal N = 2$ SQCD, the $U(2)$ theory with two fundamental matters and no Chern-Simons term. The dual theory has no gauge group and consists of free chiral multiplets only. Clearly the vortex partition function has to be trivial on both side, even though the gauged side does admit classical vortices. The 1d quiver theory does exist, yet no fully quantum BPS vortex state contributes. Nevertheless, one again finds that the vortex quiver theory  gives the correct (trivial) vortex partition functions.

These limiting cases show that  the notion of fundamental vortices emerges from low energy topological sector in the canonical theory among a 3d duality chain, yet their 1d quiver theory remains reliable throughout the duality chain, at least for the purpose of computing the vortex partition function, and, as an aside, also captures 3d particle-vortex duality faithfully.
\\

\section{Summary}
\label{sec:summary}

We conclude the note by summarizing our results.
\\

We have constructed 1d quantum mechanical systems which describe the low energy dynamics of vortices in 3d linear quiver theories preserving either $\mathcal N = 4$ or $\mathcal N = 2$ supersymmetry. The $\mathcal N = 4$ theories we consider are called $T_\rho [SU(N)]$, which are linear quiver theories with a single flavor node of the flavor symmetry rank $N$ at one end and its integer partition $\rho$ that specifies the ranks of the gauge nodes. One can realize a $T_\rho [SU(N)]$ theory as a Hanany-Witten system, in which the vortices show up as D1-brane segments ending on D3-branes. The low energy dynamics of the vortices is captured by the effective theory of those D1-branes. Inferring from this brane setup, we wrote down the 1d quiver description of the vortex quantum mechanics. We have numerically confirmed that the refined Witten indices of such vortex quantum mechanics, when appropriately combined and also augmented by 3d perturbative contributions from non-topological sectors, correctly reproduce the known partition functions of $T_\rho [SU(N)]$. The latter is computed via 3d Coulombic localization on $S^3_b$, followed by the factorization.

Unlike $T_\rho [SU(N)]$, $\mathcal N = 2$ linear quiver gauge theories in general do not have known brane setups,
so the derivation of 1d vortex theory is less straightforward. Instead, we take the following indirect route.
We first invoke the mass deformation of $T_\rho [SU(N)]$ that preserves $\mathcal N = 2$ supersymmetry, and
keep track of how this descends down to 1d vortex dynamics. As expected, this breaks the supersymmetry
of the vortex dynamics in half, and we study how the resulting 1d vortex dynamics behaves under 3d Seiberg-like
duality, and found essentially the same behavior as in $\mathcal N=4$ theories, modulo subtleties associated with
3d Chern-Simons terms.
The 1d refined Witten indices we computed are consistent with the factorization of 3d partition functions, examined in appendix \ref{sec:fact}, and can be identified as 3d vortex partition functions. Also, as another consistency check, we found their 2d limits to be perfect matches with previously known results.
\\

One main question is how these vortex quantum mechanics behave under 3d Seiberg-like dualities. As inferred from factorized 3d partition functions, the 1d quiver quantum mechanics for vortices appear by and large unaffected. The main change is how the 3d duality is realized as a sign change of some 1d FI. We have observed that 1d FI constants are related to $1/g_{YM}^2$, for $\mathcal N=4$ theories realized as Hanany-Witten, which approaches zero in the original theory under the D-brane motion that emulates the Seiberg-like duality. As such, the 1d FI constants for vortices are the natural parameter that interpolate between a dual pair and the 3d duality manifest merely as a 1d wall-crossing of the same vortex quiver quantum mechanics. If one computes refined Witten indices of the latter via JK-residue, with the
natural choice of the auxiliary Lie Algebra vector $\eta=\zeta$, such a change will affect which subsets of residues should be picked up. For simplest of theories, these choices can be seen quite naturally and generally reflected
in the UV field content of the dual pair, even for those examples where the sign change does not
result in 1d wall-crossing.

In those cases where 1d wall-crossing induces a discontinuity of the 1d refined Witten indices,
vortex contributions to the factorized 3d partition function differ between the dual pair.
Miraculously, however, such wall-crossing discontinuities of an infinite number of quiver
quantum mechanics, differing by ranks, sum up neatly into a 3d determinant factor of neutral
chiral multiplets. The invariance of the 3d partition function is then restored by the
extra neutral chiral multiplets on the dual side, whose determinant cancels this wall-crossing
discontinuity. Duality of this type has been long known as the Aharony duality.
As an example, we first considered $\mathcal N = 2$ SQCD-like theories which admit Aharony duals. We computed the wall-crossing
indices of corresponding vortex quantum mechanics for different 1d FI parameters: $\zeta > 0$
and $\zeta < 0$. These indices were shown to fit the factorized partition functions of the
3d dual pairs in the above sense, respectively, supporting the interpretation of the Aharony
duality as the wall-crossing of the vortex quantum mechanics. We also listed several 3d
linear quiver theories where this phenomenon occurs, say $T_\rho [SU(N)]$ theories
for general 3d FI parameters. For $\mathcal N=2$ linear quiver examples, the wall-crossing
interpretation goes further than this; it actually constrains and fine-tunes the vortex
quantum mechanics.

Accepting this correspondence between the 3d duality and the vortex wall-crossing, our
computation can be also regarded as a proof of the Aharony duality at the level of vortex
partition functions. Since the vortex partition function is a building block of diverse
supersymmetric partition functions that accept rotational isometry, the vortex partition
function identity we obtained can be used to prove the agreements of the supersymmetric
partition functions under the Aharony duality. We provide explicit proofs of this
for several classes of 3d partition functions in appendix \ref{sec:identity}; in
particular, the proof for topological twisted $S^2 \times S^1$ with the angular momentum refinement is a new result.
\\

\acknowledgments

We thank Sciarappa Antonio, Hee-Cheol Kim, Hyungchul Kim, Joonho Kim, Sungjay Lee, Jaemo Park, Rak-Kyeong Seong for useful comments and conversations.
\\

\appendix

\section{Factorization for 3d $\mathcal N = 2$ linear quiver gauge theories}
\label{sec:fact}

For 3d supersymmetric gauge theories, various supersymmetric partition functions on curved 3-manifolds have been studied, e.g., \cite{Kim:2009wb,Kapustin:2009kz,Jafferis:2010un,Hama:2010av,Imamura:2011su,Hama:2011ea,Kapustin:2011jm,Benini:2011nc,Imamura:2011wg,Alday:2012au,Benini:2015noa,Benini:2016hjo,Closset:2016arn}. Especially if one considers spaces of $S^1$-bundle over $S^2$, the partition functions on those spaces (with appropriate twistings) are written in the following factorized form \cite{Krattenthaler:2011da,Dimofte:2011ju,Pasquetti:2011fj,Beem:2012mb,Hwang:2012jh,Taki:2013opa,Cecotti:2013mba,Fujitsuka:2013fga,Benini:2013yva,Benini:2015noa}:\footnote{This is explicitly worked out for maximally chiral \cite{Benini:2011mf} theories.}
\begin{align}
Z = \sum_\text{vacua} Z^\text{pert} Z^\text{vort} Z^\text{antiv}
\end{align}
where $Z^\text{vort}$ is the vortex partition function on $\Omega$-deformed $\mathbb R^2_\Omega \times S^1$. This shows that various supersymmetric partition functions are built upon the same building block, the vortex partition function. The only differences are the perturbative contribution, $Z^\text{pert}$, and the parameter identifications. Thus, the study of the vortex partition function enlarges our understandings of other supersymmetric partition functions as well. Moreover, one can reverse the idea such that the factorization can be used as a method of obtaining the vortex partition function of a given theory. In both directions the study of the factorizations of supersymmetric partition functions plays an important role in understanding 3d supersymmetric gauge theories and their partition functions.
\\

In this appendix, we work out the factorization of a supersymmetric partition function, especially the (angular momentum refined) topologically twisted $S^2 \times S^1$ indices \cite{Benini:2015noa} of 3d $\mathcal N = 2$ linear quiver gauge theories that we examined in section \ref{sec:N=2}. In this way, we obtain the vortex partition functions of the $\mathcal N = 2$ linear quiver gauge theories, which agree with the results obtained from the quantum mechanics computation in section \ref{sec:N=2}.

The topologically twisted index of a 3d $\mathcal N = 2$ theory on $S^2 \times S^1$ is defined as follows:
\begin{align}
\label{eq:tind}
I = \mathrm{Tr}_{\mathcal H} (-1)^F e^{-\beta H} e^{i J_f A_f} \zeta^{2 L_\phi}
\end{align}
with a topological twist on $S^2$ \cite{Benini:2015noa}. $\zeta$, the fugacity for the angular momentum $L_\phi$ on $S^2$, is also included. One can compute the twisted index using the supersymmetric localization. The result is written in terms of the Jeffrey-Kirwan residue \cite{1993alg.geom..7001J} as follows:
\begin{align}
I = \frac{1}{|\mathsf W|} \sum_{m \in \Gamma_{\mathfrak h}} \sum_{u_* \in \mathfrak M_\text{sing}^*} \text{JK-Res}_{u = u_*} (Q_{u_*},\eta) Z_\text{int} (m,u)+\text{boundary contribution}.
\end{align}
$u$ denotes the set of bosonic zero-modes, which are elements in $\mathfrak M = H \times \mathfrak h$ where $H$ is the maximal torus in gauge group $G$ and $\mathfrak h$ is the corresponding Cartan subalgebra. $\mathfrak M_\text{sing}^*$ is a set of singular points in $\mathfrak M$. $m$ is the magnetic flux on $S^2$ living in the co-root lattice, $\Gamma_{\mathfrak h}$. Due to gauge equivalent configurations, the result is divided by the Weyl group order, $|\mathsf W|$. The integrand, $Z_\text{int}$, is given by the product of the classical contributions of (mixed) Chern-Simons actions and the 1-loop determinants of chiral/vector multiplets:
\begin{align}
\begin{aligned}
\label{eq:CS}
Z^\text{CS} &= \prod_{\rho \in F} (x^\rho)^{\kappa \rho(m)}, \\
Z^\text{BF} &= x_1^{\kappa_{12} m_2} x_2^{\kappa_{12} m_1},
\end{aligned}
\end{align}
\begin{align}
\begin{aligned}
Z^\text{chiral} &= \prod_{\rho \in R} \frac{\left(x^\rho\right)^{\frac{B}{2}}}{\left(x^\rho \zeta^{1-B};\zeta^2\right)_B}, \qquad B = \rho(m)-q+1, \\
Z^\text{vector} &= \zeta^{-\sum_{\alpha > 0} |\alpha(m)|} \prod_{\alpha} \left(1-x^\alpha \zeta^{|\alpha(m)|}\right) (i du)^r
\end{aligned}
\end{align}
where we denote $e^{i u}$ by $x$. $\rho$ is a weight of representation $R$ of a chiral multiplet (in particular we denote the fundamental representation by $F$) and $\alpha$ is a root of the gauge group. $q$ is the R-charge of a chiral multiplet. A more detailed explanation for each component can be found in \cite{Benini:2015noa}.

The boundary pieces are given by the residues at the asymptotic regions, whose contributions to the index are determined by the effective CS levels. For a single gauge node case, the $U(N)_\kappa$ theory with $(N_f,N_a)$ flavors, the boundary contribution can be avoided if $|\kappa| \leq \frac{|N_f-N_a|}{2}$, which is called maximally chiral \cite{Benini:2011mf}. Such a theory only has Higgs vacua if we choose FI parameter of sign $\mathrm{sgn}(N_f-N_a)$. For general cases with multiple gauge nodes, on the other hand, the condition of the vanishing boundary condition is not manifest and should be worked out case by case because of possible complications due to various $U(1)$ CS/BF interactions. It is plausible to expect that a criterion would be whether the theory has Higgs vacua only or not. In the following computations, we assume that the CS levels and the number of matters are chosen such that the boundary contribution vanishes. More discussions about the boundary contribution to the twisted index can be found in \cite{Benini:2015noa,Benini:2016hjo,Closset:2016arn}.
\\

For the 3d $\mathcal N = 2$ linear quiver theory in figure \ref{fig:def T[SU(N)]}, the twisted index is written as follows:
\begin{align}
\label{eq:N=2 tind}
I &= \frac{1}{\left(\prod_{l = 1}^L N_l !\right)}\sum_{m^{(l)}_i} \oint \frac{dx^{(l)}_i}{2 \pi i x^{(l)}_i} \left(\prod_{l = 1}^L \xi_l^{\sum_i m^{(l)}_i}\right) Z^\text{CS} Z^\text{BF} \zeta^{-\sum_l \sum_{i < j} |m^{(l)}_i-m^{(l)}_j|} \nonumber \\
&\quad \times \left(\prod_{l = 1}^L \prod_{i \neq j}^{N_l} \left(1-x^{(l)}_i {x^{(l)}_j}^{-1} \zeta^{|m^{(l)}_i-m^{(l)}_j|}\right)\right) \nonumber \\
&\quad \times \left(\prod_{l = 1}^L \prod_{i = 1}^{N_l} \prod_{j = 1}^{N_{l+1}} \frac{\left(x^{(l)}_i {x^{(l+1)}_j}^{-1}\right)^{\frac{1}{2} (m^{(l)}_i-m^{(l+1)}_j+1)}}{\left(x^{(l)}_i {x^{(l+1)}_j}^{-1} \zeta^{-m^{(l)}_i+m^{(l+1)}_j};\zeta^2\right)_{m^{(l)}_i-m^{(l+1)}_j+1}}\right)
\end{align}
where $x^{(L+1)}_i = y_i$ and $m^{(L+1)}_i = 0$. The fugacities for global symmetries $\prod_{l = 1}^L U(1)_T^{(l)} \times U(N_{L+1})$ are denoted by $\xi_l$ for $l = 1,\ldots,L$ and $y_i$ respectively. We have set the R-charges of the chiral multiplets zero for simplicity. The CS/BF contributions, $Z^\text{CS}$ and $Z^\text{BF}$, are determined by \eqref{eq:CS}. The integration contour is determined by the JK-residue rule and encloses the following poles if we choose the auxiliary JK-vector $\eta = \vec 1$:
\begin{align}
x^{(l)}_i &= y_{k^{L-l+1}(i)} \zeta^{m^{(l)}_i-2 \sum_{a = l}^L p^{(a)}_{k^{a-l}(i)}}, \\
&\equiv y_{k^{(l)}_i} \zeta^{m^{(l)}_i-2 \mathsf p^{(l)}_i}, \label{eq:pole}
\end{align}
which are the intersections of hyperplanes
\begin{align}
x^{(l)}_i = x^{(l+1)}_{k(i)} \zeta^{m^{(l)}_i-m^{(l+1)}_{k(i)}-2 p^{(l)}_i}, \qquad 0 \leq p^{(l)}_i \leq m^{(l)}_i-m^{(l+1)}_{k(i)}.
\end{align}
The contributing poles are classified by integers $p^{(l)}_i$ and functions $k_{(l)}:I^{(l)} \rightarrow I^{(l+1)}$ where $I^{(l)}$ is the set of gauge indices of the $l$-th gauge node (or the flavor indices if $l = L+1$). In the above equations and the followings, we omit subscript $(l)$ of function $k_{(l)}$ unless it is necessary. Note that the residue vanishes if $m^{(l)}_i-m^{(l+1)}_{k(i)} < 0$.

Now we evaluate each component in \eqref{eq:N=2 tind} at the pole \eqref{eq:pole}. First, the 1-loop determinant of the vector multiplet is evaluated as follows:
\begin{align}
& \zeta^{-\sum_l \sum_{i < j} |m^{(l)}_i-m^{(l)}_j|} \prod_{l = 1}^L \prod_{i \neq j}^{N_l} \left(1-x^{(l)}_i {x^{(l)}_j}^{-1} \zeta^{|m^{(l)}_i-m^{(l)}_j|}\right) = \nonumber \\
& \zeta^{-\sum_l \sum_{i < j} (m^{(l)}_i-m^{(l)}_j)} \prod_{l = 1}^L \prod_{i < j}^{N_l} \left(1-y_{k^{(l)}_i} y_{k^{(l)}_j}^{-1} \zeta^{2 m^{(l)}_i-2 m^{(l)}_j-2 \mathsf p^{(l)}_i+2 \mathsf p^{(l)}_j}\right) \left(1-y_{k^{(l)}_j} y_{k^{(l)}_i}^{-1} \zeta^{-2 \mathsf p^{(l)}_j+2 \mathsf p^{(l)}_i}\right)
\end{align}
where we have assumed $m^{(l)}_1 \geq \ldots \geq m^{(l)}_{N_l}$ without loss of generality. $m$ in a different order gives the same contribution. Now we define new variables $n^{(l)}_i$ and $\bar n^{(l)}_i$ in terms of $m^{(l)}_i$ and $p^{(l)}_i$:
\begin{gather}
\begin{gathered}
n^{(l)}_i = m^{(l)}_i-m^{(l+1)}_{k(i)}-p^{(l)}_i, \qquad \bar n^{(l)}_i = p^{(l)}_i, \\
0 \leq n^{(l)}_i, \bar n^{(l)}_i \leq m^{(l)}_i-m^{(l+1)}_{k(i)},
\end{gathered}
\end{gather}
which correspond to the vorticities. The vector multiplet determinant is written in terms of $n^{(l)}_i$ and $\bar n^{(l)}_i$:
\begin{align}
\label{eq:vector}
\zeta^{-\sum_l \sum_{i < j} (\mathsf n^{(l)}_i-\mathsf n^{(l)}_j+\mathsf{\bar n}^{(l)}_i-\mathsf{\bar n}^{(l)}_j)} \prod_{l = 1}^L \prod_{i < j}^{N_l} \left(1-y_{k^{(l)}_i} y_{k^{(l)}_j}^{-1} \zeta^{2 \mathsf n^{(l)}_i-2 \mathsf n^{(l)}_j}\right) \left(1-y_{k^{(l)}_j} y_{k^{(l)}_i}^{-1} \zeta^{-2 \mathsf{\bar n}^{(l)}_j+2 \mathsf{\bar n}^{(l)}_i}\right)
\end{align}
where $\mathsf n^{(l)}_i = \sum_{a = l}^L n^{(a)}_{k^{a-l}(i)} = m^{(l)}_i-\mathsf p^{(l)}_i$ and $\mathsf{\bar n}^{(l)}_i = \sum_{a = l}^L \bar n^{(a)}_{k^{a-l}(i)} = \mathsf p^{(l)}_i$. Likewise the 1-loop determinants of chiral multiplets are evaluated as follows:
\begin{align}
\label{eq:chiral}
& \prod_{l = 1}^L \prod_{i = 1}^{N_l} \prod_{j = 1}^{N_{l+1}} \frac{\left(x^{(l)}_i {x^{(l+1)}_j}^{-1}\right)^{\frac{1}{2} (m^{(l)}_i-m^{(l+1)}_j+1)}}{\left(x^{(l)}_i {x^{(l+1)}_j}^{-1} \zeta^{-m^{(l)}_i+m^{(l+1)}_j};\zeta^2\right)_{m^{(l)}_i-m^{(l+1)}_j+1}} \nonumber \\
%
%
&= \prod_{l = 1}^L \prod_{i = 1}^{N_l} \prod_{j = 1}^{N_{l+1}} \frac{\left(y_{k^{(l)}_i} y_{k^{(l+1)}_j}^{-1} \zeta^{\mathsf n^{(l)}_i-\mathsf n^{(l+1)}_j-\mathsf{\bar n}^{(l)}_i+\mathsf{\bar n}^{(l+1)}_j}\right)^{\frac{1}{2} (\mathsf n^{(l)}_i-\mathsf n^{(l+1)}_j+\mathsf{\bar n}^{(l)}_i-\mathsf{\bar n}^{(l+1)}_j+1)}}{\left(y_{k^{(l)}_i} y_{k^{(l+1)}_j}^{-1} \zeta^{-2 \mathsf{\bar n}^{(l)}_i+2 \mathsf{\bar n}^{(l+1)}_j};\zeta^2\right)_{\mathsf n^{(l)}_i-\mathsf n^{(l+1)}_j+\mathsf{\bar n}^{(l)}_i-\mathsf{\bar n}^{(l+1)}_j+1}}.
\end{align}
\eqref{eq:chiral} is nothing but a product of multiple q-Pochhammer symbols up to some monomial factor. Using the following identity of q-Pochhammer symbol:
\begin{align}
\label{eq:q-Pochh ident}
\left(y \zeta^{-2 \bar n};\zeta^2\right)_{n+\bar n+1} = (1-y) \left(y \zeta^2;\zeta^2\right)_n \left(y \zeta^{-2};\zeta^{-2}\right)_{\bar n},
\end{align}
one can distinguish the $\mathsf n^{(l)}_i$-dependent part and the $\mathsf{\bar n}^{(l)}_i$-dependent part (as well as the part independent both of $\mathsf n^{(l)}_i$ and of $\mathsf{\bar n}^{(l)}_i$):
\begin{align}
& \left(y_{k^{(l)}_i} y_{k^{(l+1)}_j}^{-1} \zeta^{-2 \mathsf{\bar n}^{(l)}_i+2 \mathsf{\bar n}^{(l+1)}_j};\zeta^2\right)_{\mathsf n^{(l)}_i-\mathsf n^{(l+1)}_j+\mathsf{\bar n}^{(l)}_i-\mathsf{\bar n}^{(l+1)}_j+1} \nonumber \\
&= \left(1-y_{k^{(l)}_i} y_{k^{(l+1)}_j}^{-1}\right) \left(y_{k^{(l)}_i} y_{k^{(l+1)}_j}^{-1} \zeta^2;\zeta^2\right)_{\mathsf n^{(l)}_i-\mathsf n^{(l+1)}_j} \left(y_{k^{(l)}_i} y_{k^{(l+1)}_j}^{-1} \zeta^{-2};\zeta^{-2}\right)_{\mathsf{\bar n}^{(l)}_i-\mathsf{\bar n}^{(l+1)}_j}.
\end{align}
Moreover, \eqref{eq:vector} can be also massaged using the same identity \eqref{eq:q-Pochh ident}. For example, the second factor in \eqref{eq:vector} is written in the following way:
\begin{align}
\left(1-y_{k^{(l)}_i} y_{k^{(l)}_j}^{-1} \zeta^{2 \mathsf n^{(l)}_i-2 \mathsf n^{(l)}_j}\right)
%
%
&= \left(-y_{k^{(l)}_i} y_{k^{(l)}_j}^{-1} \zeta^{\mathsf n^{(l)}_i-\mathsf n^{(l)}_j+1}\right)^{\mathsf n^{(l)}_i-\mathsf n^{(l)}_j} \left(1-y_{k^{(l)}_i} y_{k^{(l)}_j}^{-1}\right) \nonumber \\
&\quad \times \left(y_{k^{(l)}_i}^{-1} y_{k^{(l)}_j};\zeta^2\right)_{-\mathsf n^{(l)}_i+\mathsf n^{(l)}_j}^{-1} \left(y_{k^{(l)}_i} y_{k^{(l)}_j}^{-1};\zeta^2\right)_{\mathsf n^{(l)}_i-\mathsf n^{(l)}_j}^{-1}.
\end{align}
Next we evaluate the classical action contribution:
\begin{align}
\label{eq:cl}
\left(\prod_{l = 1}^L \xi_l^{\sum_i m^{(l)}_i}\right) Z^\text{CS} Z^\text{BF}
\end{align}
where the last two factors are determined by \eqref{eq:CS}. The first factor of \eqref{eq:cl} is the FI term contribution. The latter two factors are gauge CS/BF term contributions. More explicitly, we turn on $SU(N)$ CS level $\kappa^{(l)}$ and $U(1)$ CS level $\kappa^{(l)}+\Delta \kappa_{U(1)}^{(l)}$ for each gauge node as well as BF level $\kappa_{U(1)}^{(l,l+1)}$ for each pair of adjacent gauge nodes. At the pole \eqref{eq:pole} each factor of \eqref{eq:cl} is evaluated as follows:
\begin{align}
\prod_{l = 1}^L \xi_l^{\sum_i m^{(l)}_i} &= \prod_{l = 1}^L \xi_l^{\sum_i (\mathsf n^{(l)}_i+\mathsf{\bar n}^{(l)}_i)}, \\
Z^\text{CS} &= \prod_{l = 1}^L \left(\prod_{i = 1}^{N_l} y_{k^{(l)}_i}^{\kappa^{(l)} (\mathsf n^{(l)}_i+\mathsf{\bar n}^{(l)}_i)} \zeta^{\kappa^{(l)} (\mathsf n^{(l)}_i{}^2-\mathsf{\bar n}^{(l)}_i{}^2)}\right) \nonumber \\
&\quad \times \left(\prod_{i = 1}^{N_l} y_{k^{(l)}_i}\right)^{\Delta \kappa_{U(1)}^{(l)} \sum_i (\mathsf n^{(l)}_i+\mathsf{\bar n}^{(l)}_i)}  \zeta^{\Delta \kappa_{U(1)}^{(l)} [(\sum_i \mathsf n^{(l)}_i)^2-(\sum_i \mathsf{\bar n}^{(l)}_i)^2]}, \label{eq:CS1}\\
Z^\text{BF} &= \prod_{l = 1}^{L-1} \left(\prod_{i = 1}^{N_l} y_{k^{(l)}_i}\right)^{\kappa_{U(1)}^{(l,l+1)} \sum_i (\mathsf n^{(l+1)}_i+\mathsf{\bar n}^{(l+1)}_i)} \left(\prod_{i = 1}^{N_{l+1}} y_{k^{(l+1)}_i}\right)^{\kappa_{U(1)}^{(l,l+1)} \sum_i (\mathsf n^{(l)}_i+\mathsf{\bar n}^{(l)}_i)} \nonumber \\
&\quad \times \zeta^{2 \kappa_{U(1)}^{(l,l+1)} [(\sum_i \mathsf n^{(l)}_i) (\sum_i \mathsf n^{(l+1)}_i)-(\sum_i \mathsf{\bar n}^{(l)}_i) (\sum_i \mathsf{\bar n}^{(l+1)}_i)]}. \label{eq:CS2}
\end{align}
Due to the parity anomaly, those CS/BF levels satisfy the quantization conditions:
\begin{gather}
\kappa^{(l)}+\frac{N_{l+1}-N_{l-1}}{2} \in \mathbb Z, \\
\Delta \kappa_{U(1)}^{(l)} \in \mathbb Z, \\
\kappa_{U(1)}^{(l,l+1)} \in \mathbb Z-\frac{1}{2}.
\end{gather}
\\

Collecting all these results, the twisted index is reorganized in the following way:
\begin{align}
I = \frac{1}{\left(\prod_{l = 1}^L N_l!\right)} \sum_{m^{(l)}_i} \sum_{k} \sum_{p^{(l)}_i} \left(\prod_{l = 1}^L \tilde \xi_l^{\sum_i (\mathsf n^{(l)}_i+\mathsf{\bar n}^{(l)}_i)}\right) Z^\text{CS} Z^\text{BF} Z^0 Z^+ Z^-,
\end{align}
\begin{align}
Z^0 &= \left(\prod_{l = 1}^L \prod_{i \neq j}^{N_l} \left[y_{k^{(l)}_i} y_{k^{(l)}_j}^{-1}\right]\right) \left(\prod_{l = 1}^L \prod_{i = 1}^{N_l} \prod_{j = 1}^{N_{l+1}} \left[y_{k^{(l)}_i} y_{k^{(l+1)}_j}^{-1}\right]\right)^{-1}, \\
Z^+ &= \left(\prod_{l = 1}^L \prod_{i \neq j}^{N_l} \left[y_{k^{(l)}_i} y_{k^{(l)}_j}^{-1};\zeta^2\right]_{\mathsf n^{(l)}_i-\mathsf n^{(l)}_j}\right)^{-1} \left(\prod_{l = 1}^L \prod_{i = 1}^{N_l} \prod_{j = 1}^{N_{l+1}} \left[y_{k^{(l)}_i} y_{k^{(l+1)}_j}^{-1} \zeta^2;\zeta^2\right]_{\mathsf n^{(l)}_i-\mathsf n^{(l+1)}_j}\right)^{-1}, \label{eq:vort} \\
Z^- &= \left.Z^+\right|_{\zeta \rightarrow \zeta^{-1}}
\end{align}
where $\tilde \xi_l = (-1)^{N_l+1} \xi_l$ and we have defined the shifted q-Pochhammer symbol $[a;q]_n$:
\begin{align}
[a;q]_n = a^{-\frac{1}{2} n} q^{-\frac{1}{4} n (n-1)} (a;q)_n
\end{align}
and $[a] \equiv [a;q]_1$. $Z^\text{CS}$ and $Z^\text{BF}$ are given by \eqref{eq:CS1} and \eqref{eq:CS2} respectively.

We need to recast the summations over $m^{(l)}_i$ and $p^{(l)}_i$ in terms of $n^{(l)}_i$ and $\bar n^{(l)}_i$. Recall that we have aligned $m$ in the descending order. Now we restore the all possible permutations of $m$, which give the same contributions. Taking into account the vanishing residues, we can replace the $m^{(l)}_i$/$p^{(l)}_i$-summations by
\begin{align}
\sum_{m^{(l)}_i \in \mathbb Z} \sum_{p^{(l)}_i = 0}^{m^{(l)}_i-m^{(l+1)}_{k(i)}} = \sum_{n^{(l)}_i \geq 0} \sum_{\bar n^{(l)}_i \geq 0}.
\end{align}
Next let us consider the summation over $k$. Due to factor $[y_{k^{(l)}_i} y_{k^{(l)}_j}^{-1}]$ in $Z^0$, $k^{(l)}_i$ and $k^{(l)}_j$ should be different if $i \neq j$; i.e., $k$ is an injective function. Furthermore, the above expressions are independent of $k_{(l)}$ for $l = 1,\ldots,L-1$. Thus, we can fix $k_{(l)}$ such that $k_{(l)} (i) = i$ for $l = 1,\ldots,L-1$ and multiply $\prod_{l = 1}^{L-1} \frac{N_{l+1}!}{(N_{l+1}-N_l)!}$, which is the number of possible $k_{(l)}$ for $l = 1,\ldots,L-1$. The summation over $k$ is then reduced as follows:
\begin{align}
\sum_k = \frac{\left(\prod_{l = 1}^L N_l!\right)}{\left(\prod_{l = 1}^L (N_l-N_{l-1})!\right)} \sum_{k_{(L)}} = \frac{\left(\prod_{l = 1}^L N_l!\right)}{\left(\prod_{l = 1}^{L+1} (N_l-N_{l-1})!\right)} \sum_{\sigma}
\end{align}
where $\sigma$ is a permutation acting on $I^{(L+1)}$. $k^{(l)}_i$ in the previous expressions is replaced by $\sigma(i)$.

Therefore, the twisted index is finally written in the following factorized form:
\begin{align}
I = \frac{1}{\prod_{l = 1}^{L+1} (N_l-N_{l-1})!} \sum_{\sigma} Z^\text{pert} Z^\text{vort} Z^\text{antiv},
\end{align}
\begin{align}
& Z^\text{pert} = \left(\prod_{l = 1}^L \prod_{i \neq j}^{N_l} \left[y_{\sigma(i)} y_{\sigma(j)}^{-1}\right]\right) \left(\prod_{l = 1}^L \prod_{i = 1}^{N_l} \prod_{j = 1 (\neq i)}^{N_{l+1}} \left[y_{\sigma(i)} y_{\sigma(j)}^{-1}\right]\right)^{-1}, \\
& Z^\text{vort} = \sum_{n^{(l)}_i \geq 0} \left(\prod_{l = 1}^L \tilde \xi_l^{\sum_i \mathsf n^{(l)}_i}\right) Z^\text{cl} \nonumber \\
& \times \left(\prod_{l = 1}^L \prod_{i \neq j}^{N_l} \left[y_{\sigma(i)} y_{\sigma(j)}^{-1};\zeta^2\right]_{\mathsf n^{(l)}_i-\mathsf n^{(l)}_j}\right)^{-1} \left(\prod_{l = 1}^L \prod_{i = 1}^{N_l} \prod_{j = 1}^{N_{l+1}} \left[y_{\sigma(i)} y_{\sigma(j)}^{-1} \zeta^2;\zeta^2\right]_{\mathsf n^{(l)}_i-\mathsf n^{(l+1)}_j}\right)^{-1}, \\
& Z^\text{antiv} = \left.Z^\text{vort}\right|_{\zeta \rightarrow \zeta^{-1}}
\end{align}
where $\mathsf n^{(l)}_i = \sum_{a = l}^L n^{(a)}_{k^{a-l}(i)}$. The classical action contribution, $Z^\text{cl}$, is given by
\begin{align}
Z^\text{cl} &= \prod_{l = 1}^L \left(\prod_{i = 1}^{N_l} y_{\sigma(i)}^{\kappa^{(l)} \mathsf n^{(l)}_i} \zeta^{\kappa^{(l)} \mathsf n^{(l)}_i{}^2}\right) \left(\prod_{i = 1}^{N_l} y_{\sigma(i)}\right)^{\Delta \kappa_{U(1)}^{(l)} \sum_i \mathsf n^{(l)}_i}  \zeta^{\Delta \kappa_{U(1)}^{(l)} (\sum_i \mathsf n^{(l)}_i)^2} \nonumber \\
&\quad \times \left(\prod_{i = 1}^{N_l} y_{\sigma(i)}\right)^{\kappa_{U(1)}^{(l,l+1)} \sum_i \mathsf n^{(l+1)}_i} \left(\prod_{i = 1}^{N_{l+1}} y_{\sigma(i)}\right)^{\kappa_{U(1)}^{(l,l+1)} \sum_i \mathsf n^{(l)}_i} \zeta^{2 \kappa_{U(1)}^{(l,l+1)} (\sum_i \mathsf n^{(l)}_i) (\sum_i \mathsf n^{(l+1)}_i)}.
\end{align}
Sometimes it is convenient to replace $\sum_{n^{(l)}_i \geq 0}$ by $\sum_{\mathsf n^{(l)}_i \geq 0}$ because $Z^\text{vort}$ is completely written in terms of $\mathsf n^{(l)}_i$. This replacement is possible because $Z^\text{vort}$ vanishes if $n^{(l)}_i < 0$ although $\mathsf n^{(l)}_i \geq 0$. In section \ref{sec:N=2}, we show that $Z^\text{vort}$ obtained here agrees with the refined Witten index of vortex quantum mechanics in figure \ref{fig:def vortex}.
\\

\begin{figure}[tbp]
\centering 
\includegraphics[width=.5\textwidth]{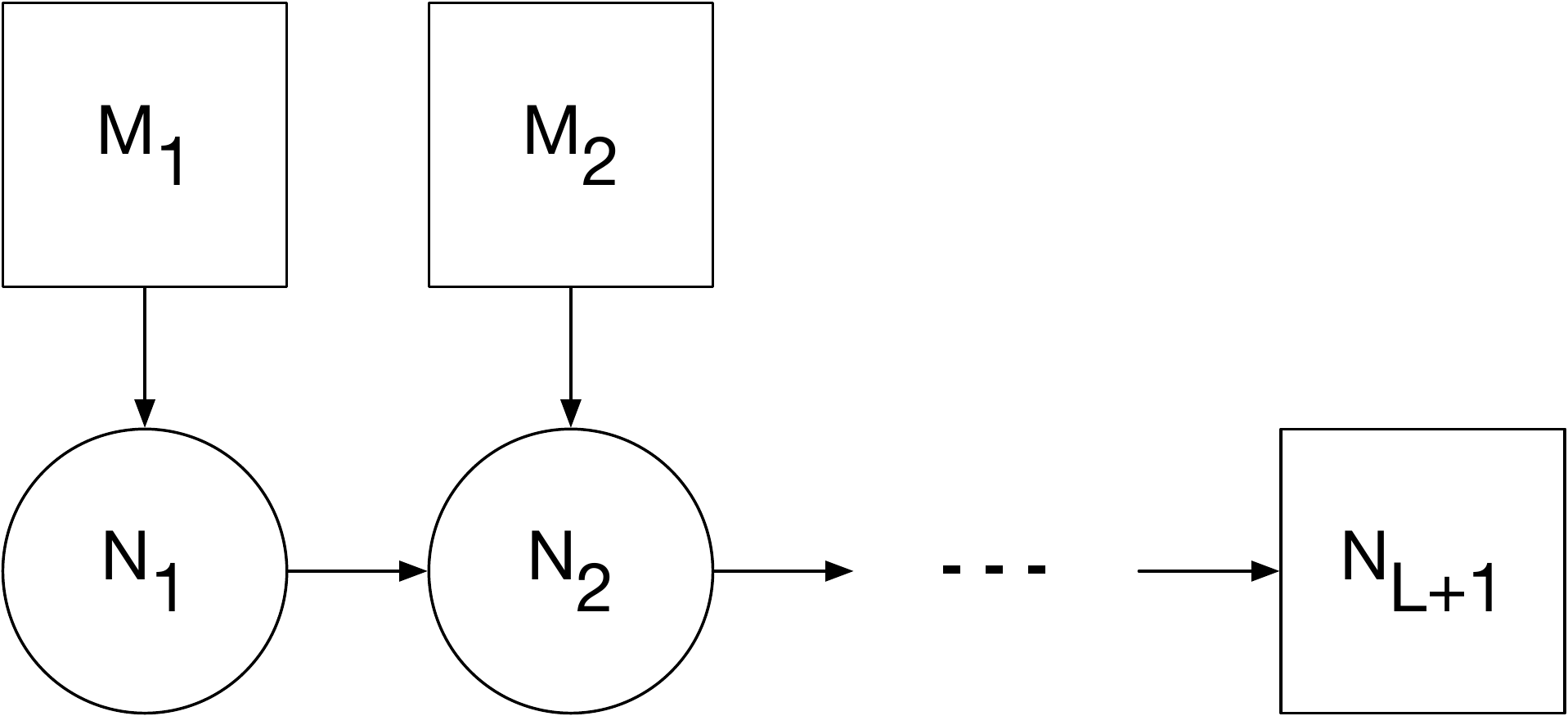}
\caption{\label{fig:flavored} More flavor nodes are attached to the quiver diagram in figure \ref{fig:def T[SU(N)]}.}
\end{figure}

We comment that one can also attach a flavor node to each gauge node, which is illustrated in figure \ref{fig:flavored}.
The 1-loop determinants of the new bi-fundamental chiral multiplets are as follows:
\begin{align}
\label{eq:flavors}
\left(\prod_{l = 1}^L \prod_{i = 1}^{M_l} \prod_{j = 1}^{N_l} \frac{\left(\tilde y^{(l)}_i {x^{(l)}_j}^{-1}\right)^{\frac{1}{2} (-m^{(l)}_j+1)}}{\left(\tilde y^{(l)}_i {x^{(l)}_j}^{-1} \zeta^{m^{(l)}_j};\zeta^2\right)_{-m^{(l)}_j+1}}\right).
\end{align}
We have additional global symmetries $\prod_{l = 1}^L U(M_l)$, whose fugacities are denoted by $\tilde y^{(l)}_i$. Because the new bi-fundamentals are heading downward, their 1-loop determinants do not develop new contributing poles.\footnote{We still assume that the CS/BF levels and the number of matters are chosen such that the boundary contribution to the index vanishes.} Thus, we only need to evaluate the contribution of \eqref{eq:flavors} at the pole \eqref{eq:pole}.
%
The resulting perturbative part and vortex part are as follows:
\begin{align}
Z^\text{pert} &= \left(\prod_{l = 1}^L \prod_{i \neq j}^{N_l} \left[y_{\sigma(i)} y_{\sigma(j)}^{-1}\right]\right) \left(\prod_{l = 1}^L \prod_{i = 1}^{N_l} \prod_{j = 1 (\neq i)}^{N_{l+1}} \left[y_{\sigma(i)} y_{\sigma(j)}^{-1}\right]\right)^{-1} \left(\prod_{l = 1}^L \prod_{i = 1}^{M_l} \prod_{j = 1}^{N_l} \left[\tilde y^{(l)}_i y_{\sigma(j)}^{-1}\right]\right)^{-1}, \\
Z^\text{vort} &= \sum_{n^{(l)}_i \geq 0} \left(\prod_{l = 1}^L \tilde \xi_l^{\sum_i \mathsf n^{(l)}_i}\right) \left(\prod_{l = 1}^L \prod_{i \neq j}^{N_l} \left[y_{\sigma(i)} y_{\sigma(j)}^{-1};\zeta^2\right]_{\mathsf n^{(l)}_i-\mathsf n^{(l)}_j}\right)^{-1} \nonumber \\
& \quad \times \left(\prod_{l = 1}^L \prod_{i = 1}^{N_l} \prod_{j = 1}^{N_{l+1}} \left[y_{\sigma(i)} y_{\sigma(j)}^{-1} \zeta^2;\zeta^2\right]_{\mathsf n^{(l)}_i-\mathsf n^{(l+1)}_j}\right)^{-1} \left(\prod_{l = 1}^L \prod_{i = 1}^{M_l} \prod_{j = 1}^{N_{l}} \left[\tilde y^{(l)}_i y_{\sigma(j)}^{-1};\zeta^{-2}\right]_{\mathsf n^{(l)}_j}\right) \label{eq:flavored vortex}.
\end{align}
If we are restricted to $L = 1$, this model corresponds to SQCD examined in section \ref{sec:Aharony}. One can see that \eqref{eq:flavored vortex} agrees with the vortex partition function \eqref{eq:Aharony vort} for $L = 1$.
\\

\section{Aharony duality for supersymmetric partition functions}
\label{sec:identity}

We have discussed that the vortex partition is a building block of various 3d supersymmetric partition functions. This is explicitly shown in appendix \ref{sec:fact} for the topologically twisted index. In this regard, the vortex partition function identity we have proven in section \ref{sec:Aharony} can be a key ingredient to exhibit the Aharony duality in terms of 3d partition functions. In this appendix, we show that indeed many identities of 3d supersymmetric partition functions can be proven using the vortex partition function identity \eqref{eq:ident} in section \ref{sec:Aharony}.
\\

For simplicity we focus on the original Aharony duality \cite{Aharony:1997gp}, i.e., $\kappa = N_f-N_a = 0$. The first example is the superconformal index, a supersymmetric partition function on $S^2 \times S^1$ \cite{Bhattacharya:2008zy,Bhattacharya:2008bja}. The superconformal index is defined by
\begin{align}
I = \mathrm{Tr}~ (-1)^F e^{-\beta' \{Q,S\}} x^{E+j} \prod_i t_i^{F_i}.
\end{align}
$Q$ is a supercharge of quantum numbers $E = \frac{1}{2}$, $j = -\frac{1}{2}$ and $R = 1$ where those quantum numbers are the Cartans of the bosonic subgroup of the 3d $\mathcal N = 2$ superconformal group, $SO(2,3) \times SO(2)$. $S$ is another supercharge satisfying $S = Q^\dagger$. Only the BPS states saturating
\begin{align}
\{Q,S\} = E-R-j \geq 0
\end{align}
contribute to the superconformal index. $t_i$'s are fugacities for global symmetries commuting with $Q$ and $S$.

The superconformal index can be computed exactly using the supersymmetric localization \cite{Kim:2009wb,Imamura:2011su}. It is also shown that the superconformal index is factorized into copies of the vortex partition functions as well as the perturbative part \cite{Krattenthaler:2011da,Hwang:2012jh,Fujitsuka:2013fga,Benini:2013yva,Hwang:2015wna}. For our interest, the $\mathcal N = 2$ $U(N_c)$ gauge theory with $N_f$ pairs of fundamental and anti-fundamental chiral multiplets, the factorization of the superconformal index is worked out in \cite{Hwang:2012jh}. They also examine the Aharony duality for the factorized index. The equality of the perturbative part is explicitly proven while that of the vortex part is numerically tested. Indeed, their conjectural identity for the vortex part is exactly a special case of our equation \eqref{eq:ident}, which we explicitly prove using the vortex wall-crossing interpretation of the Aharony duality. Therefore, combined with their result, \eqref{eq:ident} also proves the superconformal index equality for the Aharony duality.
\\

The second example is the topologically twisted index on $S^2 \times S^1$, which is defined by \eqref{eq:tind}. Its factorization is examined in appendix \ref{sec:fact}. For the $\mathcal N = 2$ $U(N_c)$ theory with $N_f$ flavors, the (angular momentum refined) twisted index is written in the following factorized form:
\begin{align}
I = \frac{1}{N_c! (N_f-N_c)!} \sum_{\sigma} Z^\text{pert} Z^\text{vort} Z^\text{antiv},
\end{align}
\begin{align}
Z^\text{pert} &= \left(\tilde \xi^{N_c r}\right) \left(\prod_{i = 1}^{N_c} \prod_{j = N_c+1}^{N_f} \left[y_{\sigma(i)} y_{\sigma(j)}^{-1}\right]\right)^{-1} \left(\prod_{i = 1}^{N_f} \prod_{j = 1}^{N_c} \left[\tilde y^{(l)}_i y_{\sigma(j)}^{-1} \tau^{-2} \zeta^{2 r};\zeta^2\right]_{-2 r+1}\right)^{-1}, \\
Z^\text{vort} &= \sum_{n_i \geq 0} \left(\tilde \xi^{\sum_i \mathsf n_i}\right) \left(\prod_{i \neq j}^{N_c} \left[y_{\sigma(i)} y_{\sigma(j)}^{-1};\zeta^2\right]_{\mathsf n_i-\mathsf n_j}\right)^{-1} \nonumber \\
& \quad \times \left(\prod_{i = 1}^{N_c} \prod_{j = 1}^{N_f} \left[y_{\sigma(i)} y_{\sigma(j)}^{-1} \zeta^2;\zeta^2\right]_{\mathsf n_i}\right)^{-1} \left(\prod_{i = 1}^{N_f} \prod_{j = 1}^{N_c} \left[\tilde y_i y_{\sigma(j)}^{-1} \tau^{-2} \zeta^{-2 r};\zeta^{-2}\right]_{\mathsf n_j}\right), \\
Z^\text{antiv} &= \left.Z^\text{vort}\right|_{\zeta \rightarrow \zeta^{-1}}
\end{align}
where we have kept $r$, the R-charge of the chiral multiplets, unfixed. $r$ should be an integer due to the topological twist. Also recall that we have defined
\begin{align}
[a;q]_n = a^{-\frac{1}{2} n} q^{-\frac{1}{4} n (n-1)} (a;q)_n.
\end{align}

Now we show how each component is identified with the counterpart in the dual theory. For the perturbative part, $Z^\text{pert}$, we first observe that the last factor in $Z^\text{pert}$ can be rewritten, using
\begin{align}
\left[a;q\right]_{-n} = \left[a q^{-n};q\right]_n^{-1},
\end{align}
as follows:
\begin{align}
\label{eq:tpert}
& \left(\prod_{i = 1}^{N_f} \prod_{j = 1}^{N_c} \left[\tilde y^{(l)}_i y_{\sigma(j)}^{-1} \tau^{-2} \zeta^{2 r};\zeta^2\right]_{-2 r+1}\right)^{-1} \nonumber \\
%
%
&= \left(\prod_{i = 1}^{N_f} \prod_{j = 1}^{N_f} \left[\tilde y^{(l)}_i y_{\sigma(j)}^{-1} \tau^{-2} \zeta^{2 r};\zeta^2\right]_{-2 r+1}\right)^{-1} \left(\prod_{i = 1}^{N_f} \prod_{j = N_c+1}^{N_f} \left[\tilde y^{(l)}_i y_{\sigma(j)}^{-1} \tau^{-2} \zeta^{2-2 r};\zeta^2\right]_{2 r-1}\right)^{-1}.
\end{align}
Note that the first factor is exactly the 1-loop determinant of $N_f^2$ gauge singlet chirals $M_a^{\tilde b}$ in the dual theory. The second factor corresponds to the contribution of the fundamental chirals in the dual theory. The factor correctly reflects the fact that those fundamental chirals in the dual theory have the R-charge $1-r$. Together with the second factor in $Z^\text{pert}$, which can be identified as the contribution of the anti-fundamental chirals in the dual theory, \eqref{eq:tpert} forms the perturbative part of the dual theory up to $\tilde \xi$-dependent factors. The $\tilde \xi$-dependent factors will be fixed after the vortex parts are taken into account.

Next, let us move on to the vortex parts. Firstly, using
\begin{align}
[a;q]_{n-m} = \frac{\left[a q^{-m};q\right]_n}{\left[a q^{-m};q\right]_m},
\end{align}
we rewrite $Z^\text{vort}$ in the following form:
\begin{align}
\label{eq:tvort}
Z^\text{vort} &= \sum_{n_i \geq 0} \left(\tilde \xi^{\sum_i \mathsf n_i}\right) \left(\prod_{i \neq j}^{N_c} \left[y_{\sigma(i)} y_{\sigma(j)}^{-1} \zeta^{2 \mathsf n_i};\zeta^{-2}\right]_{\mathsf n_j}\right)^{-1} \nonumber \\
& \quad \times \left(\prod_{i = N_c+1}^{N_f} \prod_{j = 1}^{N_c} \left[y_{\sigma(i)} y_{\sigma(j)}^{-1} \zeta^{-2};\zeta^{-2}\right]_{\mathsf n_j}\right)^{-1} \left(\prod_{i = 1}^{N_f} \prod_{j = 1}^{N_c} \left[\tilde y_i y_{\sigma(j)}^{-1} \tau^{-2} \zeta^{-2 r};\zeta^{-2}\right]_{\mathsf n_j}\right).
\end{align}
Since \eqref{eq:tvort} is nothing but \eqref{eq:Aharony vort}, we can use the identity \eqref{eq:ident} with identification $\mathsf w = -\tilde \xi$. From \eqref{eq:ident}, $Z^\text{vort}$ and $Z^\text{antiv}$ have the following relations with the dual vortex parts, $\bar Z^\text{vort}$ and $\bar Z^\text{antiv}$:
\begin{align}
Z^\text{vort} &= \bar Z^\text{vort} \times \mathrm{PE}\left[\frac{\tau^{-N_f} \zeta^{-N_c+N_f (1-r)+1}-\tau^{N_f} \zeta^{N_c-N_f (1-r)+1}}{1-\zeta^2} \mathsf w \right], \\
Z^\text{antiv} &= \bar Z^\text{antiv} \times \mathrm{PE}\left[\frac{\tau^{-N_f} \zeta^{N_c-N_f (1-r)-1}-\tau^{N_f} \zeta^{-N_c+N_f (1-r)-1}}{1-\zeta^{-2}} \mathsf w \right].
\end{align}
The product of $Z^\text{vort}$ and $Z^\text{antiv}$ is then written as
\begin{align}
& Z^\text{vort} Z^\text{antiv} = \bar Z^\text{vort} \bar Z^\text{antiv} \times \mathrm{PE}\left[\left(\mathsf w \tau^{-N_f} \zeta^{\Delta_V}+\mathsf w \tau^{N_f} \zeta^{\Delta_V}\right) \frac{1-\zeta^{-2 \Delta_V+2}}{1-\zeta^2}\right]
\end{align}
where $\Delta_V \equiv N_f (1-r)-N_c+1$ is the R-charge of gauge singlet chirals $V_\pm$ in the dual theory. Again $\Delta_V$ should be an integer due to the topological twist. One can rewrite the last Plethystic exponential factor as follows:
\begin{align}
\label{eq:tV}
& \mathrm{PE}\left[\left(\mathsf w \tau^{-N_f} \zeta^{\Delta_V}+\mathsf w \tau^{N_f} \zeta^{\Delta_V}\right) \frac{1-\zeta^{-2 \Delta_V+2}}{1-\zeta^2}\right] \nonumber \\
&= \left(-\mathsf w\right)^{\Delta_V-1} \left[\mathsf w \tau^{-N_f} \zeta^{\Delta_V};\zeta^2\right]_{-\Delta_V+1}^{-1} \left[\mathsf w^{-1} \tau^{-N_f} \zeta^{\Delta_V};\zeta^2\right]_{-\Delta_V+1}^{-1}.
\end{align}
The last two factors are exactly the 1-loop determinants of $V_\pm$ in the dual theory. Moreover, the first monomial factor $(-\mathsf w)^{\Delta_V-1} = \tilde \xi^{\Delta_V-1}$ is combined with the first factor in $Z^\text{pert}$ such that we have
\begin{align}
\label{eq:tmon}
\tilde \xi^{N_c r+\Delta_V-1} = \tilde \xi^{(N_f-N_c) (1-r)},
\end{align}
which is the correct monomial factor that should be included in the dual perturbative part $\bar Z^\text{pert}$. Thus, \eqref{eq:tV} and \eqref{eq:tmon} correctly reproduce the $\tilde \xi$-dependent factors in $\bar Z^\text{pert}$, which implies $Z^\text{pert} Z^\text{vort} Z^\text{antiv} = \bar Z^\text{pert} \bar Z^\text{vort} \bar Z^\text{antiv}$. Therefore, the twisted index equality for the Aharony duality is proven using the vortex partition function identity, \eqref{eq:ident}.
\\

Our last example is the supersymmetric partition function on the squashed 3-sphere, $S^3_b$. Using the supersymmetric localization, especially the Higgs branch localization, the partition function is given by the following factorized form \cite{Pasquetti:2011fj,Taki:2013opa,Fujitsuka:2013fga,Benini:2013yva}:\footnote{Especially we adopt the parameters used in \cite{Benini:2013yva}.}
\begin{align}
Z = \frac{1}{N_c! (N_f-N_c)!} \sum_{\sigma \in S_{N_c}} Z^\text{pert} Z^\text{vort} Z^\text{antiv},
\end{align}
\begin{align}
& Z^\text{pert} = e^{2 \pi i \xi \sum_{i = 1}^{N_c} (m_i+\mu)} \nonumber \\
& \times \left(\prod_{i < j}^{N_c} 4 \sinh [\pi b (m_i-m_j)] \sinh [\pi b^{-1} (m_i-m_j)]\right) \frac{\prod_{i = 1}^{N_c} \prod_{j = 1 (\neq i)}^{N_f} s_b(\frac{i Q}{2}+m_j-m_i)}{\prod_{i = 1}^{N_c} \prod_{j = 1}^{N_f} s_b(-\frac{i Q}{2}+\tilde m_j-m_i-2 \mu)}, \\
& Z^\text{vort} = \left.Z^\text{antiv}\right|_{b \rightarrow b^{-1}}, \\
& Z^\text{antiv} = \sum_{n_i \geq 0} \left((-1)^{N_f-N_c} e^{2 \pi b \xi}\right)^{\sum_{i = 1}^{N_c} n_i} \nonumber \\
& \times \prod_{i = 1}^{N_c} \prod_{l = 1}^{n_i} \frac{\prod_{j = 1}^{N_f} 2 \sinh \pi b (\tilde m_j-m_i-2 \mu+(l-1) i b)}{\left(\prod_{j = 1}^{N_c} 2 \sinh \pi b (m_j-m_i+(l-1-n_j) i b)\right) \left(\prod_{j = N_c+1}^{N_f} 2 \sinh \pi b (m_j-m_i+l i b)\right)}
\end{align}
where $s_b (z)$ is the double sine function, whose definition can be found in, e.g., \cite{Hama:2011ea}. $\xi$ is the FI parameter and $m_j+\mu$, $\tilde m_j-\mu$ are mass parameters.\footnote{From now on mass parameters have slightly different normalization. The new mass parameters are identified with the old ones as follows: $2 \pi b^{-1} m_\text{new} = m_\text{old}$ where $b$ is the squashing parameter.} $b$ is the squashing parameter of $S^3_b$ and $Q = b+b^{-1}$. $\sigma$ is a permutation acting on mass parameters $m_j$'s.

One should note that the $S^3_b$ partition function also allows a matrix integral form \cite{Hama:2011ea}, which is obtained by the Coulomb branch localization. In that case, the Aharony duality is realized as an integral identity of double sine functions \cite{Willett:2011gp}, which is proven in \cite{Bult:2007}. Here we recast the proof of the identity using the vortex partition function.

Firstly we examine the perturbative part. Using double-sine function identities
\begin{gather}
s_b(z) s_b(-z) = 1, \\
s_b \left(\frac{i Q}{2}+z\right) = \frac{s_b \left(\frac{i Q}{2}+z-i b\right)}{2 i \sinh \pi b z},
\end{gather}
one can rewrite $Z^\text{pert}$ as follows:
\begin{align}
\label{eq:spert1}
& Z^\text{pert} = e^{2 \pi i \xi \sum_{i = 1}^{N_c} (m_i+\mu)} \times \frac{\prod_{i = 1}^{N_c} \prod_{j = N_c+1}^{N_f} s_b(\frac{i Q}{2}+m_j-m_i)}{\prod_{i = 1}^{N_c} \prod_{j = 1}^{N_f} s_b(-\frac{i Q}{2}+\tilde m_j-m_i-2 \mu)}.
\end{align}
The numerator of the second factor appears in the dual perturbative part as it is while the denominator requires a little massage:
\begin{align}
\label{eq:spert2}
& \prod_{i = 1}^{N_c} \prod_{j = 1}^{N_f} s_b \left(-\frac{i Q}{2}+\tilde m_j-m_i-2 \mu\right) \nonumber \\
&= \left(\prod_{i = 1}^{N_f} \prod_{j = 1}^{N_f} s_b \left(\frac{i Q}{2}-\tilde m_j+m_i+2 \mu\right)\right)^{-1} \left(\prod_{i = N_c+1}^{N_f} \prod_{j = 1}^{N_f} s_b \left(\frac{i Q}{2}-\tilde m_j+m_i+2 \mu\right)\right).
\end{align}
\eqref{eq:spert1} combined with \eqref{eq:spert2} produces the perturbative part of the dual theory up to $\xi$-dependent factors, which will be fixed later. Note that the first factor on the right hand side of \eqref{eq:spert2} is the 1-loop determinant of $M_a^{\tilde b}$ in the dual theory.

Next, we examine the vortex parts. Again, from \eqref{eq:ident}, we have the following relations:
\begin{align}
\begin{aligned}
\label{eq:S3 vort}
Z^\text{vort} &= \bar Z^\text{vort} \times \mathrm{PE}\left[\frac{\tau^{-N_f} \zeta^{-N_c+N_f+1}-\tau^{N_f} \zeta^{N_c-N_f+1}}{1-\zeta^2} \mathsf w \right], \\
Z^\text{antiv} &= \bar Z^\text{antiv} \times \left.\mathrm{PE}\left[\frac{\tau^{-N_f} \zeta^{-N_c+N_f+1}-\tau^{N_f} \zeta^{N_c-N_f+1}}{1-\zeta^2} \mathsf w \right]\right|_{b \rightarrow b^{-1}}
\end{aligned}
\end{align}
where $\tau = e^{2 \pi b^{-1} \mu}$, $\zeta = e^{-\pi i b^{-2}}$ and $\mathsf w = (-1)^{N_f-N_c+1} e^{2 \pi b^{-1} \xi}$. Since the double sine function is written in terms of the Plethystic exponential as follows:
\begin{align}
s_b \left(z+\frac{i Q}{2}\right) = e^{-\frac{i \pi}{2} (z+\frac{i Q}{2})^2} \times \mathrm{PE}\left[\frac{e^{2 \pi b z}}{1-e^{-2 i \pi b^2}}+\frac{e^{2 \pi b^{-1} z}}{1-e^{-2 i \pi b^{-2}}}\right],
\end{align}
the two Plethystic exponentials in \eqref{eq:S3 vort} together form the following double sine function expression:
\begin{align}
\label{eq:sV}
e^{-2 \pi i \xi (N_f \mu+\frac{i Q}{2} (N_f-N_c))} \frac{s_b(\xi-N_f \mu-\frac{i Q}{2} (N_f-N_c))}{s_b(\xi+N_f \mu+\frac{i Q}{2} (N_f-N_c))}.
\end{align}
Note that the second factor is exactly the 1-loop determinant of $V_\pm$ in the dual theory. Combining the monomial factors in \eqref{eq:spert2} and in \eqref{eq:sV}, we have
\begin{align}
\label{eq:smon}
e^{-2 \pi i \xi \sum_{i = N_c+1}^{N_f} (m_i+\mu+\frac{i Q}{2})}.
\end{align}
Taking this monomial factor into account, \eqref{eq:sV} correctly reproduces the $\xi$-dependent factors in the dual perturbative part. Thus, we have proven the identity of the partition functions on $S^3_b$ for the Aharony duality.
\\

We have shown that three kinds of partition function identities for the Aharony duality can be proven using the identity \eqref{eq:ident} of the vortex partition function. Except the $S^3_b$ partition function, for which the Aharony duality is proven using an integral identity of the hyperbolic gamma function \cite{Bult:2007,Willett:2011gp,Benini:2011mf,Amariti:2014lla}, analytic proofs of the other partition function identities were only available for specific examples with the fixed gauge rank and the fixed number of flavors \cite{Krattenthaler:2011da,Hwang:2012jh} or for specific fugacities \cite{Closset:2016arn}.\footnote{The twisted index identity was proven without the angular-momentum refinement, i.e., $\zeta = 1$, in \cite{Closset:2016arn}.} Our proof, on the other hand, is for the arbitrary gauge rank and the arbitrary number of flavors with full generality of fugacities. The key of our proof is the vortex partition function identity, which is a building block of various 3d partition functions. Thus, our method is not sensitive to the type of a partition function. Furthermore, since those partition functions allow integral expressions, our results can be regarded as proofs of integral identities of special functions.

Our approach is based on the observation of the physical phenomenon, the wall-crossing of vortex quantum mechanics. This shows that a understanding of a physical phenomenon even suggests a new way to prove a nontrivial mathematical identity. Indeed, from various Seiberg-like dualities, there are many conjectural identities beyond what we have proven in this appendix, e.g., \cite{Bashkirov:2011vy,Hwang:2011qt,Hwang:2011ht,Kapustin:2011vz,Cheon:2012be,Kim:2013cma,Aharony:2013dha,Park:2013wta,Aharony:2013kma,Gahramanov:gka,Amariti:2014iza,Gahramanov:2014ona,Amariti:2015vwa}. A better understanding of physics behind those dualities may give new understandings of various integral identities of special functions. More related discussions are also found in \cite{Gahramanov:2013rda,Gahramanov:2016wxi}.
\\


\bibliographystyle{JHEP}
\bibliography{vortices}

\providecommand{\href}[2]{#2}\begingroup\raggedright\begin{thebibliography}{10}

\bibitem{Jockers:2012dk}
H.~Jockers, V.~Kumar, J.~M. Lapan, D.~R. Morrison and M.~Romo,
  \emph{{Two-Sphere Partition Functions and Gromov-Witten Invariants}},
  \href{http://dx.doi.org/10.1007/s00220-013-1874-z}{\emph{Commun. Math. Phys.}
  {\bfseries 325} (2014) 1139--1170},
  [\href{https://arxiv.org/abs/1208.6244}{{\ttfamily 1208.6244}}].

\bibitem{Gomis:2012wy}
J.~Gomis and S.~Lee, \emph{{Exact Kahler Potential from Gauge Theory and Mirror
  Symmetry}}, \href{http://dx.doi.org/10.1007/JHEP04(2013)019}{\emph{JHEP}
  {\bfseries 04} (2013) 019},
  [\href{https://arxiv.org/abs/1210.6022}{{\ttfamily 1210.6022}}].

\bibitem{Benini:2012ui}
F.~Benini and S.~Cremonesi, \emph{{Partition Functions of ${\mathcal{N}=(2,2)}$
  Gauge Theories on S$^{2}$ and Vortices}},
  \href{http://dx.doi.org/10.1007/s00220-014-2112-z}{\emph{Commun. Math. Phys.}
  {\bfseries 334} (2015) 1483--1527},
  [\href{https://arxiv.org/abs/1206.2356}{{\ttfamily 1206.2356}}].

\bibitem{Doroud:2012xw}
N.~Doroud, J.~Gomis, B.~Le~Floch and S.~Lee, \emph{{Exact Results in D=2
  Supersymmetric Gauge Theories}},
  \href{http://dx.doi.org/10.1007/JHEP05(2013)093}{\emph{JHEP} {\bfseries 05}
  (2013) 093}, [\href{https://arxiv.org/abs/1206.2606}{{\ttfamily 1206.2606}}].

\bibitem{Benini:2013xpa}
F.~Benini, R.~Eager, K.~Hori and Y.~Tachikawa, \emph{{Elliptic Genera of 2d
  ${\mathcal{N}}$ = 2 Gauge Theories}},
  \href{http://dx.doi.org/10.1007/s00220-014-2210-y}{\emph{Commun. Math. Phys.}
  {\bfseries 333} (2015) 1241--1286},
  [\href{https://arxiv.org/abs/1308.4896}{{\ttfamily 1308.4896}}].

\bibitem{Hori:2014tda}
K.~Hori, H.~Kim and P.~Yi, \emph{{Witten Index and Wall Crossing}},
  \href{http://dx.doi.org/10.1007/JHEP01(2015)124}{\emph{JHEP} {\bfseries 01}
  (2015) 124}, [\href{https://arxiv.org/abs/1407.2567}{{\ttfamily 1407.2567}}].

\bibitem{Intriligator:2013lca}
K.~Intriligator and N.~Seiberg, \emph{{Aspects of 3d N=2 Chern-Simons-Matter
  Theories}}, \href{http://dx.doi.org/10.1007/JHEP07(2013)079}{\emph{JHEP}
  {\bfseries 07} (2013) 079},
  [\href{https://arxiv.org/abs/1305.1633}{{\ttfamily 1305.1633}}].

\bibitem{Krattenthaler:2011da}
C.~Krattenthaler, V.~P. Spiridonov and G.~S. Vartanov, \emph{{Superconformal
  indices of three-dimensional theories related by mirror symmetry}},
  \href{http://dx.doi.org/10.1007/JHEP06(2011)008}{\emph{JHEP} {\bfseries 06}
  (2011) 008}, [\href{https://arxiv.org/abs/1103.4075}{{\ttfamily 1103.4075}}].

\bibitem{Dimofte:2011ju}
T.~Dimofte, D.~Gaiotto and S.~Gukov, \emph{{Gauge Theories Labelled by
  Three-Manifolds}},
  \href{http://dx.doi.org/10.1007/s00220-013-1863-2}{\emph{Commun. Math. Phys.}
  {\bfseries 325} (2014) 367--419},
  [\href{https://arxiv.org/abs/1108.4389}{{\ttfamily 1108.4389}}].

\bibitem{Pasquetti:2011fj}
S.~Pasquetti, \emph{{Factorisation of N = 2 Theories on the Squashed
  3-Sphere}}, \href{http://dx.doi.org/10.1007/JHEP04(2012)120}{\emph{JHEP}
  {\bfseries 04} (2012) 120},
  [\href{https://arxiv.org/abs/1111.6905}{{\ttfamily 1111.6905}}].

\bibitem{Beem:2012mb}
C.~Beem, T.~Dimofte and S.~Pasquetti, \emph{{Holomorphic Blocks in Three
  Dimensions}}, \href{http://dx.doi.org/10.1007/JHEP12(2014)177}{\emph{JHEP}
  {\bfseries 12} (2014) 177},
  [\href{https://arxiv.org/abs/1211.1986}{{\ttfamily 1211.1986}}].

\bibitem{Hwang:2012jh}
C.~Hwang, H.-C. Kim and J.~Park, \emph{{Factorization of the 3d superconformal
  index}}, \href{http://dx.doi.org/10.1007/JHEP08(2014)018}{\emph{JHEP}
  {\bfseries 08} (2014) 018},
  [\href{https://arxiv.org/abs/1211.6023}{{\ttfamily 1211.6023}}].

\bibitem{Taki:2013opa}
M.~Taki, \emph{{Holomorphic Blocks for 3d Non-abelian Partition Functions}},
  \href{https://arxiv.org/abs/1303.5915}{{\ttfamily 1303.5915}}.

\bibitem{Cecotti:2013mba}
S.~Cecotti, D.~Gaiotto and C.~Vafa, \emph{{$tt^*$ geometry in 3 and 4
  dimensions}}, \href{http://dx.doi.org/10.1007/JHEP05(2014)055}{\emph{JHEP}
  {\bfseries 05} (2014) 055},
  [\href{https://arxiv.org/abs/1312.1008}{{\ttfamily 1312.1008}}].

\bibitem{Fujitsuka:2013fga}
M.~Fujitsuka, M.~Honda and Y.~Yoshida, \emph{{Higgs branch localization of 3d
  ?? = 2 theories}}, \href{http://dx.doi.org/10.1093/ptep/ptu158}{\emph{PTEP}
  {\bfseries 2014} (2014) 123B02},
  [\href{https://arxiv.org/abs/1312.3627}{{\ttfamily 1312.3627}}].

\bibitem{Benini:2013yva}
F.~Benini and W.~Peelaers, \emph{{Higgs branch localization in three
  dimensions}}, \href{http://dx.doi.org/10.1007/JHEP05(2014)030}{\emph{JHEP}
  {\bfseries 05} (2014) 030},
  [\href{https://arxiv.org/abs/1312.6078}{{\ttfamily 1312.6078}}].

\bibitem{Benini:2015noa}
F.~Benini and A.~Zaffaroni, \emph{{A topologically twisted index for
  three-dimensional supersymmetric theories}},
  \href{http://dx.doi.org/10.1007/JHEP07(2015)127}{\emph{JHEP} {\bfseries 07}
  (2015) 127}, [\href{https://arxiv.org/abs/1504.03698}{{\ttfamily
  1504.03698}}].

\bibitem{Nekrasov:2002qd}
N.~A. Nekrasov, \emph{{Seiberg-Witten prepotential from instanton counting}},
  \href{http://dx.doi.org/10.4310/ATMP.2003.v7.n5.a4}{\emph{Adv. Theor. Math.
  Phys.} {\bfseries 7} (2003) 831--864},
  [\href{https://arxiv.org/abs/hep-th/0206161}{{\ttfamily hep-th/0206161}}].

\bibitem{Hanany:2003hp}
A.~Hanany and D.~Tong, \emph{{Vortices, instantons and branes}},
  \href{http://dx.doi.org/10.1088/1126-6708/2003/07/037}{\emph{JHEP} {\bfseries
  07} (2003) 037}, [\href{https://arxiv.org/abs/hep-th/0306150}{{\ttfamily
  hep-th/0306150}}].

\bibitem{Weinberg:1982ev}
E.~J. Weinberg, \emph{{Fundamental Monopoles in Theories With Arbitrary
  Symmetry Breaking}},
  \href{http://dx.doi.org/10.1016/0550-3213(82)90324-8}{\emph{Nucl. Phys.}
  {\bfseries B203} (1982) 445--471}.

\bibitem{Weinberg:2006rq}
E.~J. Weinberg and P.~Yi, \emph{{Magnetic Monopole Dynamics, Supersymmetry, and
  Duality}}, \href{http://dx.doi.org/10.1016/j.physrep.2006.11.002}{\emph{Phys.
  Rept.} {\bfseries 438} (2007) 65--236},
  [\href{https://arxiv.org/abs/hep-th/0609055}{{\ttfamily hep-th/0609055}}].

\bibitem{Denef:2002ru}
F.~Denef, \emph{{Quantum quivers and Hall / hole halos}},
  \href{http://dx.doi.org/10.1088/1126-6708/2002/10/023}{\emph{JHEP} {\bfseries
  10} (2002) 023}, [\href{https://arxiv.org/abs/hep-th/0206072}{{\ttfamily
  hep-th/0206072}}].

\bibitem{Aharony:1997gp}
O.~Aharony, \emph{{IR duality in d = 3 N=2 supersymmetric USp(2N(c)) and
  U(N(c)) gauge theories}},
  \href{http://dx.doi.org/10.1016/S0370-2693(97)00530-3}{\emph{Phys. Lett.}
  {\bfseries B404} (1997) 71--76},
  [\href{https://arxiv.org/abs/hep-th/9703215}{{\ttfamily hep-th/9703215}}].

\bibitem{Jafferis:2010un}
D.~L. Jafferis, \emph{{The Exact Superconformal R-Symmetry Extremizes Z}},
  \href{http://dx.doi.org/10.1007/JHEP05(2012)159}{\emph{JHEP} {\bfseries 05}
  (2012) 159}, [\href{https://arxiv.org/abs/1012.3210}{{\ttfamily 1012.3210}}].

\bibitem{Aharony:1997bx}
O.~Aharony, A.~Hanany, K.~A. Intriligator, N.~Seiberg and M.~J. Strassler,
  \emph{{Aspects of N=2 supersymmetric gauge theories in three-dimensions}},
  \href{http://dx.doi.org/10.1016/S0550-3213(97)00323-4}{\emph{Nucl. Phys.}
  {\bfseries B499} (1997) 67--99},
  [\href{https://arxiv.org/abs/hep-th/9703110}{{\ttfamily hep-th/9703110}}].

\bibitem{Giveon:2008zn}
A.~Giveon and D.~Kutasov, \emph{{Seiberg Duality in Chern-Simons Theory}},
  \href{http://dx.doi.org/10.1016/j.nuclphysb.2008.09.045}{\emph{Nucl. Phys.}
  {\bfseries B812} (2009) 1--11},
  [\href{https://arxiv.org/abs/0808.0360}{{\ttfamily 0808.0360}}].

\bibitem{Hanany:2015via}
A.~Hanany, C.~Hwang, H.~Kim, J.~Park and R.-K. Seong, \emph{{Hilbert Series for
  Theories with Aharony Duals}},
  \href{http://dx.doi.org/10.1007/JHEP11(2015)132,
  10.1007/JHEP04(2016)064}{\emph{JHEP} {\bfseries 11} (2015) 132},
  [\href{https://arxiv.org/abs/1505.02160}{{\ttfamily 1505.02160}}].

\bibitem{Cremonesi:2015dja}
S.~Cremonesi, \emph{{The Hilbert series of 3d ${\boldsymbol{\mathcal{N}}}=2$
  Yang?Mills theories with vectorlike matter}},
  \href{http://dx.doi.org/10.1088/1751-8113/48/45/455401}{\emph{J. Phys.}
  {\bfseries A48} (2015) 455401},
  [\href{https://arxiv.org/abs/1505.02409}{{\ttfamily 1505.02409}}].

\bibitem{Cremonesi:2016nbo}
S.~Cremonesi, N.~Mekareeya and A.~Zaffaroni, \emph{{The moduli spaces of 3d $
  \mathcal{N}\ge 2 $ Chern-Simons gauge theories and their Hilbert series}},
  \href{http://dx.doi.org/10.1007/JHEP10(2016)046}{\emph{JHEP} {\bfseries 10}
  (2016) 046}, [\href{https://arxiv.org/abs/1607.05728}{{\ttfamily
  1607.05728}}].

\bibitem{Witten:1999ds}
E.~Witten, \emph{{Supersymmetric index of three-dimensional gauge theory}},
  \href{https://arxiv.org/abs/hep-th/9903005}{{\ttfamily hep-th/9903005}}.

\bibitem{Benini:2011mf}
F.~Benini, C.~Closset and S.~Cremonesi, \emph{{Comments on 3d Seiberg-like
  dualities}}, \href{http://dx.doi.org/10.1007/JHEP10(2011)075}{\emph{JHEP}
  {\bfseries 10} (2011) 075},
  [\href{https://arxiv.org/abs/1108.5373}{{\ttfamily 1108.5373}}].

\bibitem{Hanany:1996ie}
A.~Hanany and E.~Witten, \emph{{Type IIB superstrings, BPS monopoles, and
  three-dimensional gauge dynamics}},
  \href{http://dx.doi.org/10.1016/S0550-3213(97)00157-0,
  10.1016/S0550-3213(97)80030-2}{\emph{Nucl. Phys.} {\bfseries B492} (1997)
  152--190}, [\href{https://arxiv.org/abs/hep-th/9611230}{{\ttfamily
  hep-th/9611230}}].

\bibitem{Peskin1978122}
M.~E. Peskin, \emph{Mandelstam-'t hooft duality in abelian lattice models},
  \href{http://dx.doi.org/http://dx.doi.org/10.1016/0003-4916(78)90252-X}{\emph{Annals
  of Physics} {\bfseries 113} (1978) 122 -- 152}.

\bibitem{PhysRevLett.47.1556}
C.~Dasgupta and B.~I. Halperin, \emph{Phase transition in a lattice model of
  superconductivity},
  \href{http://dx.doi.org/10.1103/PhysRevLett.47.1556}{\emph{Phys. Rev. Lett.}
  {\bfseries 47} (Nov, 1981) 1556--1560}.

\bibitem{Son:2015xqa}
D.~T. Son, \emph{{Is the Composite Fermion a Dirac Particle?}},
  \href{http://dx.doi.org/10.1103/PhysRevX.5.031027}{\emph{Phys. Rev.}
  {\bfseries X5} (2015) 031027},
  [\href{https://arxiv.org/abs/1502.03446}{{\ttfamily 1502.03446}}].

\bibitem{2015PhRvX...5d1031W}
C.~{Wang} and T.~{Senthil}, \emph{{Dual Dirac Liquid on the Surface of the
  Electron Topological Insulator}},
  \href{http://dx.doi.org/10.1103/PhysRevX.5.041031}{\emph{Physical Review X}
  {\bfseries 5} (Oct., 2015) 041031},
  [\href{https://arxiv.org/abs/1505.05141}{{\ttfamily 1505.05141}}].

\bibitem{Metlitski:2015eka}
M.~A. Metlitski and A.~Vishwanath, \emph{{Particle-vortex duality of
  two-dimensional Dirac fermion from electric-magnetic duality of
  three-dimensional topological insulators}},
  \href{http://dx.doi.org/10.1103/PhysRevB.93.245151}{\emph{Phys. Rev.}
  {\bfseries B93} (2016) 245151},
  [\href{https://arxiv.org/abs/1505.05142}{{\ttfamily 1505.05142}}].

\bibitem{Intriligator:1996ex}
K.~A. Intriligator and N.~Seiberg, \emph{{Mirror symmetry in three-dimensional
  gauge theories}},
  \href{http://dx.doi.org/10.1016/0370-2693(96)01088-X}{\emph{Phys. Lett.}
  {\bfseries B387} (1996) 513--519},
  [\href{https://arxiv.org/abs/hep-th/9607207}{{\ttfamily hep-th/9607207}}].

\bibitem{Kapustin:1999ha}
A.~Kapustin and M.~J. Strassler, \emph{{On mirror symmetry in three-dimensional
  Abelian gauge theories}},
  \href{http://dx.doi.org/10.1088/1126-6708/1999/04/021}{\emph{JHEP} {\bfseries
  04} (1999) 021}, [\href{https://arxiv.org/abs/hep-th/9902033}{{\ttfamily
  hep-th/9902033}}].

\bibitem{Karch:2016sxi}
A.~Karch and D.~Tong, \emph{{Particle-Vortex Duality from 3d Bosonization}},
  \href{http://dx.doi.org/10.1103/PhysRevX.6.031043}{\emph{Phys. Rev.}
  {\bfseries X6} (2016) 031043},
  [\href{https://arxiv.org/abs/1606.01893}{{\ttfamily 1606.01893}}].

\bibitem{Murugan:2016zal}
J.~Murugan and H.~Nastase, \emph{{Particle-vortex duality in topological
  insulators and superconductors}},
  \href{https://arxiv.org/abs/1606.01912}{{\ttfamily 1606.01912}}.

\bibitem{Seiberg:2016gmd}
N.~Seiberg, T.~Senthil, C.~Wang and E.~Witten, \emph{{A Duality Web in 2+1
  Dimensions and Condensed Matter Physics}},
  \href{http://dx.doi.org/10.1016/j.aop.2016.08.007}{\emph{Annals Phys.}
  {\bfseries 374} (2016) 395--433},
  [\href{https://arxiv.org/abs/1606.01989}{{\ttfamily 1606.01989}}].

\bibitem{Kachru:2016rui}
S.~Kachru, M.~Mulligan, G.~Torroba and H.~Wang, \emph{{Bosonization and Mirror
  Symmetry}}, \href{http://dx.doi.org/10.1103/PhysRevD.94.085009}{\emph{Phys.
  Rev.} {\bfseries D94} (2016) 085009},
  [\href{https://arxiv.org/abs/1608.05077}{{\ttfamily 1608.05077}}].

\bibitem{Kachru:2016aon}
S.~Kachru, M.~Mulligan, G.~Torroba and H.~Wang, \emph{{The many faces of mirror
  symmetry}},
  \href{http://dx.doi.org/10.1103/PhysRevLett.118.011602}{\emph{Phys. Rev.
  Lett.} {\bfseries 118} (2017) 011602},
  [\href{https://arxiv.org/abs/1609.02149}{{\ttfamily 1609.02149}}].

\bibitem{Gaiotto:2008ak}
D.~Gaiotto and E.~Witten, \emph{{S-Duality of Boundary Conditions In N=4 Super
  Yang-Mills Theory}},
  \href{http://dx.doi.org/10.4310/ATMP.2009.v13.n3.a5}{\emph{Adv. Theor. Math.
  Phys.} {\bfseries 13} (2009) 721--896},
  [\href{https://arxiv.org/abs/0807.3720}{{\ttfamily 0807.3720}}].

\bibitem{Kapustin:2010mh}
A.~Kapustin, B.~Willett and I.~Yaakov, \emph{{Tests of Seiberg-like Duality in
  Three Dimensions}},  \href{https://arxiv.org/abs/1012.4021}{{\ttfamily
  1012.4021}}.

\bibitem{Kim:2012uz}
H.-C. Kim, J.~Kim, S.~Kim and K.~Lee, \emph{{Vortices and 3 dimensional
  dualities}},  \href{https://arxiv.org/abs/1204.3895}{{\ttfamily 1204.3895}}.

\bibitem{Yaakov:2013fza}
I.~Yaakov, \emph{{Redeeming Bad Theories}},
  \href{http://dx.doi.org/10.1007/JHEP11(2013)189}{\emph{JHEP} {\bfseries 11}
  (2013) 189}, [\href{https://arxiv.org/abs/1303.2769}{{\ttfamily 1303.2769}}].

\bibitem{Gaiotto:2013bwa}
D.~Gaiotto and P.~Koroteev, \emph{{On Three Dimensional Quiver Gauge Theories
  and Integrability}},
  \href{http://dx.doi.org/10.1007/JHEP05(2013)126}{\emph{JHEP} {\bfseries 05}
  (2013) 126}, [\href{https://arxiv.org/abs/1304.0779}{{\ttfamily 1304.0779}}].

\bibitem{Witten:1982df}
E.~Witten, \emph{{Constraints on Supersymmetry Breaking}},
  \href{http://dx.doi.org/10.1016/0550-3213(82)90071-2}{\emph{Nucl. Phys.}
  {\bfseries B202} (1982) 253}.

\bibitem{AlvarezGaume:1986nm}
L.~Alvarez-Gaume, \emph{{SUPERSYMMETRY AND INDEX THEORY}},  in \emph{{1984 NATO
  ASI on Supersymmetry Bonn, Germany, August 20-31, 1984}}, pp.~1--44, 1986.

\bibitem{Lee:2016dbm}
S.-J. Lee and P.~Yi, \emph{{Witten Index for Noncompact Dynamics}},
  \href{http://dx.doi.org/10.1007/JHEP06(2016)089}{\emph{JHEP} {\bfseries 06}
  (2016) 089}, [\href{https://arxiv.org/abs/1602.03530}{{\ttfamily
  1602.03530}}].

\bibitem{Hwang:2014uwa}
C.~Hwang, J.~Kim, S.~Kim and J.~Park, \emph{{General instanton counting and 5d
  SCFT}}, \href{http://dx.doi.org/10.1007/JHEP07(2015)063,
  10.1007/JHEP04(2016)094}{\emph{JHEP} {\bfseries 07} (2015) 063},
  [\href{https://arxiv.org/abs/1406.6793}{{\ttfamily 1406.6793}}].

\bibitem{Cordova:2014oxa}
C.~Cordova and S.-H. Shao, \emph{{An Index Formula for Supersymmetric Quantum
  Mechanics}},  \href{https://arxiv.org/abs/1406.7853}{{\ttfamily 1406.7853}}.

\bibitem{1993alg.geom..7001J}
L.~C. {Jeffrey} and F.~C. {Kirwan}, \emph{{Localization for nonabelian group
  actions}},  in \emph{eprint arXiv:alg-geom/9307001}, July, 1993.

\bibitem{1999math......3178B}
M.~{Brion} and M.~{Vergne}, \emph{{Arrangements of hyperplanes I: Rational
  functions and Jeffrey-Kirwan residue}}, {\emph{ArXiv Mathematics e-prints}
  (Mar., 1999) }, [\href{https://arxiv.org/abs/math/9903178}{{\ttfamily
  math/9903178}}].

\bibitem{2004InMat.158..453S}
A.~{Szenes} and M.~{Vergne}, \emph{{Toric reduction and a conjecture of Batyrev
  and Materov}},
  \href{http://dx.doi.org/10.1007/s00222-004-0375-2}{\emph{Inventiones
  Mathematicae} {\bfseries 158} (June, 2004) 453--495},
  [\href{https://arxiv.org/abs/math/0306311}{{\ttfamily math/0306311}}].

\bibitem{Aganagic:2014oia}
M.~Aganagic, N.~Haouzi and S.~Shakirov, \emph{{$A_n$-Triality}},
  \href{https://arxiv.org/abs/1403.3657}{{\ttfamily 1403.3657}}.

\bibitem{Bullimore:2016hdc}
M.~Bullimore, T.~Dimofte, D.~Gaiotto, J.~Hilburn and H.-C. Kim, \emph{{Vortices
  and Vermas}},  \href{https://arxiv.org/abs/1609.04406}{{\ttfamily
  1609.04406}}.

\bibitem{Bullimore:2014awa}
M.~Bullimore, H.-C. Kim and P.~Koroteev, \emph{{Defects and Quantum
  Seiberg-Witten Geometry}},
  \href{http://dx.doi.org/10.1007/JHEP05(2015)095}{\emph{JHEP} {\bfseries 05}
  (2015) 095}, [\href{https://arxiv.org/abs/1412.6081}{{\ttfamily 1412.6081}}].

\bibitem{Nawata:2014nca}
S.~Nawata, \emph{{Givental J-functions, Quantum integrable systems, AGT
  relation with surface operator}},
  \href{http://dx.doi.org/10.4310/ATMP.2015.v19.n6.a4}{\emph{Adv. Theor. Math.
  Phys.} {\bfseries 19} (2015) 1277--1338},
  [\href{https://arxiv.org/abs/1408.4132}{{\ttfamily 1408.4132}}].

\bibitem{Finkelberg2014}
M.~Finkelberg and L.~Rybnikov, \emph{{Quantization of Drinfeld Zastava in type
  A}}, {\emph{J. Eur. Math. Soc. (JEMS)} {\bfseries 16} (2014) 235--271},
  [\href{https://arxiv.org/abs/1009.0676}{{\ttfamily 1009.0676}}].

\bibitem{Venugopalan2013}
S.~Venugopalan and C.~T. Woodward, \emph{{Classification of affine vortices}},
  {\emph{Duke. Math. J} {\bfseries 165} (2016) 1695--1751},
  [\href{https://arxiv.org/abs/1301.7052}{{\ttfamily 1301.7052}}].

\bibitem{Givental2003}
A.~Givental and Y.-P. Lee, \emph{{Quantum K-theory on flag manifolds,
  finite-difference Toda lattices and quantum groups}}, {\emph{Inventiones
  Mathematicae.} {\bfseries 151} (2003) 193?219},
  [\href{https://arxiv.org/abs/math/0108105}{{\ttfamily math/0108105}}].

\bibitem{Dimofte:2010tz}
T.~Dimofte, S.~Gukov and L.~Hollands, \emph{{Vortex Counting and Lagrangian
  3-manifolds}}, \href{http://dx.doi.org/10.1007/s11005-011-0531-8}{\emph{Lett.
  Math. Phys.} {\bfseries 98} (2011) 225--287},
  [\href{https://arxiv.org/abs/1006.0977}{{\ttfamily 1006.0977}}].

\bibitem{Hwang:2015wna}
C.~Hwang and J.~Park, \emph{{Factorization of the 3d superconformal index with
  an adjoint matter}},
  \href{http://dx.doi.org/10.1007/JHEP11(2015)028}{\emph{JHEP} {\bfseries 11}
  (2015) 028}, [\href{https://arxiv.org/abs/1506.03951}{{\ttfamily
  1506.03951}}].

\bibitem{Willett:2011gp}
B.~Willett and I.~Yaakov, \emph{{N=2 Dualities and Z Extremization in Three
  Dimensions}},  \href{https://arxiv.org/abs/1104.0487}{{\ttfamily 1104.0487}}.

\bibitem{Amariti:2014lla}
A.~Amariti and C.~Klare, \emph{{Chern-Simons and RG Flows: Contact with
  Dualities}}, \href{http://dx.doi.org/10.1007/JHEP08(2014)144}{\emph{JHEP}
  {\bfseries 08} (2014) 144},
  [\href{https://arxiv.org/abs/1405.2312}{{\ttfamily 1405.2312}}].

\bibitem{Bult:2007}
F.~van~de Bult, \emph{{Hyperbolic Hypergeometric Functions}}, {\emph{Thesis}
  (2007) }.

\bibitem{Closset:2016arn}
C.~Closset and H.~Kim, \emph{{Comments on twisted indices in 3d supersymmetric
  gauge theories}},
  \href{http://dx.doi.org/10.1007/JHEP08(2016)059}{\emph{JHEP} {\bfseries 08}
  (2016) 059}, [\href{https://arxiv.org/abs/1605.06531}{{\ttfamily
  1605.06531}}].

\bibitem{Kim:2009wb}
S.~Kim, \emph{{The Complete superconformal index for N=6 Chern-Simons theory}},
  \href{http://dx.doi.org/10.1016/j.nuclphysb.2012.07.015,
  10.1016/j.nuclphysb.2009.06.025}{\emph{Nucl. Phys.} {\bfseries B821} (2009)
  241--284}, [\href{https://arxiv.org/abs/0903.4172}{{\ttfamily 0903.4172}}].

\bibitem{Kapustin:2009kz}
A.~Kapustin, B.~Willett and I.~Yaakov, \emph{{Exact Results for Wilson Loops in
  Superconformal Chern-Simons Theories with Matter}},
  \href{http://dx.doi.org/10.1007/JHEP03(2010)089}{\emph{JHEP} {\bfseries 03}
  (2010) 089}, [\href{https://arxiv.org/abs/0909.4559}{{\ttfamily 0909.4559}}].

\bibitem{Hama:2010av}
N.~Hama, K.~Hosomichi and S.~Lee, \emph{{Notes on SUSY Gauge Theories on
  Three-Sphere}}, \href{http://dx.doi.org/10.1007/JHEP03(2011)127}{\emph{JHEP}
  {\bfseries 03} (2011) 127},
  [\href{https://arxiv.org/abs/1012.3512}{{\ttfamily 1012.3512}}].

\bibitem{Imamura:2011su}
Y.~Imamura and S.~Yokoyama, \emph{{Index for three dimensional superconformal
  field theories with general R-charge assignments}},
  \href{http://dx.doi.org/10.1007/JHEP04(2011)007}{\emph{JHEP} {\bfseries 04}
  (2011) 007}, [\href{https://arxiv.org/abs/1101.0557}{{\ttfamily 1101.0557}}].

\bibitem{Hama:2011ea}
N.~Hama, K.~Hosomichi and S.~Lee, \emph{{SUSY Gauge Theories on Squashed
  Three-Spheres}}, \href{http://dx.doi.org/10.1007/JHEP05(2011)014}{\emph{JHEP}
  {\bfseries 05} (2011) 014},
  [\href{https://arxiv.org/abs/1102.4716}{{\ttfamily 1102.4716}}].

\bibitem{Kapustin:2011jm}
A.~Kapustin and B.~Willett, \emph{{Generalized Superconformal Index for Three
  Dimensional Field Theories}},
  \href{https://arxiv.org/abs/1106.2484}{{\ttfamily 1106.2484}}.

\bibitem{Benini:2011nc}
F.~Benini, T.~Nishioka and M.~Yamazaki, \emph{{4d Index to 3d Index and 2d
  TQFT}}, \href{http://dx.doi.org/10.1103/PhysRevD.86.065015}{\emph{Phys. Rev.}
  {\bfseries D86} (2012) 065015},
  [\href{https://arxiv.org/abs/1109.0283}{{\ttfamily 1109.0283}}].

\bibitem{Imamura:2011wg}
Y.~Imamura and D.~Yokoyama, \emph{{N=2 supersymmetric theories on squashed
  three-sphere}},
  \href{http://dx.doi.org/10.1103/PhysRevD.85.025015}{\emph{Phys. Rev.}
  {\bfseries D85} (2012) 025015},
  [\href{https://arxiv.org/abs/1109.4734}{{\ttfamily 1109.4734}}].

\bibitem{Alday:2012au}
L.~F. Alday, M.~Fluder and J.~Sparks, \emph{{The Large N limit of M2-branes on
  Lens spaces}}, \href{http://dx.doi.org/10.1007/JHEP10(2012)057}{\emph{JHEP}
  {\bfseries 10} (2012) 057},
  [\href{https://arxiv.org/abs/1204.1280}{{\ttfamily 1204.1280}}].

\bibitem{Benini:2016hjo}
F.~Benini and A.~Zaffaroni, \emph{{Supersymmetric partition functions on
  Riemann surfaces}},  \href{https://arxiv.org/abs/1605.06120}{{\ttfamily
  1605.06120}}.

\bibitem{Bhattacharya:2008zy}
J.~Bhattacharya, S.~Bhattacharyya, S.~Minwalla and S.~Raju, \emph{{Indices for
  Superconformal Field Theories in 3,5 and 6 Dimensions}},
  \href{http://dx.doi.org/10.1088/1126-6708/2008/02/064}{\emph{JHEP} {\bfseries
  02} (2008) 064}, [\href{https://arxiv.org/abs/0801.1435}{{\ttfamily
  0801.1435}}].

\bibitem{Bhattacharya:2008bja}
J.~Bhattacharya and S.~Minwalla, \emph{{Superconformal Indices for N = 6 Chern
  Simons Theories}},
  \href{http://dx.doi.org/10.1088/1126-6708/2009/01/014}{\emph{JHEP} {\bfseries
  01} (2009) 014}, [\href{https://arxiv.org/abs/0806.3251}{{\ttfamily
  0806.3251}}].

\bibitem{Bashkirov:2011vy}
D.~Bashkirov, \emph{{Aharony duality and monopole operators in three
  dimensions}},  \href{https://arxiv.org/abs/1106.4110}{{\ttfamily 1106.4110}}.

\bibitem{Hwang:2011qt}
C.~Hwang, H.~Kim, K.-J. Park and J.~Park, \emph{{Index computation for 3d
  Chern-Simons matter theory: test of Seiberg-like duality}},
  \href{http://dx.doi.org/10.1007/JHEP09(2011)037}{\emph{JHEP} {\bfseries 09}
  (2011) 037}, [\href{https://arxiv.org/abs/1107.4942}{{\ttfamily 1107.4942}}].

\bibitem{Hwang:2011ht}
C.~Hwang, K.-J. Park and J.~Park, \emph{{Evidence for Aharony duality for
  orthogonal gauge groups}},
  \href{http://dx.doi.org/10.1007/JHEP11(2011)011}{\emph{JHEP} {\bfseries 11}
  (2011) 011}, [\href{https://arxiv.org/abs/1109.2828}{{\ttfamily 1109.2828}}].

\bibitem{Kapustin:2011vz}
A.~Kapustin, H.~Kim and J.~Park, \emph{{Dualities for 3d Theories with Tensor
  Matter}}, \href{http://dx.doi.org/10.1007/JHEP12(2011)087}{\emph{JHEP}
  {\bfseries 12} (2011) 087},
  [\href{https://arxiv.org/abs/1110.2547}{{\ttfamily 1110.2547}}].

\bibitem{Cheon:2012be}
S.~Cheon, D.~Gang, C.~Hwang, S.~Nagaoka and J.~Park, \emph{{Duality between N=5
  and N=6 Chern-Simons matter theory}},
  \href{http://dx.doi.org/10.1007/JHEP11(2012)009}{\emph{JHEP} {\bfseries 11}
  (2012) 009}, [\href{https://arxiv.org/abs/1208.6085}{{\ttfamily 1208.6085}}].

\bibitem{Kim:2013cma}
H.~Kim and J.~Park, \emph{{Aharony Dualities for 3d Theories with Adjoint
  Matter}}, \href{http://dx.doi.org/10.1007/JHEP06(2013)106}{\emph{JHEP}
  {\bfseries 06} (2013) 106},
  [\href{https://arxiv.org/abs/1302.3645}{{\ttfamily 1302.3645}}].

\bibitem{Aharony:2013dha}
O.~Aharony, S.~S. Razamat, N.~Seiberg and B.~Willett, \emph{{3d dualities from
  4d dualities}}, \href{http://dx.doi.org/10.1007/JHEP07(2013)149}{\emph{JHEP}
  {\bfseries 07} (2013) 149},
  [\href{https://arxiv.org/abs/1305.3924}{{\ttfamily 1305.3924}}].

\bibitem{Park:2013wta}
J.~Park and K.-J. Park, \emph{{Seiberg-like Dualities for 3d N=2 Theories with
  SU(N) gauge group}},
  \href{http://dx.doi.org/10.1007/JHEP10(2013)198}{\emph{JHEP} {\bfseries 10}
  (2013) 198}, [\href{https://arxiv.org/abs/1305.6280}{{\ttfamily 1305.6280}}].

\bibitem{Aharony:2013kma}
O.~Aharony, S.~S. Razamat, N.~Seiberg and B.~Willett, \emph{{3$d$ dualities
  from 4$d$ dualities for orthogonal groups}},
  \href{http://dx.doi.org/10.1007/JHEP08(2013)099}{\emph{JHEP} {\bfseries 08}
  (2013) 099}, [\href{https://arxiv.org/abs/1307.0511}{{\ttfamily 1307.0511}}].

\bibitem{Gahramanov:gka}
I.~B. Gahramanov and G.~S. Vartanov, \emph{{Superconformal indices and
  partition functions for supersymmetric field theories}},  in \emph{{XVIIth
  Intern. Cong. Math. Phys. 695-703 (2013)}}, 2013.
\newblock \href{https://arxiv.org/abs/1310.8507}{{\ttfamily 1310.8507}}.
\newblock \href{http://dx.doi.org/10.1142/9789814449243_0076}{DOI}.

\bibitem{Amariti:2014iza}
A.~Amariti and C.~Klare, \emph{{A journey to 3d: exact relations for adjoint
  SQCD from dimensional reduction}},
  \href{http://dx.doi.org/10.1007/JHEP05(2015)148}{\emph{JHEP} {\bfseries 05}
  (2015) 148}, [\href{https://arxiv.org/abs/1409.8623}{{\ttfamily 1409.8623}}].

\bibitem{Gahramanov:2014ona}
I.~Gahramanov and H.~Rosengren, \emph{{Integral pentagon relations for 3d
  superconformal indices}},  \href{https://arxiv.org/abs/1412.2926}{{\ttfamily
  1412.2926}}.

\bibitem{Amariti:2015vwa}
A.~Amariti, \emph{{Integral identities for 3d dualities with SP(2N) gauge
  groups}},  \href{https://arxiv.org/abs/1509.02199}{{\ttfamily 1509.02199}}.

\bibitem{Gahramanov:2013rda}
I.~Gahramanov and H.~Rosengren, \emph{{A new pentagon identity for the
  tetrahedron index}},
  \href{http://dx.doi.org/10.1007/JHEP11(2013)128}{\emph{JHEP} {\bfseries 11}
  (2013) 128}, [\href{https://arxiv.org/abs/1309.2195}{{\ttfamily 1309.2195}}].

\bibitem{Gahramanov:2016wxi}
I.~Gahramanov and H.~Rosengren, \emph{{Basic hypergeometry of supersymmetric
  dualities}},
  \href{http://dx.doi.org/10.1016/j.nuclphysb.2016.10.004}{\emph{Nucl. Phys.}
  {\bfseries B913} (2016) 747--768},
  [\href{https://arxiv.org/abs/1606.08185}{{\ttfamily 1606.08185}}].

\end{thebibliography}\endgroup
\end{document}